\documentclass[prd,nofootinbib,12pt]{revtex4-1}
\usepackage{dcolumn}% Align table columns on decimal point
\usepackage{graphicx}
\usepackage{xcolor}
\usepackage{amsmath}
\usepackage{amssymb}
\usepackage{amsthm}
\usepackage{latexsym}
\usepackage{bm}
\usepackage{multirow}
\usepackage{cancel}
\usepackage{hyperref}
\usepackage{tabularx}
\usepackage{enumitem}%for making itemize compact
\usepackage{listings}%used for displaying code, should be removable in final version
\usepackage{appendix}
\usepackage[compat=1.1.0]{tikz-feynman}

\newcommand*{\gev}{{\rm GeV}}
\newcommand*{\tev}{{\rm TeV}}
\newcommand*{\abs}[1] {\lvert {#1} \rvert}
\newcommand*{\beq}{\begin{equation}}
\newcommand*{\eeq}{\end{equation}}
\newcommand*{\bea}{\begin{eqnarray}}
\newcommand*{\eea}{\end{eqnarray}}

\newcommand*{\hpm}{\ensuremath{H^\pm}}

\newcommand*{\mhpm}{\ensuremath{m_{H^\pm}}}

\newcommand*{\tanb}{\ensuremath{\tan\beta}}

\newcolumntype{Z}{>{\centering\arraybackslash}X}

\graphicspath{{./figures/}}

\begin{document}

\begin{flushright}
MAN/HEP/2018/005\\
\end{flushright}
\vspace*{1.0cm}

\title{{\LARGE Charged Higgs Bosons in Naturally Aligned} \\[3mm]
{\LARGE Two Higgs Doublet Models at the LHC}\\[1mm]
${}$}

\author{Emily Hanson$^a$, William Klemm$^a$, Roger Naranjo$^b$, Yvonne Peters$^a$, Apostolos Pilaftsis$^a$\\
${}$\vspace{-3mm}}

\affiliation{
$^a$School of Physics \& Astronomy, University of Manchester,
Manchester M13 9PL, UK\\
$^b$Deutsches Elektronen-Synchrotron DESY, Hamburg, Germany
}

\begin{abstract}
${}$

\centerline{\bf ABSTRACT} \medskip

\noindent
Measurements of a Higgs boson at the Large Hadron Collider (LHC) have become increasingly consistent with the predictions of the Standard Model~(SM). This fact puts severe constraints on many potential low-energy extensions of the Higgs sector of the SM. In the well-known Two Higgs Doublet Model (2HDM), an `alignment limit' of parameters readily furnishes one SM-like scalar, and can be achieved naturally through an underlying symmetry.  Among the other physical states of the 2HDM, a~charged scalar $H^\pm$ would provide striking evidence of new physics if observed.  We~propose a novel technique for the observation of the process ${pp\to t b H^\pm \to t\bar{t}b\bar{b}}$ in the dileptonic decay channel at the LHC. The reconstruction of events in this channel is complicated by multiple $b$-jets and unobserved neutrinos in the final state. To~determine the neutrino momenta, we implement a neutrino weighting procedure to study, for the first time, the $t\bar{t}b\bar{b}$ signature.  We further train a pair of boosted decision trees to reconstruct and classify signal events. We determine the resulting reach within the context of {\em naturally} aligned 2HDMs, such as the Maximally Symmetric Two Higgs Doublet Model (MS-2HDM). By testing at the integrated luminosity\- of 150 fb$^{-1}$ achieved in Run 2 of the LHC, we find that this channel may restrict the parameter space of a Type-II MS-2HDM with charged Higgs masses as high as~680~\gev.

\end{abstract}

\preprint{}

\maketitle
\section{Introduction}

One of the great achievements of the Large Hadron Collider (LHC) has been the discovery of a resonance around 125~\gev~\cite{Aad:2012tfa,Chatrchyan:2012xdj}, whose measured signal rates in dominant decay channels increasingly agree with that of a Standard Model (SM) Higgs boson~\cite{Khachatryan:2016vau}. The observation of a Higgs boson further opens the door for the possibility of extended Higgs sectors, with parameters constrained by measured properties.  One of the simplest such extensions is the Two Higgs Doublet Model (2HDM)~\cite{Branco:2011iw}, which introduces one additional electroweak isodoublet.  Versions of the 2HDM appear in a variety of well-motivated scenarios for new physics, both with and without supersymmetry~\cite{Haber:1984rc,Pilaftsis:1999qt,Djouadi:2005gj}, in which the additional Higgs field is either an essential ingredient or necessary byproduct in addressing issues such as the origin of dark matter, the generation of a baryon asymmetry, the gauge hierarchy problem, and the strong CP problem.

Any version of the 2HDM attempting to describe the observed 125~\gev\ state, $h$, must be able to reproduce the SM-like signals seen at the LHC.  One simple way to achieve this, known as the `decoupling limit', is to set the masses of additional scalars so high that they play a minimal role around the electroweak scale~\cite{Gunion:2002zf}.  Another possibility, which can lead to new scalars at energies accessible to the LHC, is the `alignment limit', where the parameters of the theory force one CP-even scalar to have SM-like couplings~\cite{Carena:2013ooa,Dev:2014yca,Bernon:2015qea,Ferreira:2012my, Bernon:2015wef,Pilaftsis:2016erj,Grzadkowski:2018ohf}.  While this limit can be achieved by pure conspiracy of parameters, it is more natural to consider the possibility that it arises from an underlying symmetry~\cite{Dev:2014yca,Pilaftsis:2016erj,Benakli:2018vqz,Benakli:2018vjk,Darvishi:2019ltl}. The simplest scenario, dubbed the Maximally Symmetric Two Higgs Doublet Model (MS-2HDM), has been shown to be a viable option with new states accessible at LHC energies~\cite{Dev:2014yca,Darvishi:2019ltl}.

One possible striking signature of naturally aligned 2HDMs comes from the existence of a charged scalar state, \hpm, present in some extended Higgs sectors and general 2HDMs. Collider searches to date have yielded constraints on models containing a charged Higgs, but the 2HDM parameter space still contains unexplored regions which could be accessible with continued running of the LHC~\cite{Akeroyd:2016ymd,Arbey:2017gmh}.  The decays of a sufficiently heavy charged Higgs boson are typically dominated by $H^\pm\to t b$, giving the possible $pp\to t b H^\pm\to t\bar{t}b\bar{b}$ signature\footnote{In this work, we consider both $H^+$ and $H^-$ together; quark/anti-quark assignments may be inferred.  Thus, in our notation, $\sigma(pp\to~tbH^\pm) = \sigma(pp\to t\bar{b}H^-)+\sigma(pp\to \bar{t}bH^+)$.}.  As the top quarks can decay either hadronically or leptonically, there are a few possible resulting final states, each of which poses its own challenges for reconstruction and classification.  Here we focus on the dileptonic channel, where both top quarks decay to a $b$-jet, charged lepton, and neutrino.

This article is organised as follows. In Sect.~\ref{sec:2hdm} we review the 2HDM and the naturally aligned MS-2HDM.  In Sect.~\ref{sec:LHC} we introduce a novel analysis for identifying a charged Higgs boson at the LHC in the dileptonic decay channel and determine the resulting reach for an $H^\pm$ in the MS-2HDM. Finally, the results of our analysis are summarised in Sect.~\ref{sec:disc}.

\section{\label{sec:2hdm}The Two Higgs Doublet Model}

The two complex scalar Higgs fields, transforming as isodoublets $({\bf  2},1)$ under the SM electroweak gauge group ${\rm
  SU(2)}_L\otimes  {\rm  U(1)}_Y$, may be represented as
\begin{eqnarray}
\Phi_i  =   \left(\begin{array}{c}
  \phi_i^+ \\  \phi_i^0 \end{array}\right) ,
\end{eqnarray}
 with $i=1,2$; then the most general 2HDM potential may be written as
\begin{eqnarray}
  \label{eq:potential}
V & = & -\mu_1^2(\Phi_1^\dag \Phi_1) -\mu_2^2 (\Phi_2^\dag \Phi_2)
-\left[m_{12}^2 (\Phi_1^\dag \Phi_2)+{\rm H.c.}\right]\nonumber\\
& + &\lambda_1(\Phi_1^\dag \Phi_1)^2+\lambda_2(\Phi_2^\dag \Phi_2)^2
 +\lambda_3(\Phi_1^\dag \Phi_1)(\Phi_2^\dag \Phi_2)
 +\lambda_4(\Phi_1^\dag \Phi_2)(\Phi_2^\dag \Phi_1)\nonumber \\
&
+ &\left[\frac{1}{2}\lambda_5(\Phi_1^\dag \Phi_2)^2
+\lambda_6(\Phi_1^\dag \Phi_1)(\Phi_1^\dag \Phi_2)
+\lambda_7(\Phi_1^\dag \Phi_2)(\Phi_2^\dag \Phi_2)+{\rm H.c.}\right].
\end{eqnarray}
This contains four real mass parameters $\mu_{1,2}^2$, Re$(m^2_{12})$, Im$(m^2_{12})$, and ten real quartic couplings $\lambda_{1,2,3,4}$, Re($\lambda_{5,6,7})$, and Im($\lambda_{5,6,7}$).  Of these 14 parameters, three parameters can be removed by a $U(2)$ reparameterisation of the Higgs doublets $\Phi_1$ and $\Phi_2$~\cite{Ginzburg:2004vp}.  If we assume CP conservation, which allows the SM-like Higgs to be a CP-even scalar, then the parameters in~\eqref{eq:potential} are required to be real. After electroweak symmetry breaking (EWSB), each iso\-doublet acquires a vacuum expectation value (VEV) $v_j$ such that $\sqrt{v_1^2+v_2^2}=v\approx 246~\gev$ and $\phi_j^0 = (v_j+\phi_j+i a_j)/\sqrt{2}$, where $\phi_j$ and $a_j$ are real scalar fields. Three degrees of freedom become the longitudinal modes of the electroweak gauge bosons, leaving five physical states: two CP-even scalars $h$, $H$ with $m_h < m_H$; one CP-odd pseudoscalar $A$; and two charged scalars $H^\pm$.  It is then often useful to re-express the mass parameters $\mu^2_{1,2}$ and quartic couplings $\lambda_1$--$\lambda_5$ in terms of the physical masses $m_h$, $m_H$, $m_A$, $\mhpm$, along with the ratio of VEVs, $\tan\beta=v_2/v_1$ and the neutral sector mixing term $\sin(\beta-\alpha)$.  The angles $\alpha$ and $\beta$ govern the mixing between mass eigenstates in the CP-even sector and CP-odd/charged sectors, respectively.

Each Higgs field has Yukawa interactions with SM fermions, with the quark-sector Yukawa Lagrangian given by
\begin{equation}
-{\cal L}^q_Y  =  \bar{Q}_L(h_1^u \widetilde{\Phi}_1+ h_2^u\widetilde{\Phi}_2)u_R + \bar{Q}_L(h_1^d {\Phi}_1+ h_2^d {\Phi}_2)d_R,
\end{equation}
where $Q_L=(u_L,d_L)^T$ is the $SU(2)_L$ quark doublet, $u_R$ and $d_R$ are right-handed quark singlets, and $\widetilde{\Phi}_i=i\sigma_2\Phi_i^{*}$ are the isospin conjugates of $\Phi_i$. A similar expression holds for the leptons.
Because the potential contains couplings which mix the two isodoublets, a general 2HDM will produce tree-level flavour-changing neutral currents (FCNCs). One way to suppress these FCNCs is to impose the Glashow-Weinberg condition~\cite{Glashow:1976nt}, introducing a discrete $Z_2$ symmetry under which charges are assigned to ensure that each type of fermion couples to only a single Higgs doublet. If the fields transform as
\begin{equation}
\Phi_1 \to -\Phi_1, \  \Phi_2 \to \Phi_2, \ u_R \to u_R, \ d_R \to \pm d_R
\end{equation}
then at tree level the up-type quarks acquire mass solely from $\Phi_2$ and the down-type quarks acquire mass solely from $\Phi_2$ ($+$) or from $\Phi_1$ ($-$). Including the leptons, there are four possible unique assignments; here we will focus mainly on the Type-II 2HDM, in which $\Phi_2$ couples only to up-type quarks and $\Phi_1$ couples to down-type quarks and charged leptons, as in the minimal supersymmetric standard model.

\subsection{\label{sec:ms2hdm} Natural Alignment in the 2HDM}

The couplings of $h$ and $H$ to SM gauge bosons are related to the SM value by a factor of $\sin(\beta-\alpha)$ and $\cos(\beta-\alpha)$, respectively\footnote{A different convention for $\alpha$ is sometimes chosen such that these assignments are reversed.}, such that when $\sin(\beta-\alpha) = 1$ ($0$), $h$ ($H$) has SM-like gauge couplings.  In addition, the couplings of fermions to the neutral scalars are related to the SM value for $h$ by $\cos\alpha/\sin\beta$ or $-\sin\alpha/\cos\beta$, depending on the type of fermion and $Z_2$ symmetry, and by $\sin\alpha/\sin\beta$ or $\cos\alpha/\cos\beta$ for $H$. Therefore, when $\sin(\beta-\alpha) = 1$ ($0$), $h$ ($H$) also has SM-like couplings to fermions. When this alignment condition is met, one of the neutral scalars looks identical to a SM Higgs boson in its tree-level interactions with other SM particles.

The CP-even mass matrix in the 2HDM may be expressed as~\cite{Dev:2014yca}
\begin{equation}
M_S^2 = \left( \begin{matrix} c_\beta & -s_\beta \\ s_\beta & c_\beta \end{matrix}\right )
\left( \begin{matrix} \widehat{A} & \widehat{C} \\ \widehat{C} & \widehat{B}\end{matrix}\right )
\left( \begin{matrix} c_\beta & s_\beta \\ -s_\beta & c_\beta \end{matrix}\right ),
\end{equation}
where $c_\beta=\cos\beta$, $s_\beta=\sin\beta$, and
\begin{align}
\widehat{A}&=2v^2\left[c_\beta^4\lambda_1+s_\beta^2c_\beta^2\lambda_{345}+s_\beta^4\lambda_2+2s_\beta c_\beta(c_\beta^2\lambda_6+s_\beta^2\lambda_7) \right],\\
\widehat{B}&=m_A^2+\lambda_5v^2+2v^2\left[ s_\beta^2c_\beta^2(\lambda_1+\lambda_2-\lambda_{345})-s_\beta c_\beta(c_\beta^2-s_\beta^2)(\lambda_6-\lambda_7)\right],\\
\widehat{C}&=v^2\left[s_\beta^3 c_\beta(2\lambda_2-\lambda_{345})-c_\beta^3 s_\beta(2\lambda_1-\lambda_{345}) + c_\beta^2(1-4s_\beta^2)\lambda_6+s_\beta^2(4c_\beta^2-1)\lambda_7          \right ],
\end{align}
where $\lambda_{345}=\lambda_3+\lambda_4+\lambda_5$. The pseudoscalar mass $m_A$ and charged Higgs mass $\mhpm$ are
\begin{align}
m_A^2&= \mhpm^2+\frac{v^2}{2}(\lambda_4-\lambda_5),\\
\mhpm&=\frac{m_{12}^2}{s_\beta c_\beta}-\frac{v^2}{2}(\lambda_4+\lambda_5)+\frac{v^2}{2s_\beta c_\beta}(\lambda_6c_\beta^2+\lambda_7s_\beta^2).
\end{align}
Diagonalisation of $M_S^2$ gives the CP-even mass eigenstates, $H$ and $h$,
\begin{equation}
\left(\begin{matrix}H \\ h\end{matrix}\right)=\left(\begin{matrix} c_\alpha&s_\alpha\\-s_\alpha&c_\alpha\end{matrix} \right)\left(\begin{matrix}\phi_1 \\ \phi_2 \end{matrix} \right),
\end{equation}
such that
\begin{align}\label{eq:massalign}\nonumber
\left(\begin{matrix} m_H^2 & 0 \\ 0 & m_h^2 \end{matrix}\right)=&\left(\begin{matrix}c_\alpha & s_\alpha \\ -s_\alpha & c_\alpha \end{matrix}\right)M_S^2\left(\begin{matrix}c_\alpha & -s_\alpha \\ s_\alpha & c_\alpha \end{matrix}\right)\\
=&\left(\begin{matrix}c_{\beta-\alpha} & -s_{\beta-\alpha} \\ s_{\beta-\alpha} & c_{\beta-\alpha} \end{matrix}\right)
\left( \begin{matrix} \widehat{A} & \widehat{C} \\ \widehat{C} & \widehat{B}\end{matrix}\right )
\left(\begin{matrix}c_{\beta-\alpha} & s_{\beta-\alpha} \\ -s_{\beta-\alpha} & c_{\beta-\alpha} \end{matrix}\right).
\end{align}
For the alignment condition, $\sin(\beta-\alpha)=1$ (or $\cos(\beta-\alpha)=1$), \eqref{eq:massalign} may only be satisfied if $\widehat{C}=0$.
The alignment condition in a general CP-conserving 2HDM may then be expressed as
\begin{equation}
\lambda_7 t_\beta^4-(2\lambda_2-\lambda_{345})t_\beta^3+3(\lambda_6-\lambda_7)t_\beta^2+(2\lambda_1-\lambda_{345})t_\beta-\lambda_6=0,
\label{eq:alignmentcondition}
\end{equation}
where $t_\beta=\tan\beta$ (see also~\cite{Carena:2013ooa} for an equivalent expression). To satisfy~\eqref{eq:alignmentcondition} for all values of $\tan\beta$, the coefficients of each power of $\tan\beta$ must vanish, thus yielding the conditions for {\em natural} alignment~\cite{Dev:2014yca,Pilaftsis:2016erj}:
\begin{equation}
\lambda_1=\lambda_2=\lambda_{345}/2\,,\qquad \lambda_6=\lambda_7=0\;.
\label{eq:naturalalignmentcondition}
\end{equation}
The $U(1)_Y$-invariant 2HDM potential contains 13 accidental symmetries, which have been fully classified in~\cite{Battye:2011jj,Pilaftsis:2011ed} upon extending the bilinear field formalism in~\cite{Maniatis:2007vn,Ivanov:2007de,Nishi:2007dv}\footnote{We note that only six symmetries of 13 are preserved by $U(1)_Y$ gauge interactions beyond the tree-level approximation~\cite{Ivanov:2007de,Ferreira:2009wh,Ferreira:2010yh}}.  Of these, three restrict the quartic couplings such that the {\em natural} alignment conditions of~\eqref{eq:naturalalignmentcondition} are met~\cite{Dev:2014yca,Pilaftsis:2016erj}:
\begin{align}
\label{eq:SO5parameters}
SO(5)&:               &  \lambda_1&=\lambda_2=\lambda_3/2, & \lambda_4&=\lambda_5=\lambda_6=\lambda_7=0,  & &\\
O(3)\times O(2) & :   &  \lambda_1&=\lambda_2,             & \lambda_3&=2\lambda_1-\lambda_4,            & \lambda_5&=\lambda_6=\lambda_7=0,\\
\label{eq:O22}
Z_2\times[O(2)]^2 & : &  \lambda_1&=\lambda_2,             & \lambda_3&=2\lambda_1-(\lambda_4+\lambda_5),& \lambda_6&=\lambda_7=0.
\end{align}
While all the above symmetries are exactly realized when $\mu_1^2=\mu_2^2$ and $m_{12}^2=0$, their soft-breaking by an arbitrary choice of the parameters $\mu^2_{1,2}$ and $m^2_{12}$ will still be sufficient to give rise to alignment, at least at the tree level.
For this reason, we call~\eqref{eq:SO5parameters}--\eqref{eq:O22} symmetries of {\em natural alignment}.

In the following, we will focus on the Maximally Symmetric 2HDM (MS-2HDM)~\cite{Dev:2014yca} which possesses an $SO(5)$ invariant potential in the extended bilinear field formalism~\cite{Battye:2011jj}. The para\-meters in~\eqref{eq:SO5parameters} produce one massive CP-even scalar with $m_h=2\lambda_1 v^2$. The other four physical scalars ($H$, $A$, $H^\pm$) become massless and would participate in decays of SM particles, which is inconsistent with observation. The custodial $SO(5)$ symmetry, which is violated by $U(1)_Y$ hypercharge and the Yukawa couplings, could be realised at some high scale, $\mu_X$, with the electroweak scale behaviour determined by the renormalisation group (RG) evolution of the parameters, but this alone is unable to sufficiently raise the masses~\cite{Dev:2014yca}.  However, a viable Higgs spectrum is achievable by introducing a soft breaking term $\mathrm{Re}(m_{12}^2)$, which yields
\begin{equation}
\label{eq:softlybrokenparameters}
m_h^2 = 2\lambda_1 v^2\,,\qquad m_H^2=m_A^2=\mhpm^2=\frac{\mathrm{Re}(m_{12}^2)}{s_\beta c_\beta}\;.
\end{equation}
As stated above, the alignment conditions in~\eqref{eq:naturalalignmentcondition} do not depend on the soft-breaking parameters, e.g.~$m_{12}^2$, and as such alignment will still occur at the symmetry breaking scale.  RG~evolution${}$ to the electroweak scale will introduce some misalignment, but for a wide range of $\tan\beta$ and $\mu_X$, a viable low energy theory is possible~\cite{Dev:2014yca}.  In the remainder of this work, we do not choose a particular scale $\mu_X$ to evaluate the RG evolution of parameters, but define 2HDM parameters according to~\eqref{eq:SO5parameters} and~\eqref{eq:softlybrokenparameters}, ingoring RG effects.
For the purposes\- of this study, this consideration provides an appropriate working hypothesis for our numerical analysis that follows in Section~\ref{sec:LHC}.

\subsection{Charged Higgs Bosons in the 2HDM}
\begin{figure}[ht!]
  \begin{tikzpicture}
    \begin{feynman}
      %% Gluon and top quarks
      \vertex (g1) {\(g\)};
      \vertex[below right=0.5cm and 2cm of g1] (t4);
      \vertex[below right=1.5cm and 0.8cm of t4] (t3);
      \vertex[below left=1.5cm and 0.8cm of t3] (t2);
      \vertex[below left=0.5cm and 1.8cm of t2] (g2) {\(g\)};

      \vertex[above right=0.5cm and 1.5cm of t4] (t5);
      \vertex[below right=0.75cm and 2.5cm of t2] (t1) {\(b\)};

      %% Upper top shower
      \vertex[right=1cm of t5] (f3) {\(b\)};
      \vertex[above right=0.5cm and 0.5cm of t5] (W1);
      \vertex[above right=0.8cm and 0.1cm of W1] (f1) {\(\ell^{-}\)};
      \vertex[above right=0.3cm and 0.8cm of W1] (f2) {\(\bar{\nu}_{\ell}\)};

      %% Higgs
      \vertex[right=1.5cm of t3] (H);
      \vertex[above right=0.75cm and 1cm of H] (f4);
      \vertex[below right=0.75cm and 1cm of H] (f5) {\(\bar{b}\)};

      %% H+->top shower
      \vertex[right=1cm of f4] (f6) {\(\bar{b}\)};
      \vertex[above right=0.5cm and 0.5cm of f4] (W2);
      \vertex[above right=0.8cm and 0.1cm of W2] (f7) {\(\ell^{+}\)};
      \vertex[above right=0.3cm and 0.8cm of W2] (f8) {\(\nu_{\ell}\)};

      \diagram* {
        (g1) -- [gluon] (t4),
        (g2) -- [gluon] (t2),
        {[edges=fermion]
          (f3) -- (t5)
               -- [edge label=\(\bar{t}\)] (t4)
               -- [edge label'=\(t\)] (t3)
               -- [edge label'=\(\bar{b}\)] (t2)
               -- (t1),
          (f2) -- (W1) -- (f1),
          (f5) -- (H)
               -- [edge label=\(t\)] (f4)
               -- (f6),
          (f7) -- (W2) -- (f8),
        },
        (t3) -- [scalar, edge label=\(H^{+}\)] (H),
        (t5) -- [boson, edge label=\(W^{-}\)] (W1),
        (f4) -- [boson, edge label=\(W^{+}\)] (W2),
      };
    \end{feynman}
  \end{tikzpicture}
  \hspace{1cm}
  \begin{tikzpicture}
    \centering
    \begin{feynman}
      \vertex (i1) {\(g\)};
      \vertex[below=5cm of i1] (i2) {\(\bar{b}\)};
      \vertex[below right=1cm and 2cm of i1] (gbt);
      \vertex[above right=1cm and 2cm of i2] (btH);
      \vertex[right=2cm of gbt] (t1);
      \vertex[above right=0.5cm and 0.5cm of t1] (w1);
      \vertex[above right=0.25cm and 0.75cm of w1] (f1) {\(\ell^{-}\)};
      \vertex[below right=0.25cm and 0.75cm of w1] (f2) {\(\bar{\nu_{\ell}}\)};
      \vertex[below right=0.5cm and 1cm of t1] (f3) {\(\bar{b}\)};
      \vertex[right=2cm of btH] (hc);
      \vertex[above right=0.5cm and 1cm of hc] (t2);
      \vertex[above right=0.5cm and 0.5cm of t2] (w2);
      \vertex[above right=0.25cm and 0.75cm of w2] (f4) {\(\ell^{+}\)};
      \vertex[below right=0.25cm and 0.75cm of w2] (f5) {\(\nu_{\ell}\)};
      \vertex[below right=0.5cm and 1cm of t2] (f6) {\(b\)};
      \vertex[below right=0.5cm and 1cm of hc] (f7) {\(\bar{b}\)};

      \diagram* {
        (i1) -- [gluon] (gbt),
        {[edges=fermion]
          (f3) -- (t1) -- [edge label=\(\bar{t}\)] (gbt) -- (btH) -- (i2),
          (f2) -- (w1) -- (f1),
          (f7) -- (hc) -- [edge label=\(t\)] (t2) -- (f6),
          (f4) -- (w2) -- (f5),
        },
        (btH) -- [scalar, edge label=\(H^{+}\)] (hc),
        (t2) -- [boson, edge label=\(W^{+}\)] (w2),
        (t1) -- [boson, edge label=\(W^{-}\)] (w1),
      };
    \end{feynman}
  \end{tikzpicture}
\caption{Dileptonic decay channel of charged Higgs production in association with top quarks in the four-flavour and five-flavour schemes, respectively.}
\label{fig:tbhc_diagram}
\end{figure}
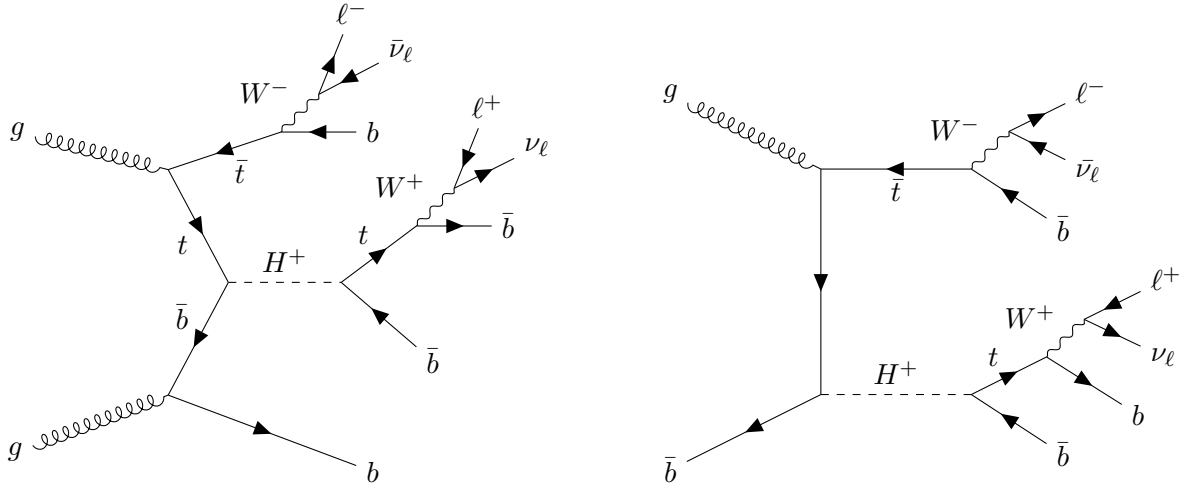

In the 2HDM, charged Higgs bosons have couplings to fermions given by

\begin{equation}
\mathcal{L}_{\hpm}=-H^+\left (\frac{\sqrt{2}V_{ud}}{v}\bar{u}(m_u X P_L+m_d Y)d+ \frac{\sqrt{2}m_\ell}{v}Z\bar{\nu}_L\ell_R \right )+\mathrm{H.c.},
\end{equation}
where terms containing $u$, $d$, and $\ell$ are summed over three generations and $V_{ud}$ is the CKM matrix.  In Type-II models, the real parameters become $X=\cot\beta$, and $Y=Z=-\tan\beta$. Because the couplings are proportional to fermion masses, the $H^\pm t b$ coupling typically dominates; in Type-II models, it is maximised at large and small $\tan\beta$.  Consequently, this coupling can play a major role in charged Higgs production. A light charged Higgs can be produced through the top quark decay $t\to H^\pm b$,
and a heavy charged Higgs can be produced as $gg\to t b H^\pm$, or in the five flavour scheme, $g b\to t H^\pm$, as seen in Fig.~\ref{fig:tbhc_diagram}.

Charged Higgs bosons $H^\pm$ may also decay through their couplings to fermions, with a preference for heavier fermions when kinematically allowed. Numerous searches have been performed\- at LEP~\cite{Abbiendi:2013hk}, Tevatron~\cite{Aaltonen:2009ke,Abazov:2009aa,Gutierrez:2010zz}, and the LHC~\cite{Aaboud:2018gjj,Khachatryan:2015qxa,Aad:2013hla,Khachatryan:2015uua,Aaboud:2018cwk} for the decays of $H^+$ to $\tau^+\nu_\tau$, $c\bar{s}$, and for sufficiently heavy $H^+$, to $t\bar{b}$. Charged Higgs bosons can, in principal, also decay to $W^\pm$-bosons and any of the neutral Higgs bosons, $h$, $H$ and $A$.  Although $H^\pm\to W^\pm h$ can be observed by taking advantage of the already-measured properties of the observed $h$~\cite{Enberg:2014pua}, the $H^\pm W^\mp h$ coupling is proportional to $\cos(\beta-\alpha)$, which vanishes in the alignment limit considered here.  Moreover, in the MS-2HDM, the near-degeneracy of $H$, $A$, and $H^\pm$ leads to a kinematic suppression of the decays $H^\pm\to W^\pm H$ or $H^\pm\to W^\pm A$.  Then $pp\to t b H^\pm\to t\bar{t}b\bar{b}$ is a natural search channel for a heavy $H^\pm$ in the MS-2HDM.  The ATLAS collaboration has recently published such a search using $36.1~\textrm{fb}^{-1}$ of data at $\sqrt{s}=13~\tev$, combining dileptonic and semi-leptonic final states to place limits on $\sigma(pp\to tbH^\pm)\times BR(H^\pm\to t b)$ ranging from 2.9 pb at $\mhpm = 200~\gev$ to 0.070 pb at $\mhpm = 2000~\gev$~\cite{Aaboud:2018cwk}.

\begin{figure}[!ht]
\includegraphics[width=0.38\textwidth]{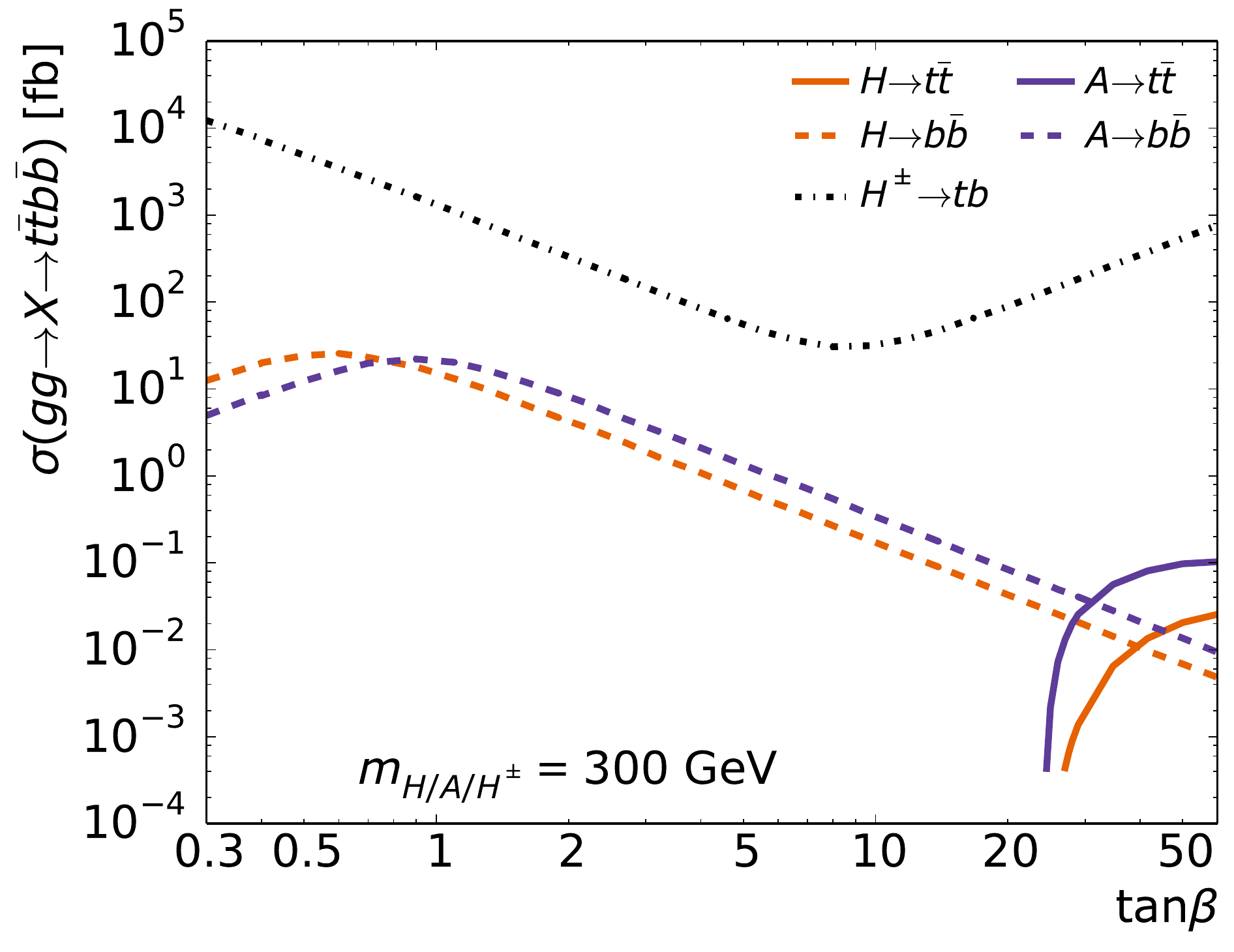}
\includegraphics[width=0.38\textwidth]{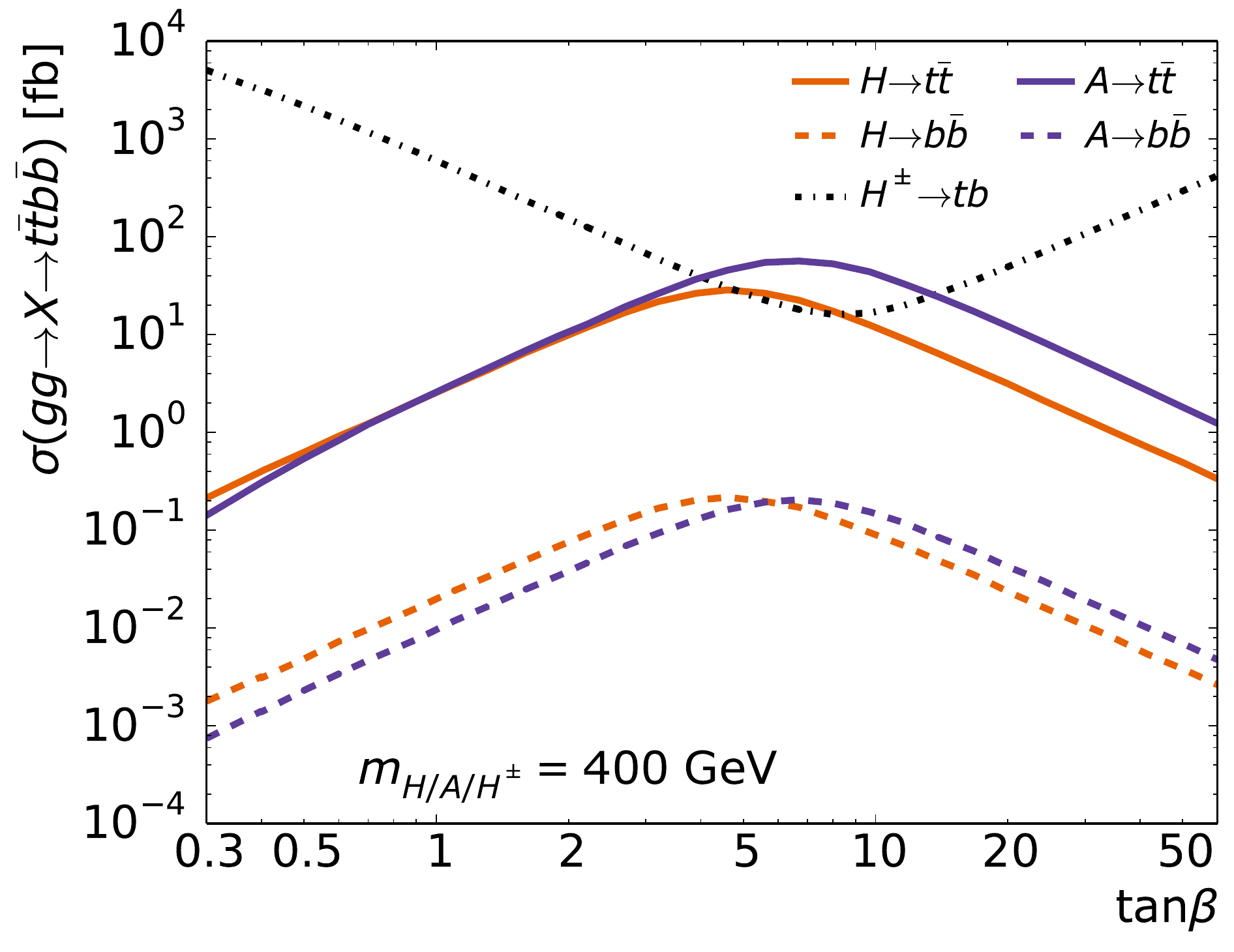}
\includegraphics[width=0.38\textwidth]{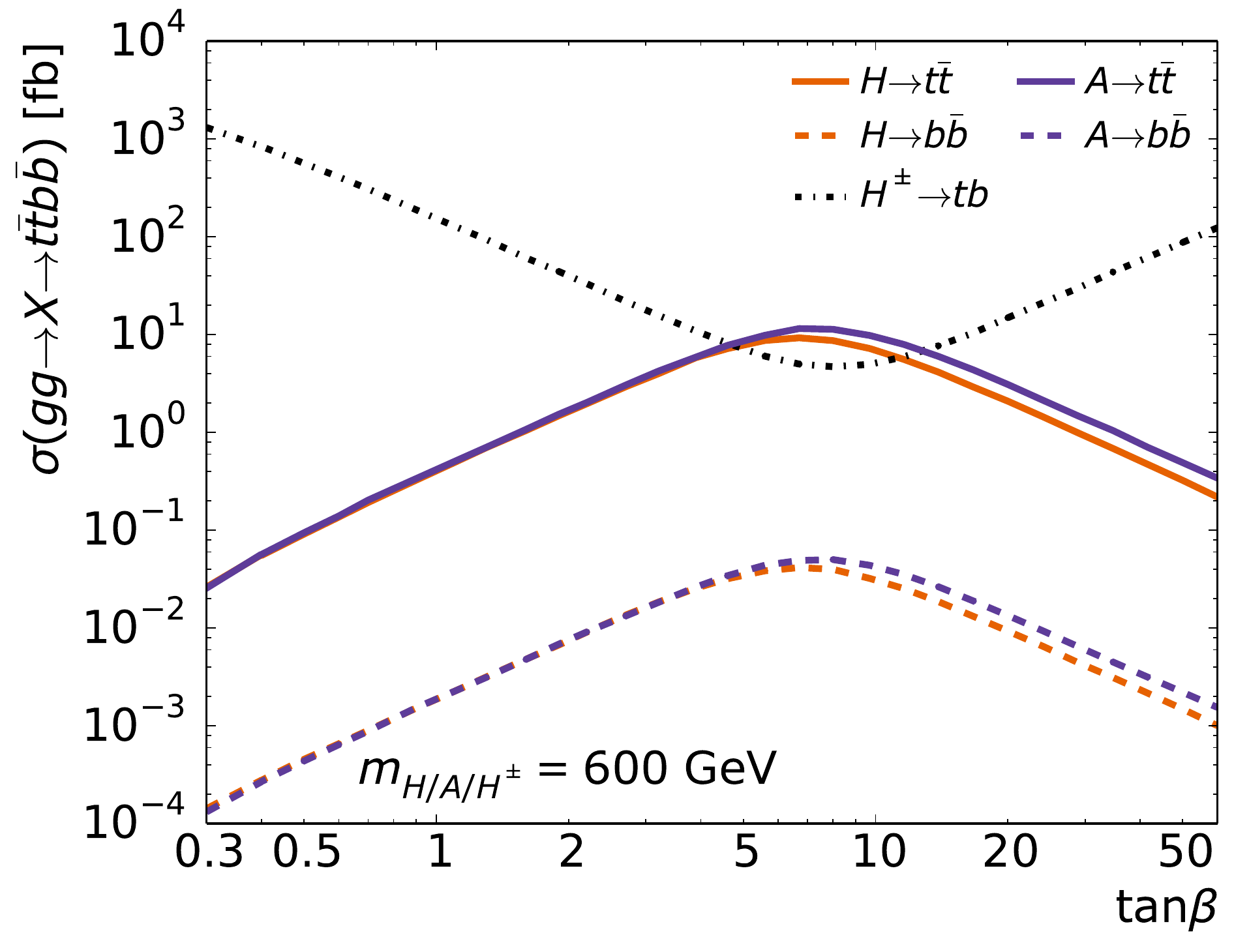}
\includegraphics[width=0.38\textwidth]{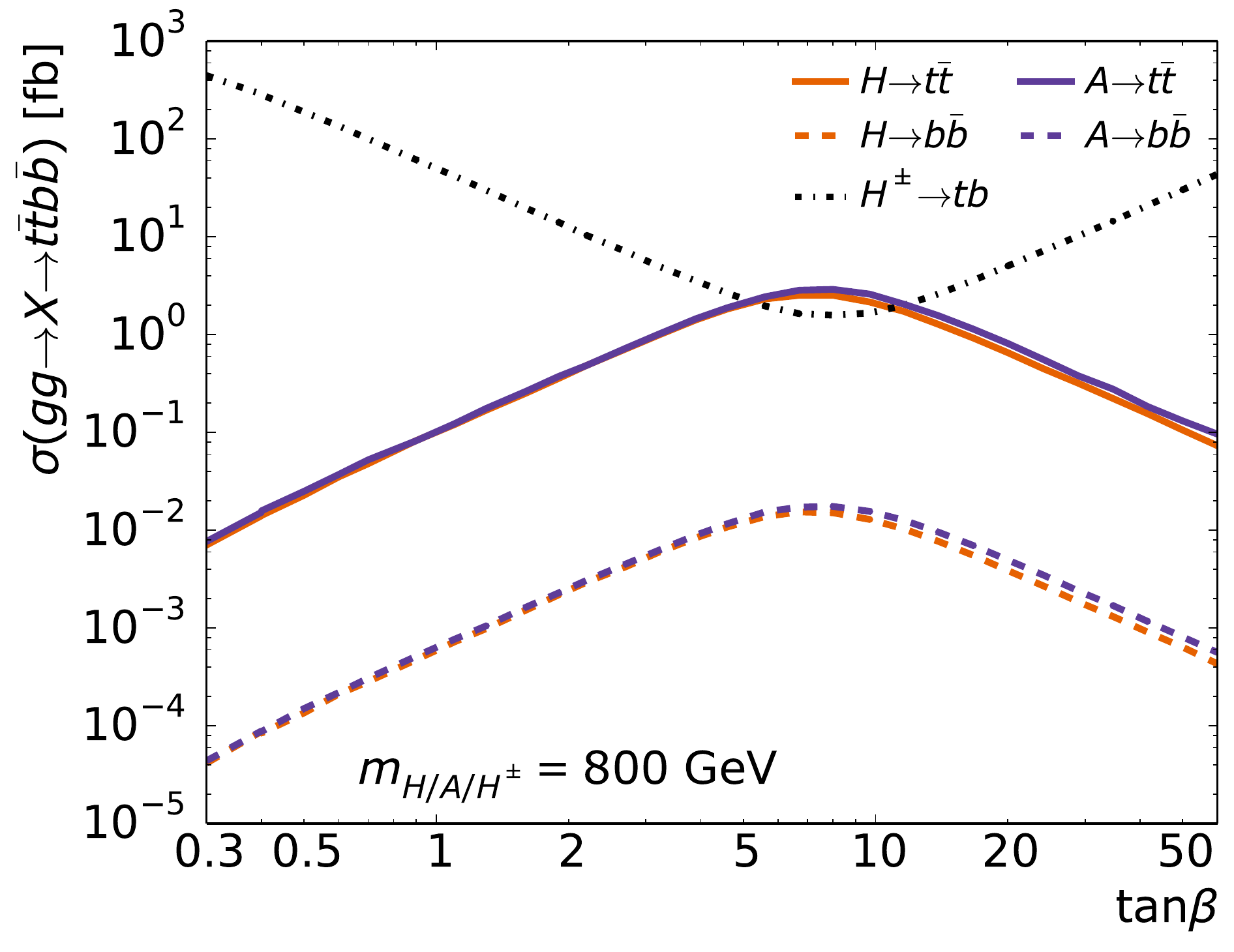}
\caption{
Cross sections for $t\bar{t}b\bar{b}$ production in the the MS-2HDM in $pp$ collisions at $\sqrt{s}=13~\tev$. The legend denotes how the heavy Higgs bosons decay. All cross sections here are calculated in the four flavour scheme with no kinematic cuts with MadGraph5\_aMC@NLO~\cite{Alwall:2014hca}.}
\label{fig:sigma_ttbb_MS2HDM}
\end{figure}

\begin{figure}[!ht]
\includegraphics[width=0.38\textwidth]{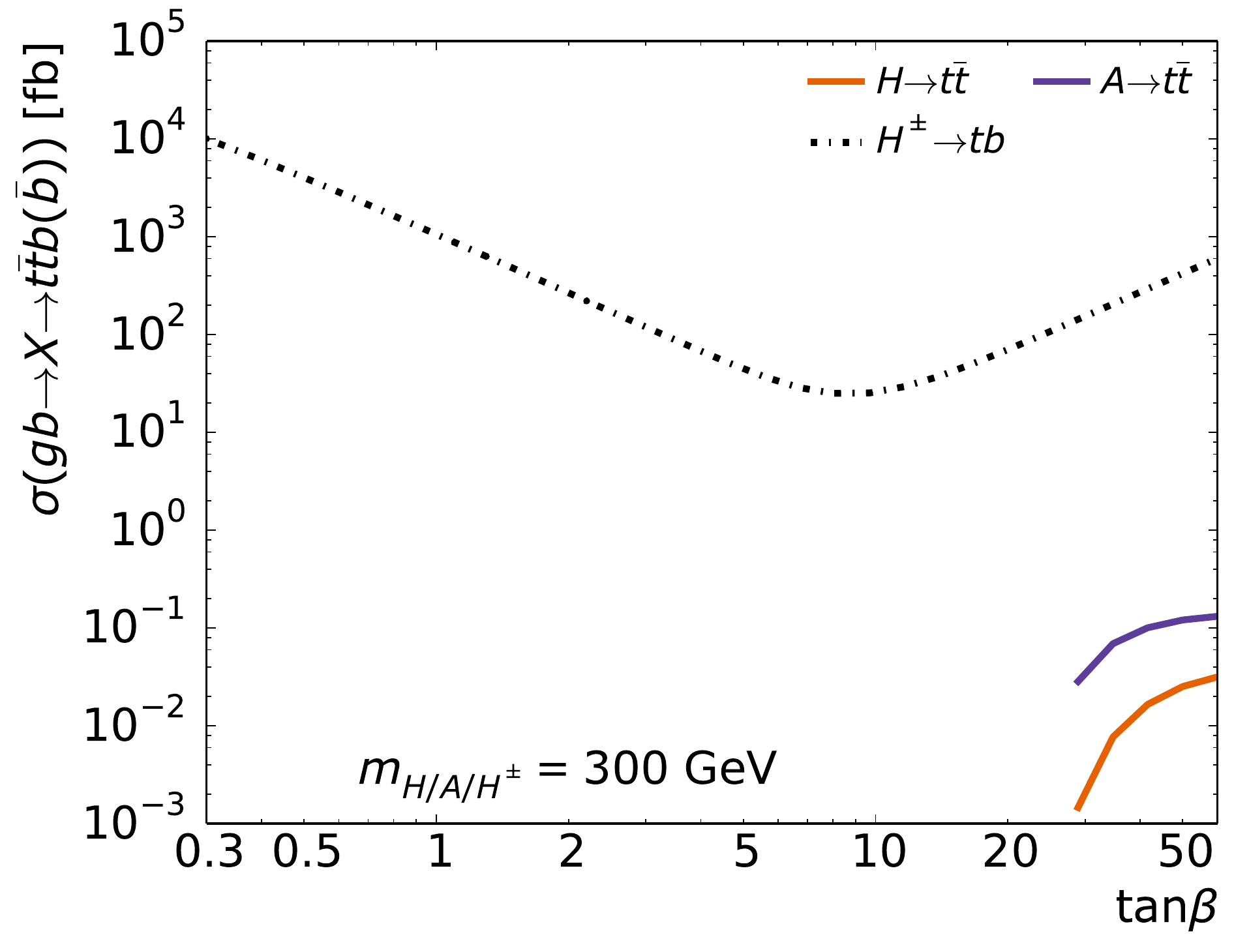}
\includegraphics[width=0.38\textwidth]{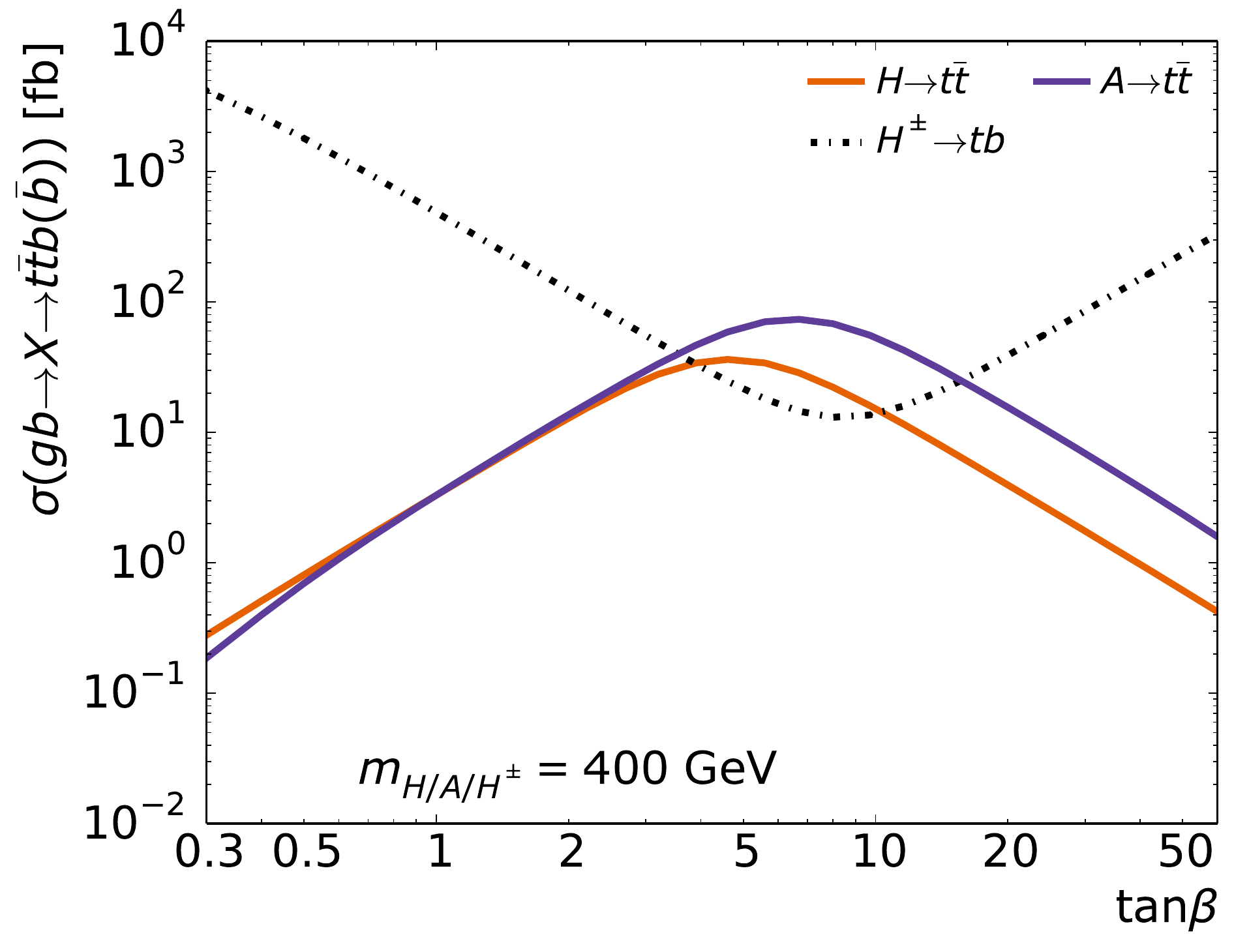}
\includegraphics[width=0.38\textwidth]{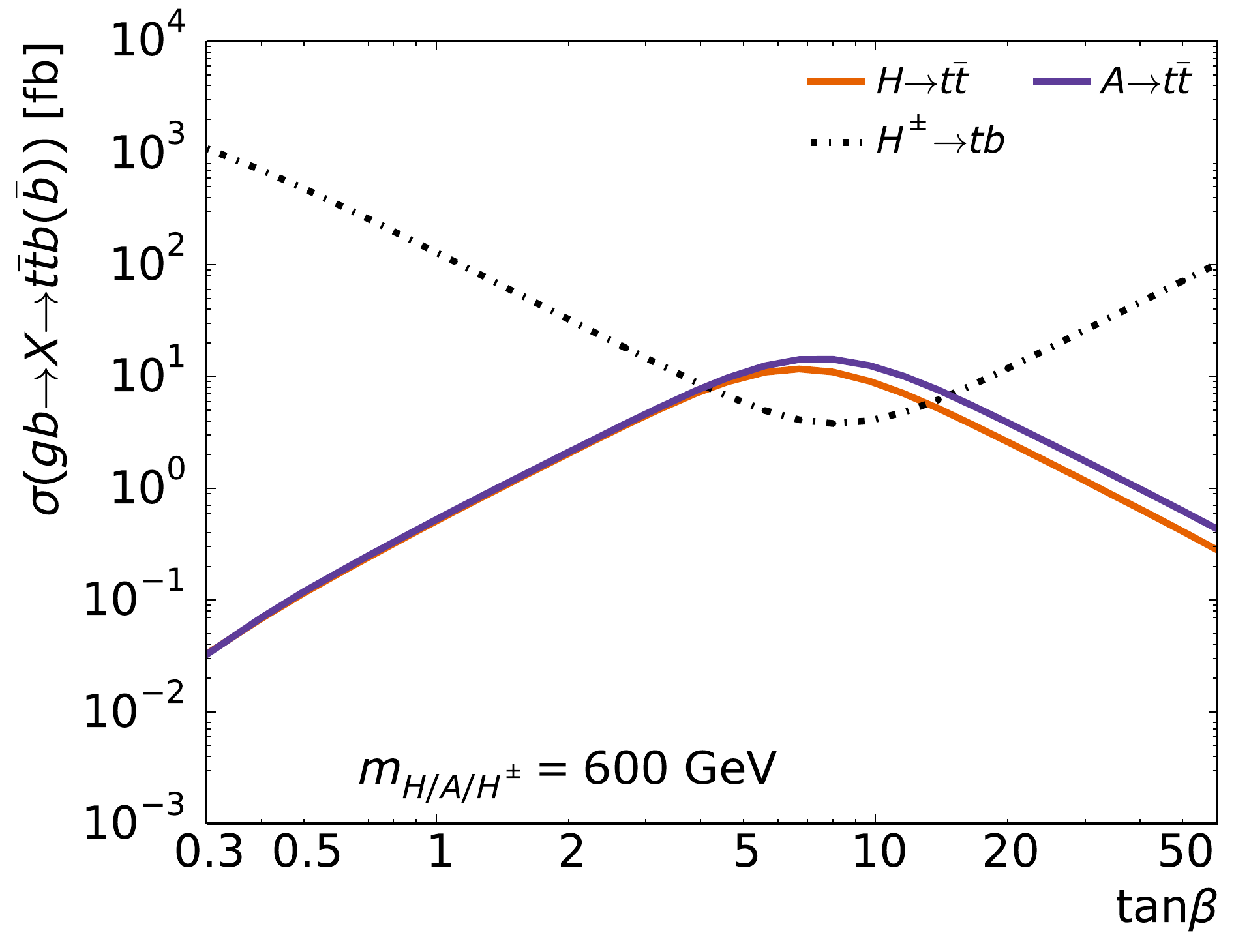}
\includegraphics[width=0.38\textwidth]{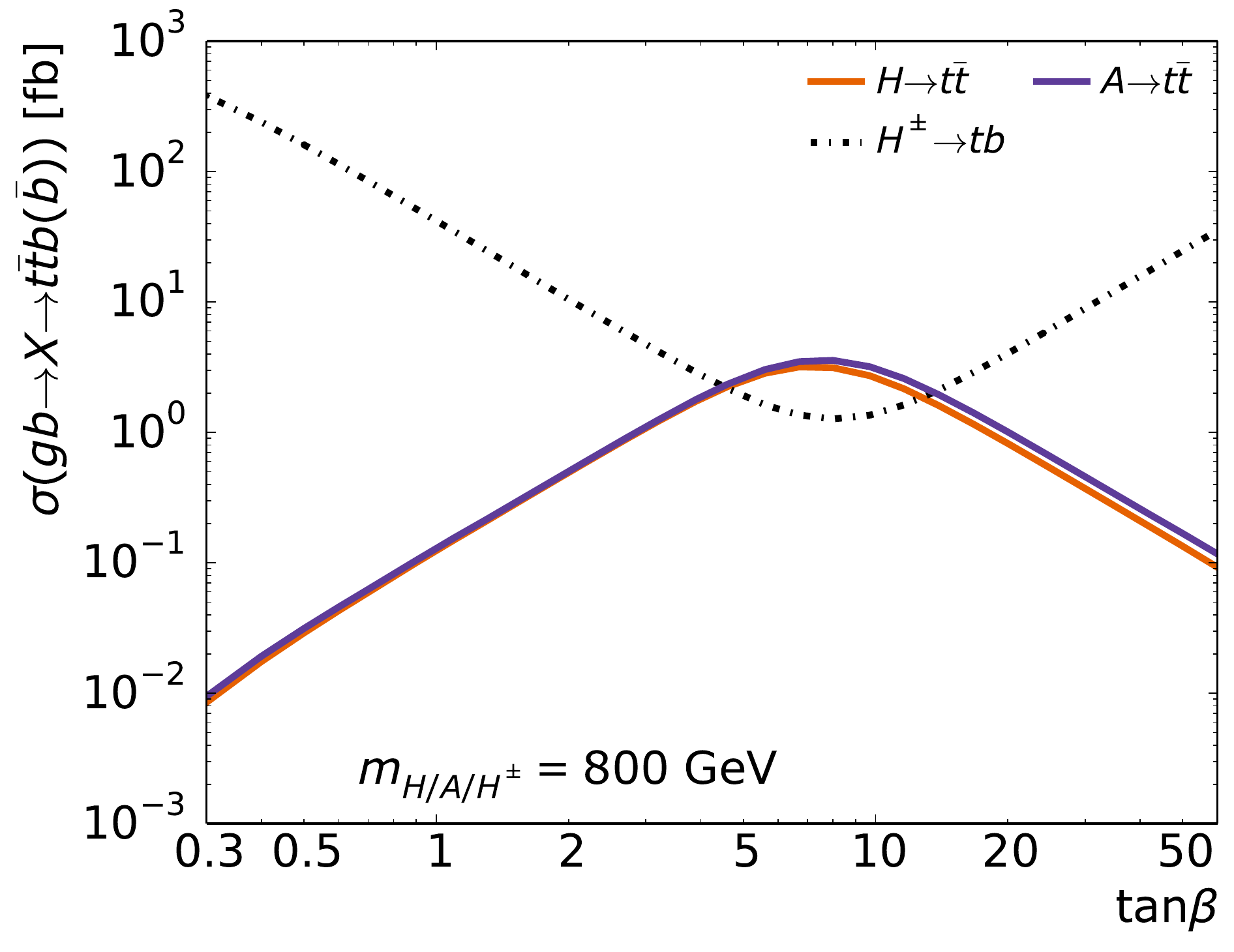}
\caption{The same as in Fig.~\ref{fig:sigma_ttbb_MS2HDM}, but in the five flavour scheme.}
\label{fig:sigma_ttbb_MS2HDM_5FS}
\end{figure}

The predicted signal cross sections for the production of heavy Higgs bosons in association with top quarks in the MS-2HDM are shown in Figs.~\ref{fig:sigma_ttbb_MS2HDM} and~\ref{fig:sigma_ttbb_MS2HDM_5FS} for the four and five flavour schemes, respectively. Like the charged Higgs bosons, the neutral Higgs bosons couple preferentially to third-generation fermions. This means that they can also mediate large $t\bar{t}b\bar{b}$ signals, also illustrated in Figs.~\ref{fig:sigma_ttbb_MS2HDM} and~\ref{fig:sigma_ttbb_MS2HDM_5FS}. However, for additional Higgs boson masses which are nearly degenerate, charged Higgs production dominates this channel for a large range of $\tan\beta$.

In addition to direct searches, charged Higgs bosons can enhance rare flavour-changing decays, in particular the B meson decay $B\to X_s \gamma$.
Precise measurement of these decay rates and calculation of their predicted values in the SM place tight constraints on Type-II models, with lower limits on $\mhpm$ in the $570-800~\gev$ range, with a high sensitivity to the exact method used~\cite{Misiak:2017bgg}. Performing direct searches for $H^\pm$ at the LHC provides a complimentary means of testing these limits while probing a larger mass range.

\section{\label{sec:LHC} Charged Higgs Signal at the LHC}

In this section, we consider the LHC production and decay of a charged Higgs boson via the process $pp\to t(b)H^\pm \to t\bar{t}b\bar{b}$, focusing on the dileptonic decay channel\footnote{See \cite{Guchait:2018nkp} for a recent proposal for the hadronic and semi-leptonic channels.} ($bb\bar{b}\bar{b}\ell^+\ell^-\nu_\ell\bar{\nu}_\ell$), as illustrated in Fig.~\ref{fig:tbhc_diagram}.
The backgrounds we consider for this channel are $t\bar{t}b\bar{b}$, $t\bar{t}c\bar{c}$, and $t\bar{t}+\textrm{light jets}(g,u,d,s)$. We generate signal and background events with MadGraph5\_aMC@NLO~\cite{Alwall:2014hca}, shower with Pythia6~\cite{Sjostrand:2006za}, and finally perform jet reconstruction and detector simulation using FastJet~\cite{Cacciari:2011ma} and DELPHES-3.4.1~\cite{deFavereau:2013fsa}, using the ATLAS configuration card. Jets are defined using the anti-$k_{t}$ algorithm with radius parameter $R=0.4$. The signal process for a Type-II 2HDM in the alignment limit is generated using the 2HDMC~\cite{Eriksson:2009ws} model for MadGraph.  For the signal, $tH^\pm$, $tH^\pm j$, and $tH^\pm b$ samples are generated in the 5-flavour scheme and matched using the MLM procedure as implemented by MadGraph and Pythia, with a matching scale of $m_{H^\pm}/4$.  The cross sections are then normalised to the Santander-matched cross sections given by the  LHC Higgs Cross Section Working Group~\cite{Degrande:2015vpa,Flechl:2014wfa,deFlorian:2016spz,PhysRevD.83.055005,PhysRevD.71.115012}. For the backgrounds, $t\bar{t}+0,1,2$ jet (five flavour) samples are generated\footnote{For practical reasons, samples are produced separately for different jet flavours accompanying the $t\bar{t}$ pair.} and matched at a scale of $80~\gev$. Other minor backgrounds, such as single-top production, are not considered; these are shown to affect the final result by, at most, 3--4.5\%, and more commonly by less than 1\%. Interferences between signal and background are also found to be negligible.
Though other scalars in the 2HDM could contribute to a $t\bar{t}b\bar{b}$ signal, we restrict ourselves here to processes containing a charged Higgs boson. We generate samples of signal events with $\tan\beta=1,2,5,10,15,30,60$ and $200~\gev \leq \mhpm \leq 1000~\gev$ in steps of $100~\gev$. Additional samples with $(\tan\beta,\mhpm)=(50,200),(40,300),(50,300),(40,400),(50,400),(50,500)$ for finer granularity in the regions with most sensitivity.

In the dileptonic channel, the final state contains four $b$ quarks, two charged leptons ($\ell^\pm=e^\pm,\mu^\pm$), and two neutrinos, on which we impose an initial selection:
\begin{itemize}
\item Exactly two leptons with transverse momenta $p_T > 20~\gev$, pseudorapidity $\abs{\eta}<2.5$, invariant dilepton mass $m_{\ell\ell}>12~\gev$ and $\abs{m_{\ell\ell}-m_Z}>10~\gev$, with separation $\Delta R_{\ell\ell}>0.4$.
\item Missing transverse energy $\cancel{E}_T>40~\gev$.
\item The event must contain at least three jets with $p_T>25~\gev$ and $\abs{\eta}<2.4$, with a leading jet $p_T>30~\gev$; at least two of these jets must be $b$-tagged. A $b$-tagged jet is one that is identified as likely to contain a $b$-hadron. The sample is split into 3-jet and $\geq$4-jet regions. After ordering by $p_T$, the first four (three for 3-jet events) $b$-tagged jets are taken if possible. If less than four jets are $b$-tagged, the highest $p_T$ non-$b$-tagged jet(s) are additionally taken to select four (three for 3-jet events) total jets. These are henceforth collectively called $b$-jets.

\end{itemize}

\begin{figure}[!ht]
\includegraphics[width=0.42\textwidth]{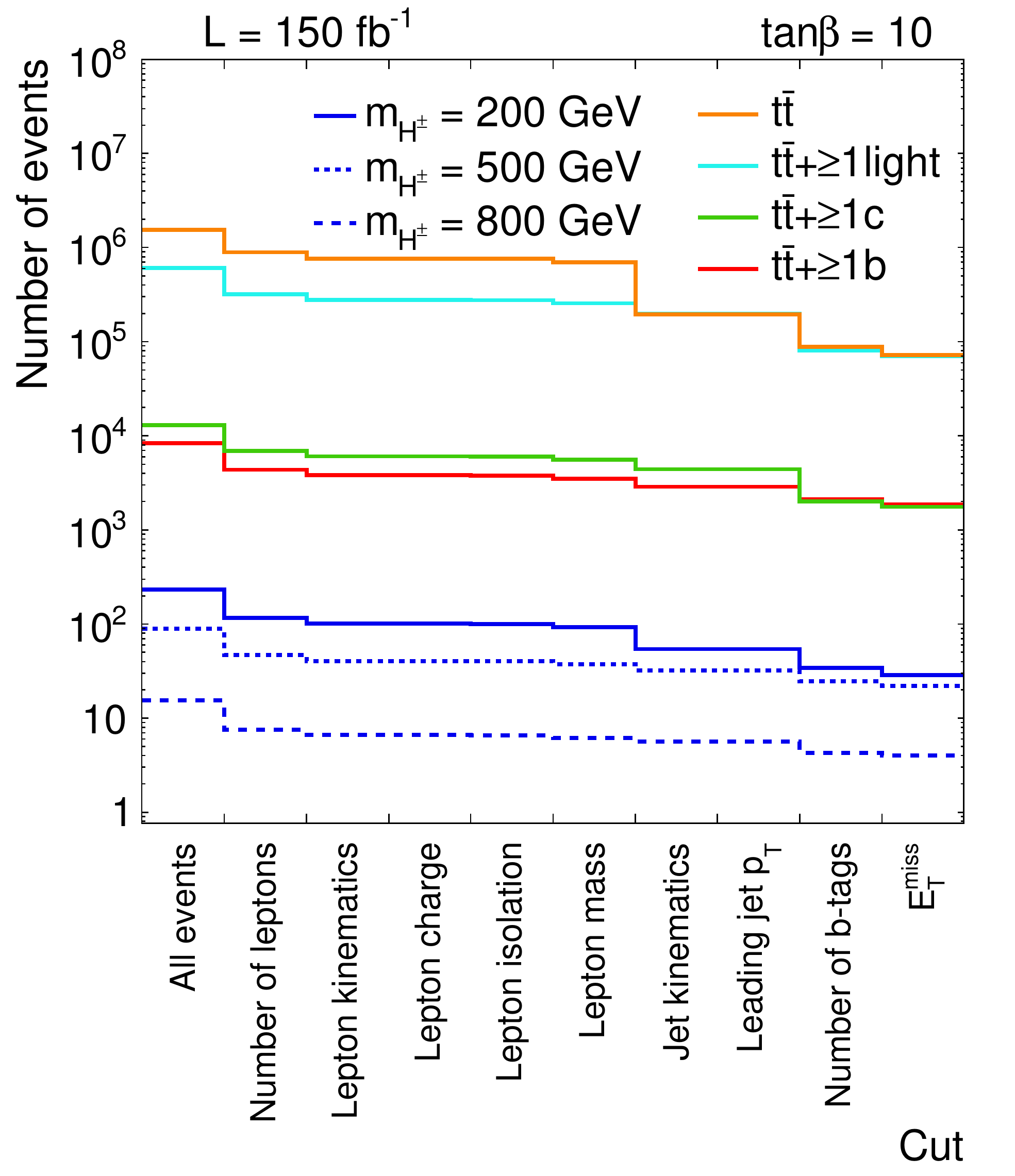}
\caption{The number of events, normalised to an integrated luminosity of 150 fb$^{-1}$, that survive after each stage of the selection process, for three signal masses at $\tanb=10$, overlaid with the SM backgrounds.}
\label{fig:cutflows}
\end{figure}

The effect of each of these requirements is shown in Fig.~\ref{fig:cutflows} for an illustrative value of $\tanb=10$ and at three masses, $\mhpm=200,500,800$~GeV. The proportion of signal events that remain after selection is approximately constant across $\tanb$ for a given mass. The effect of the selection on the SM backgrounds is also shown in Fig.~\ref{fig:cutflows}.

The reconstruction and classification of signal events in this channel present several challenges, which we address with a three step analysis utilising boosted decision trees (BDTs):
\begin{itemize}
\item A reconstruction BDT to identify the $b$-jets originating from the $b$-quarks $b_t$,$b_{tH}$,$b_H$ as defined in Fig~\ref{fig:tbhc_diagram}.
\item A neutrino weighting procedure to reconstruct the two neutrino momenta and to identify the correct $b$--$\ell^\pm$ pairings in top decays.
\item A classification BDT trained to distinguish signal and background events using the reconstruction from the first two steps. A template fit is performed on the output of this BDT to obtain limits.
\end{itemize}

\subsection{Boosted decision trees}

A BDT is a structure for classifying events by considering observables to produce a single value quantifying how signal- or background-like a given event is. This is done by training on Monte Carlo samples of signal and background. A decision tree consists of several successive layers of nodes, beginning with a single root node. At each node, the variable providing the greatest discriminating power is determined, using the training events, and an appropriate cut is applied. This causes a split into two new nodes, one expected to contain signal and one for background, and the best discriminating variables for the new nodes are determined. This processing of splitting and creating new nodes continues until a newly created node receives a subset of training events which contains less than a threshold number of events, meets a condition on purity (e.g.~mostly signal events), or reaches a maximum tree depth. This node is designated as an end node which assigns a classification corresponding to the dominant type of event in the subset of training events it received. After training, each event can thus be categorised as `signal-like' or `background-like'.

Boosting is a procedure which combines several weak classifiers into a stronger classifier. When applied to decision trees, boosting has been shown to improve both performance and stability~\cite{Hocker:2007ht}. Once an initial decision tree has been generated, events in the training sample are assigned weights. Training events which are misclassified by the initial decision tree are weighted more heavily than those which are correctly classified.  This new reweighted sample is then used to train a new decision tree, which may then be used to generate a new set of weights for the training sample to generate yet another decision tree.  This procedure repeats several times to create a set of decision trees (a `forest'); when analysing an event, each tree is queried for a classification (e.g. $-1$ for background, $+1$ for signal), and a weighted average of the responses gives a final score. Several different boosting algorithms exist with different weighting procedures for the training events and trees. In this work, we use the AdaBoost algorithm~\cite{Freund96experimentswith} with $\beta=0.5$ for the reconstruction BDT and the GradientBoost algorithm with Shrinkage=0.3 for the classification BDT.

Throughout this work we implement BDTs using the TMVA package~\cite{Hocker:2007ht} to generate forests of 400 (100) trees for the reconstruction (classification) BDT, each with a maximum depth of three layers. The cuts at each node are chosen to minimise the sum of the Gini indices of the resulting subsets of events, weighted by the fraction of events in each subset, where $Gini=p(1-p)$ for a sample with signal purity $p=N_{\textrm{signal}}/N_{\textrm{total}}$.

\subsection{Reconstruction BDT}

As shown in Fig.~\ref{fig:tbhc_diagram}, there are three $b$-jets whose origin must be determined: $b_{t}$ from the decay of the associated top, $b_H$ from the charged Higgs decay, and $b_{tH}$ from the top quark from the charged Higgs decay. An additional jet, $b_g$ is emitted from the initial gluon (or, in the five flavour scheme, produced in the parton shower), and is not considered in the reconstruction BDT. The $b$-jets are matched to parton-level (truth) $b$-quarks by determining which jet-quark pairs have the smallest separation in $\eta$-$\phi$, called $\Delta R$, providing $\Delta R\leq 0.4$. The performance of this $b$-jet to quark matching procedure is shown in the first row of Table~\ref{tab:performance_reco}. This shows that the matching is generally stable with mass, except at 200 GeV where the efficiency is lower.

From the $b$-jet to quark matching, we know the true origin of each $b$-jet. Then, we iterate through all combinations of $b$-jets, labelling them $b_{tH}$, $b_{t}$ and $b_{H}$. A permutation is, therefore, correct if all three $b$-jets have the same true origin as the label assigned to them; otherwise, it is incorrect. In order to separate the correct permutation from all of the possible incorrect permutations for a given event, we train a reconstruction BDT on each of the signal samples. This takes advantage of variations in kinematics due to model parameters. In this BDT, the correct permutation in an event serves as the `signal', and all incorrect permutations are `background'. For events where the matching procedure does not find pairings to all of $b_{tH}$, $b_{t}$, and $b_{H}$, all permutations in that event are `background'. The reconstruction BDT is trained on 57 observables:

\begin{itemize}[itemsep=1pt]
\item $\Delta R(b_{i},l^{a})$, $\Delta\eta(b_{i},l^{a})$, $\Delta\phi(b_{i},l^{a})$, $p_{T}^{b_{i}+l^{a}}$, $m(b_{i},l^{a})$,  where  $i = tH,t$ and $a = +,-$

\item $\left|m(l^{+}, b_{tH}) - m(l^{-},b_{t})\right|$ and $\left|m(l^{-}, b_{tH}) - m(l^{+},b_{t})\right|$

\item $p_{T}^{b_{j}}$ where $j=tH,H,t$

\item $\Delta R(b_{tH},b_{k})$, $\Delta\eta(b_{tH},b_{k})$, $\Delta\phi(b_{tH},b_{k})$, $p_{T}^{b_{tH}+b_{k}}$, $m(b_{tH},b_{k})$ where $k=H,t$

\item $\Delta R(t_{H^{a}},b_{H})$, $\Delta\eta(t_{H^{a}},b_{H})$, $\Delta\phi(t_{H^{a}},b_{H})$, $p_{T}^{t_{H^{a}},b_{H}}$, $m(t_{H^{a}},b_{H})$ where $a=+,-$

\item $\Delta R(t_{H^{a}},t_{c})$, $\Delta\eta(t_{H^{a}},t_{c})$, $\Delta\phi(t_{H^{a}},t_{c})$, where $(H^{a},t_{c})=(H^{+},\bar{t})$ or $(H^{-},t)$

\item $m(H^{a})-m(b_{H})$ where $a=+,-$

\item $m(H^{+})-m(\bar{t})$ and $m(H^{-})-m(t)$

\item $p_{T}^{H^{\pm}+t_{\text{other}}}$

\item $m(H^{\pm},t_{\text{other}})$
\end{itemize}
Here we define $t_{H^\pm}=b_{tH} + \ell_{H}$, $H^\pm = t_{H^\pm}+b_H$, and $t_{\text{other}} = b_t + \ell_{\text{other}}$, where $\ell_{H}$ is the charged lepton from the $H^{\pm}$ decay, and $\ell_{\text{other}}$ is the lepton not used in defining $H^\pm$.

\begin{figure}[!ht]
\includegraphics[width=0.38\textwidth]{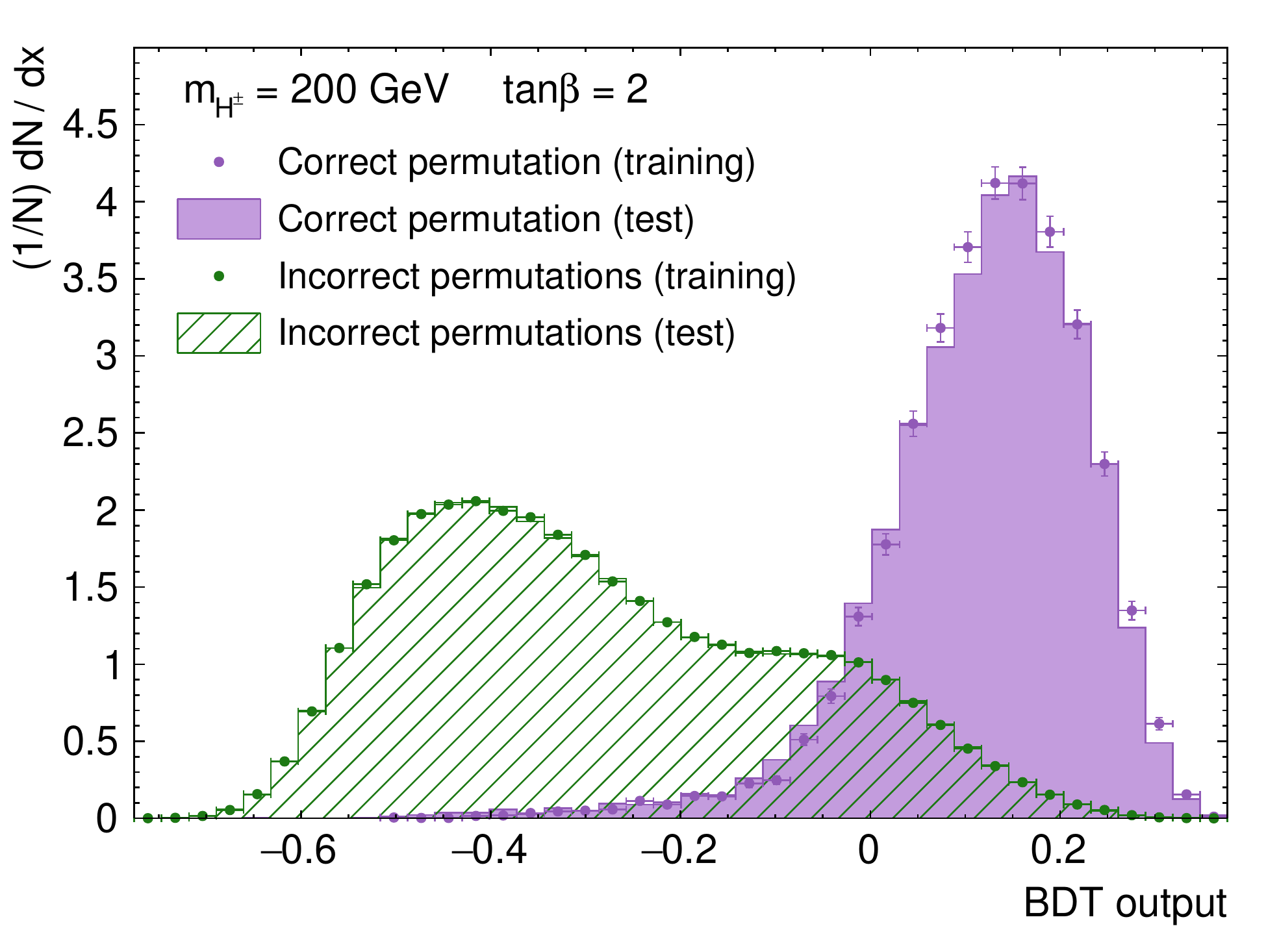}
\includegraphics[width=0.38\textwidth]{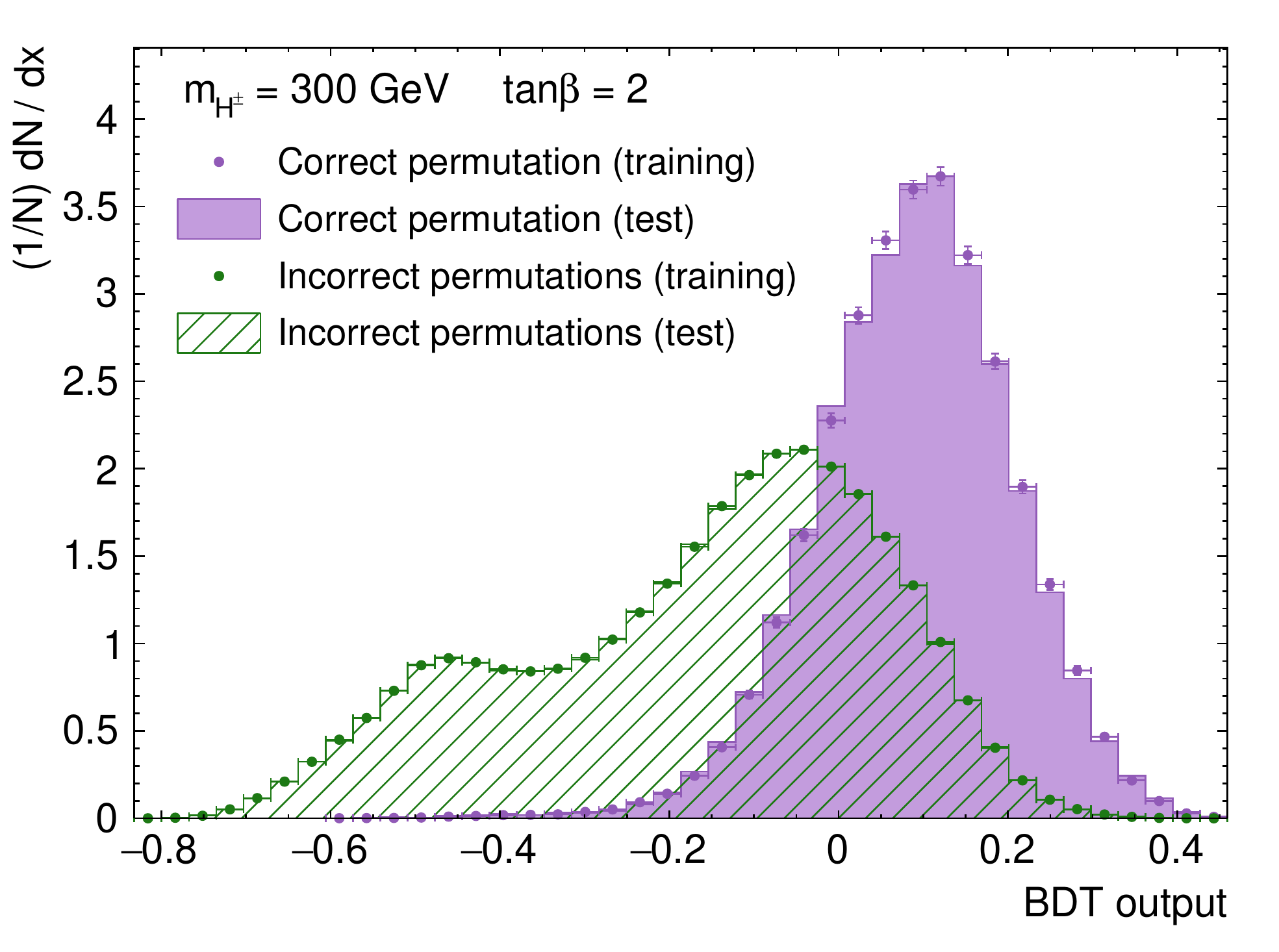}
\includegraphics[width=0.38\textwidth]{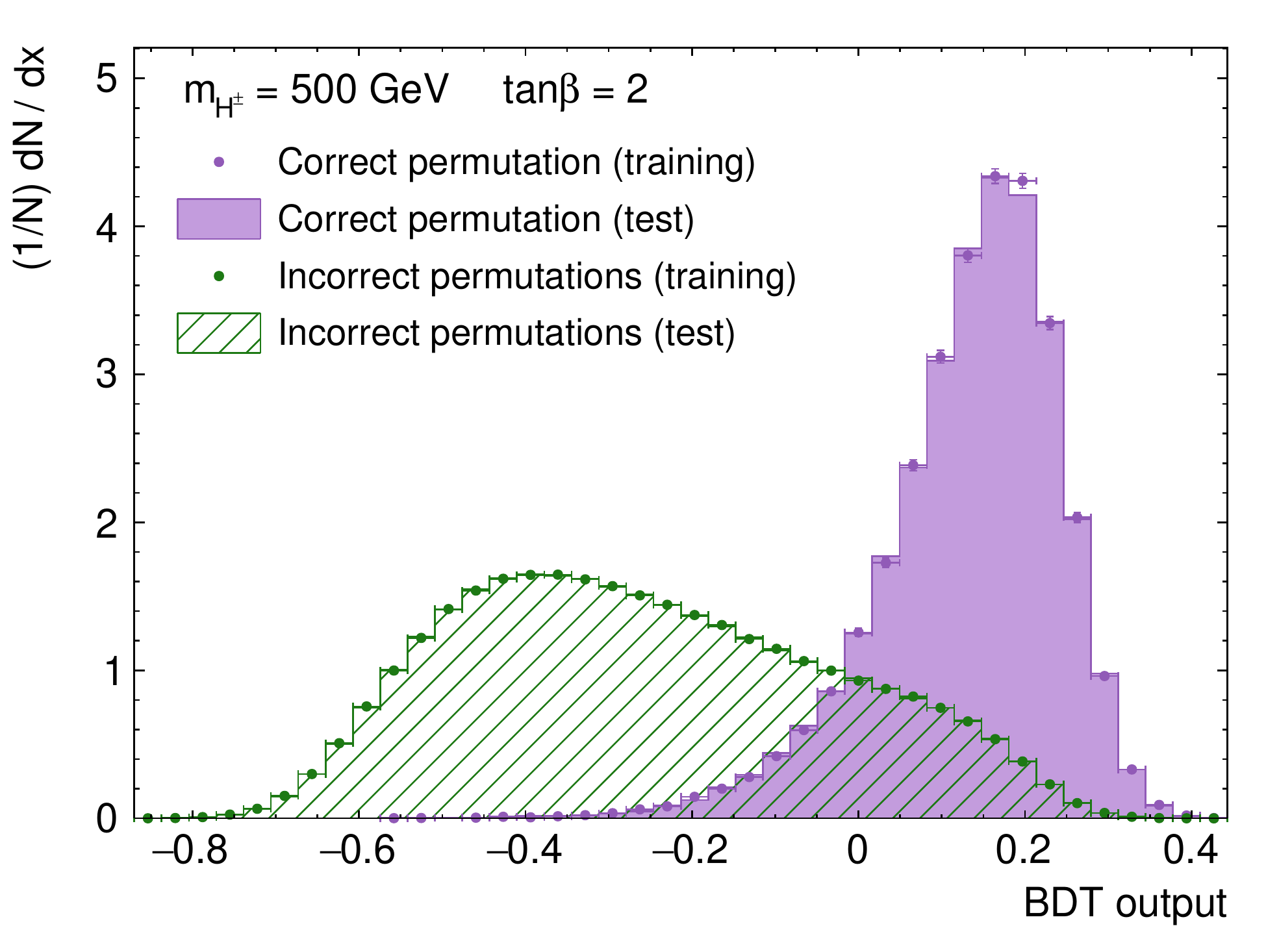}
\includegraphics[width=0.38\textwidth]{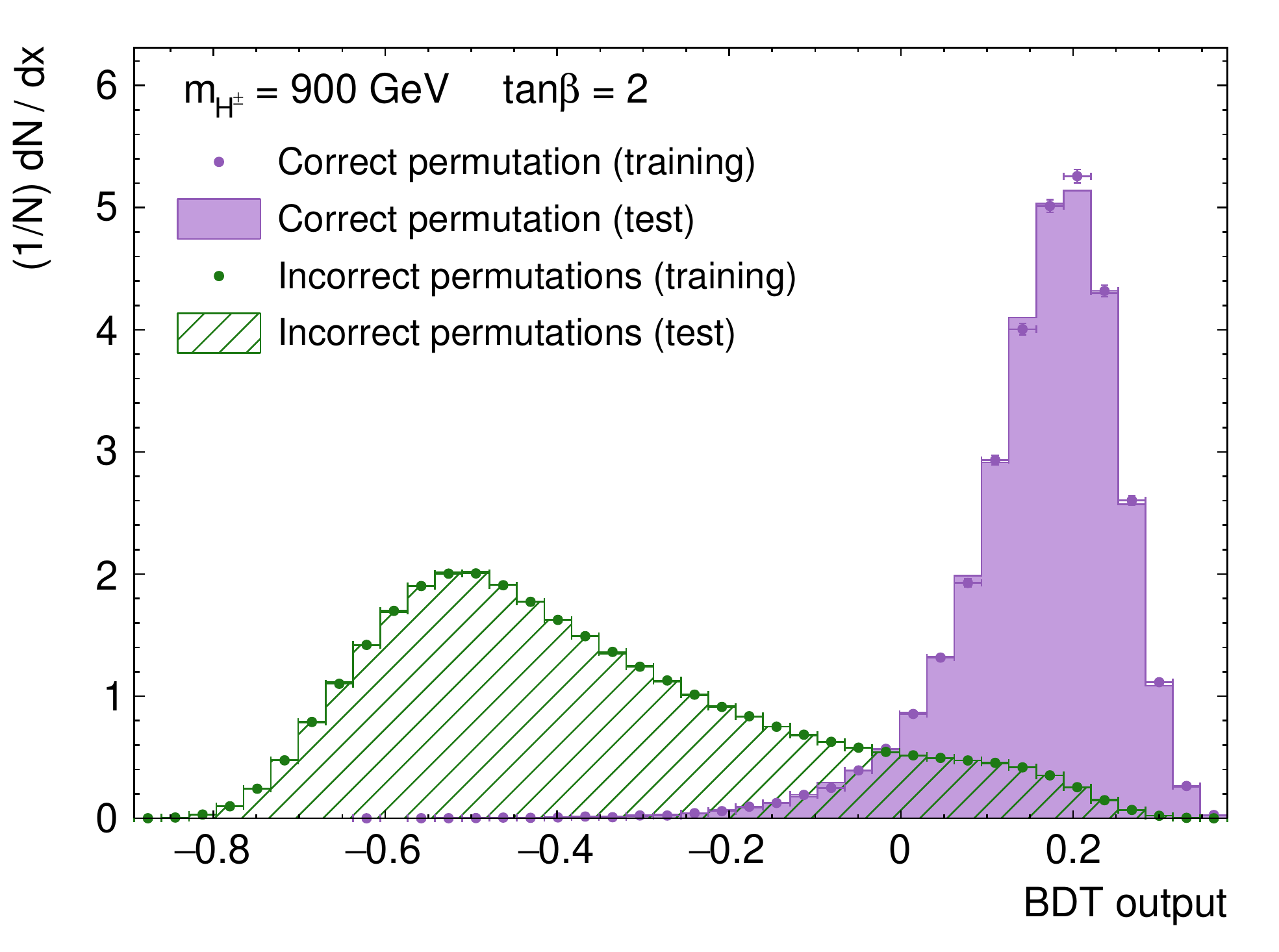}
\caption{Reconstruction BDT response for $\mhpm = 200,300,500,900~\gev$  and $\tan\beta=2$. }
\label{fig:reco_BDT}
\end{figure}

When using the reconstruction BDT to analyse an event, we obtain the BDT output for each possible arrangement of jets and select the one with the highest value for further analysis.
The BDT output distributions for correctly and incorrectly matched events is shown in Fig.~\ref{fig:reco_BDT}. There is a clear separation between `signal' and `background' that improves for large $\mhpm$, after falling off from $\mhpm=200~\gev$ to $300~\gev$.
While these distributions are shown only for $\tan\beta=2$, the performance of the reconstruction BDT for the full range of $\tan\beta=1-60$ is shown in the second and third rows of Table~\ref{tab:performance_reco}. The separation, $\langle S^2 \rangle$, is defined as
\begin{equation}
\langle S^2\rangle = \frac{1}{2}\int\frac{(\hat{y}_S(y)-\hat{y}_B(y))^2}{\hat{y}_S(y)+\hat{y}_B(y)}dy,
\end{equation}
where $y$ is the BDT response and $\hat{y}_S$ and $\hat{y}_B$ are the signal and background probability distribution functions, respectively.
The performance improves with $\mhpm$ following a steep decline from $\mhpm=200~\gev$ to $300~\gev$. Also, the correct assignments are identified in a large fraction of events, bearing in mind both the large number of incorrect combinations and events in which at least one of the relevant $b$-jets is not reconstructed or chosen in the initial selection.  We further note that the small variation in performance indicates that this step in the analysis is only mildly dependent on $\tan\beta$.

\begin{figure}
\includegraphics[width=0.32\textwidth]{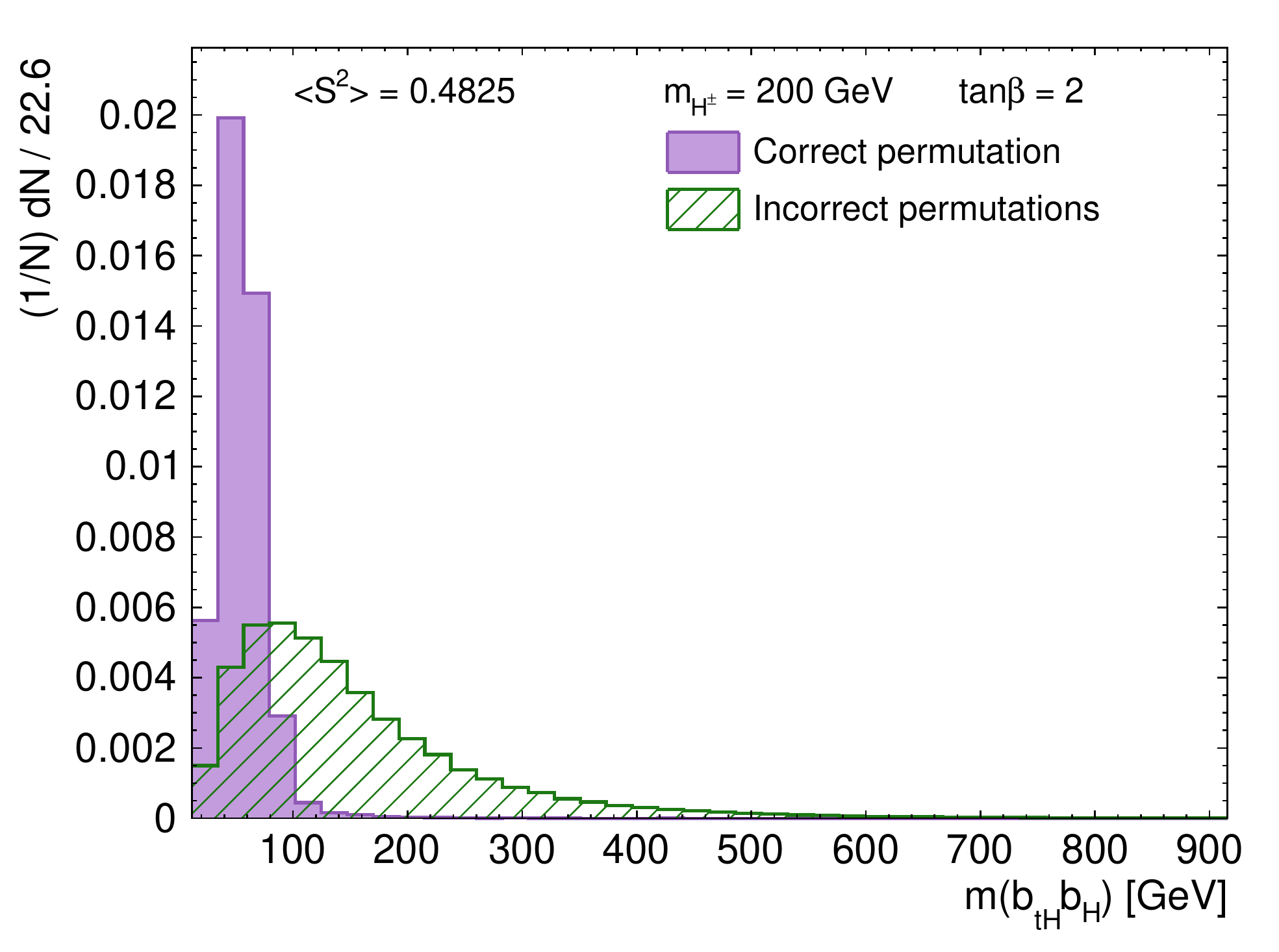}
\includegraphics[width=0.32\textwidth]{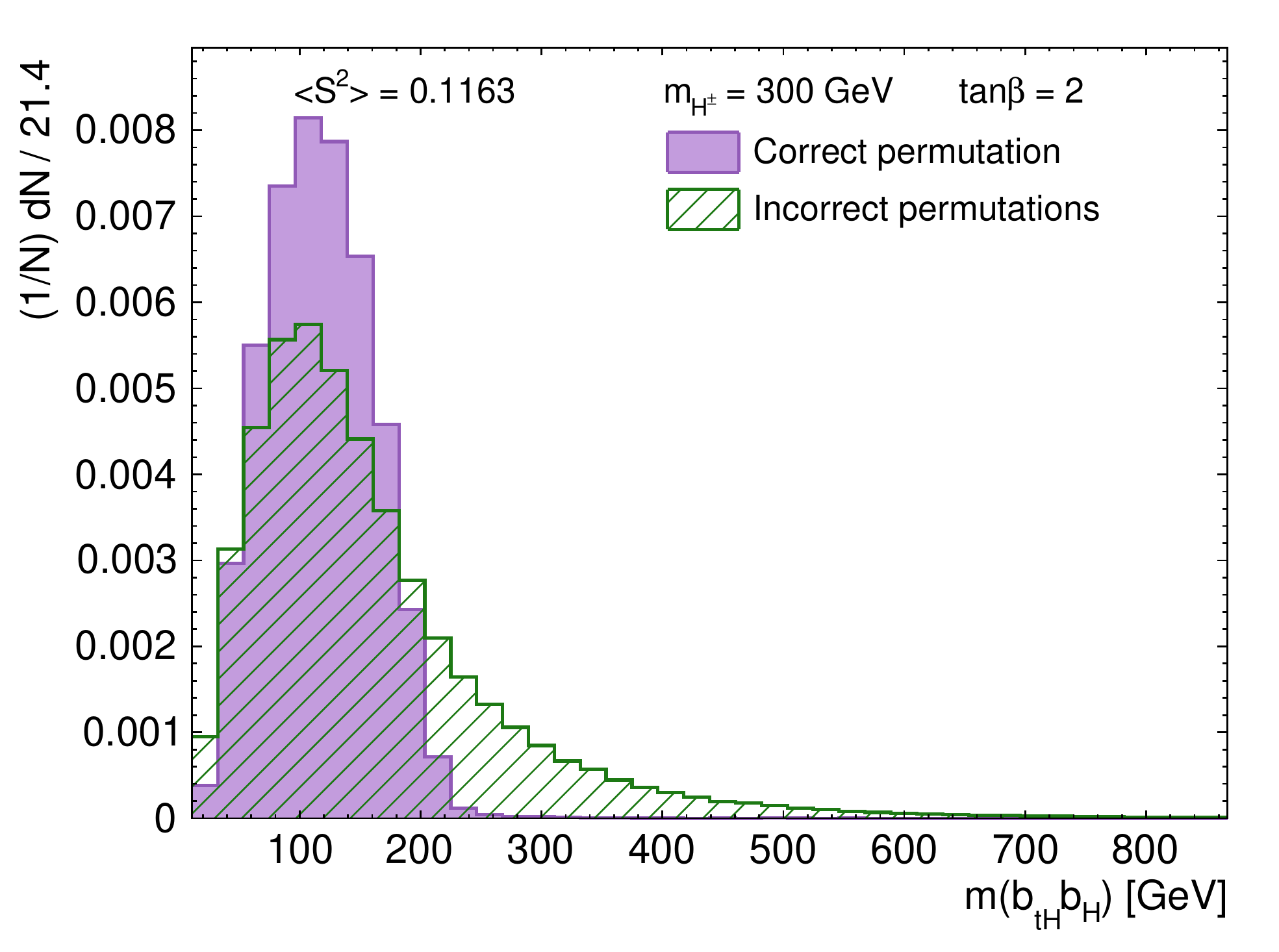}
\includegraphics[width=0.32\textwidth]{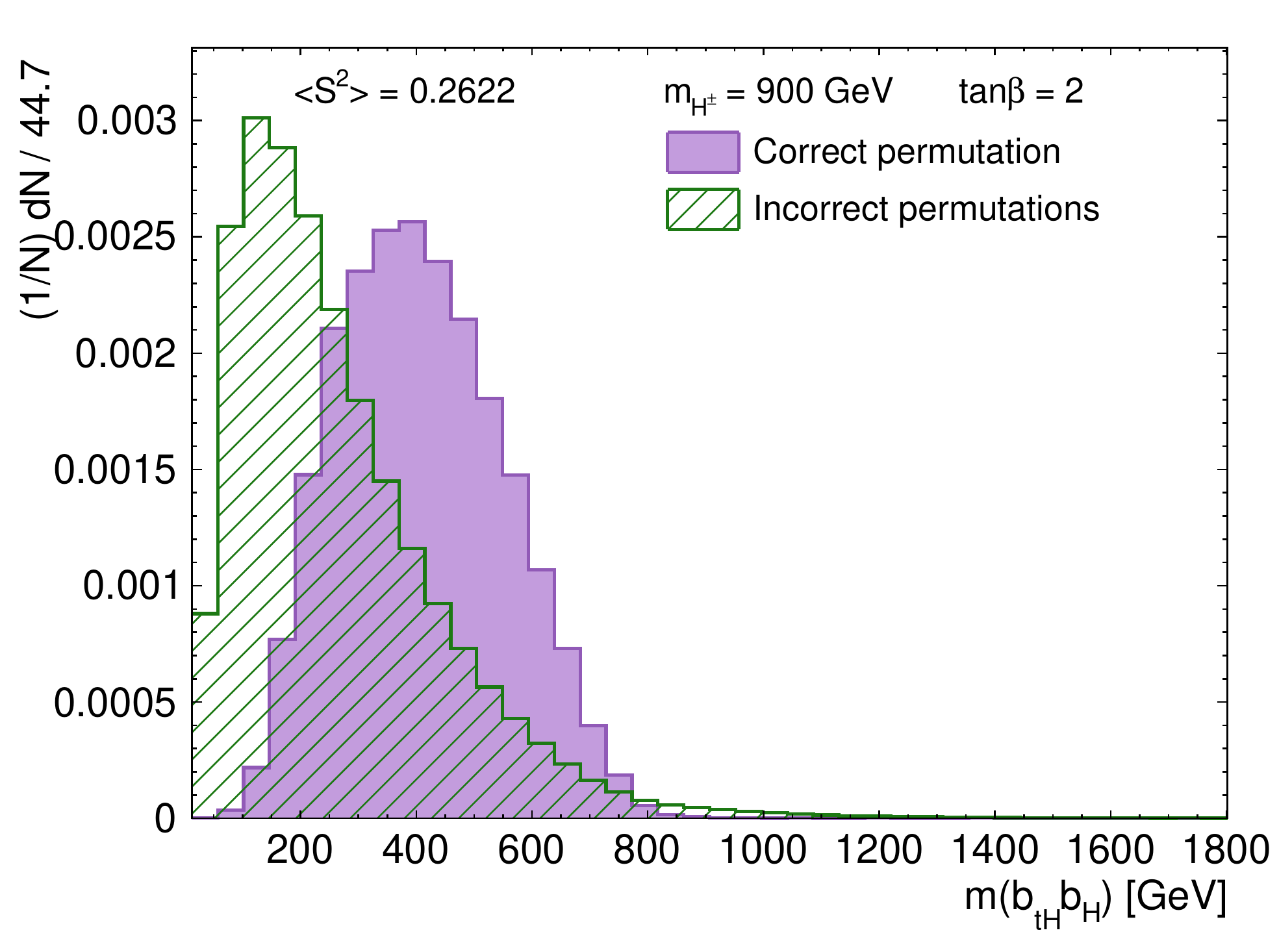}\\
\includegraphics[width=0.32\textwidth]{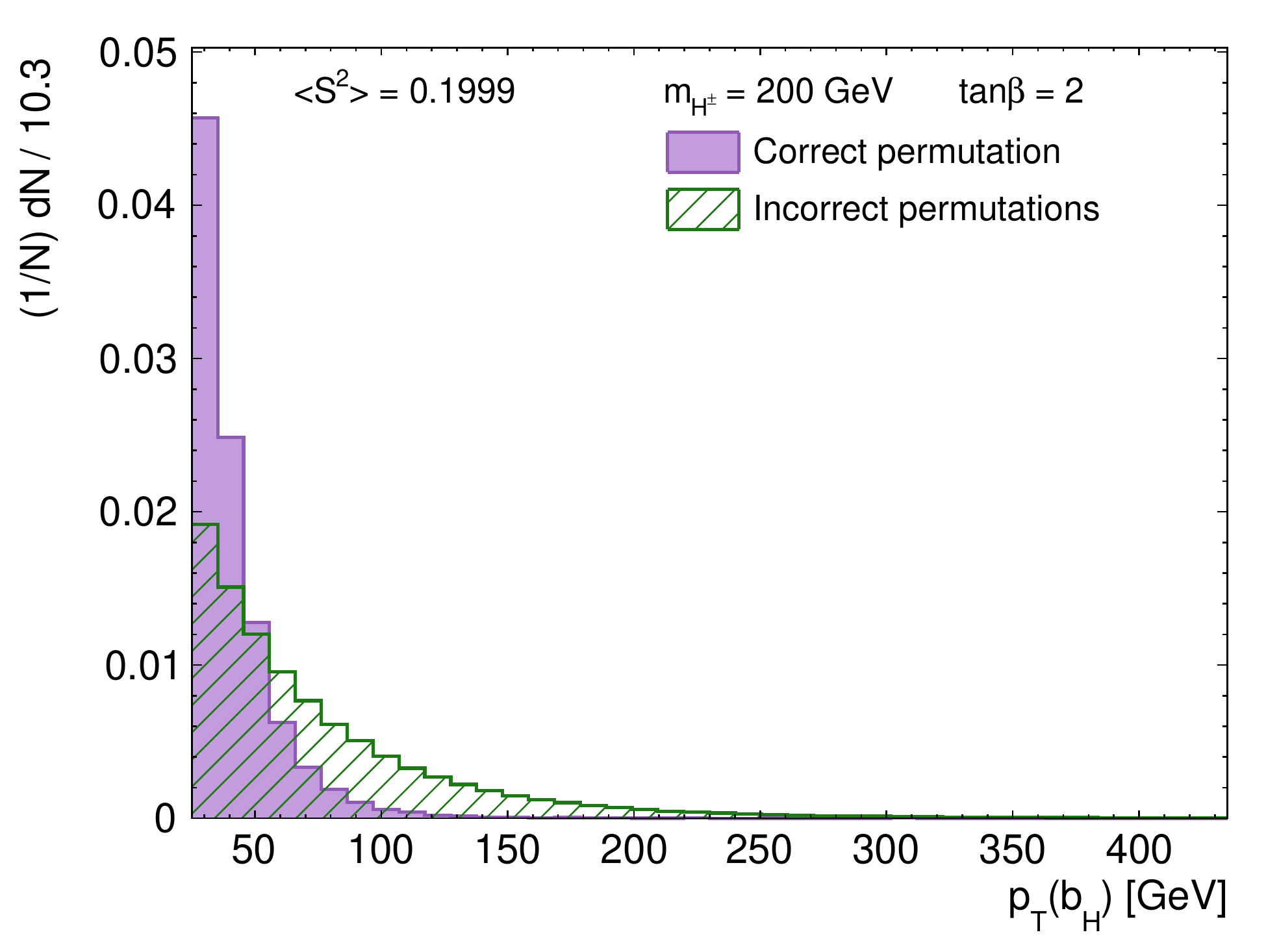}
\includegraphics[width=0.32\textwidth]{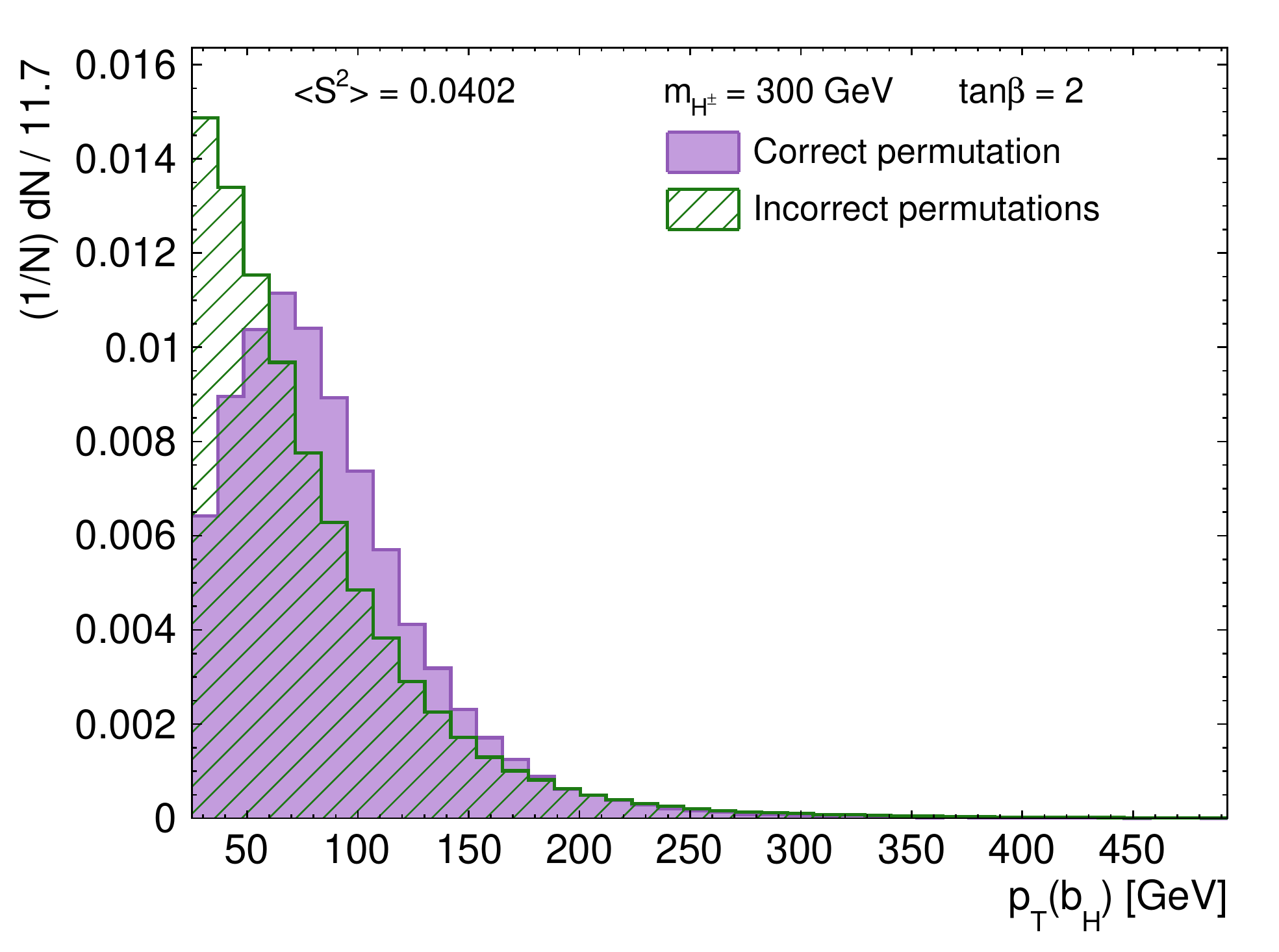}
\includegraphics[width=0.32\textwidth]{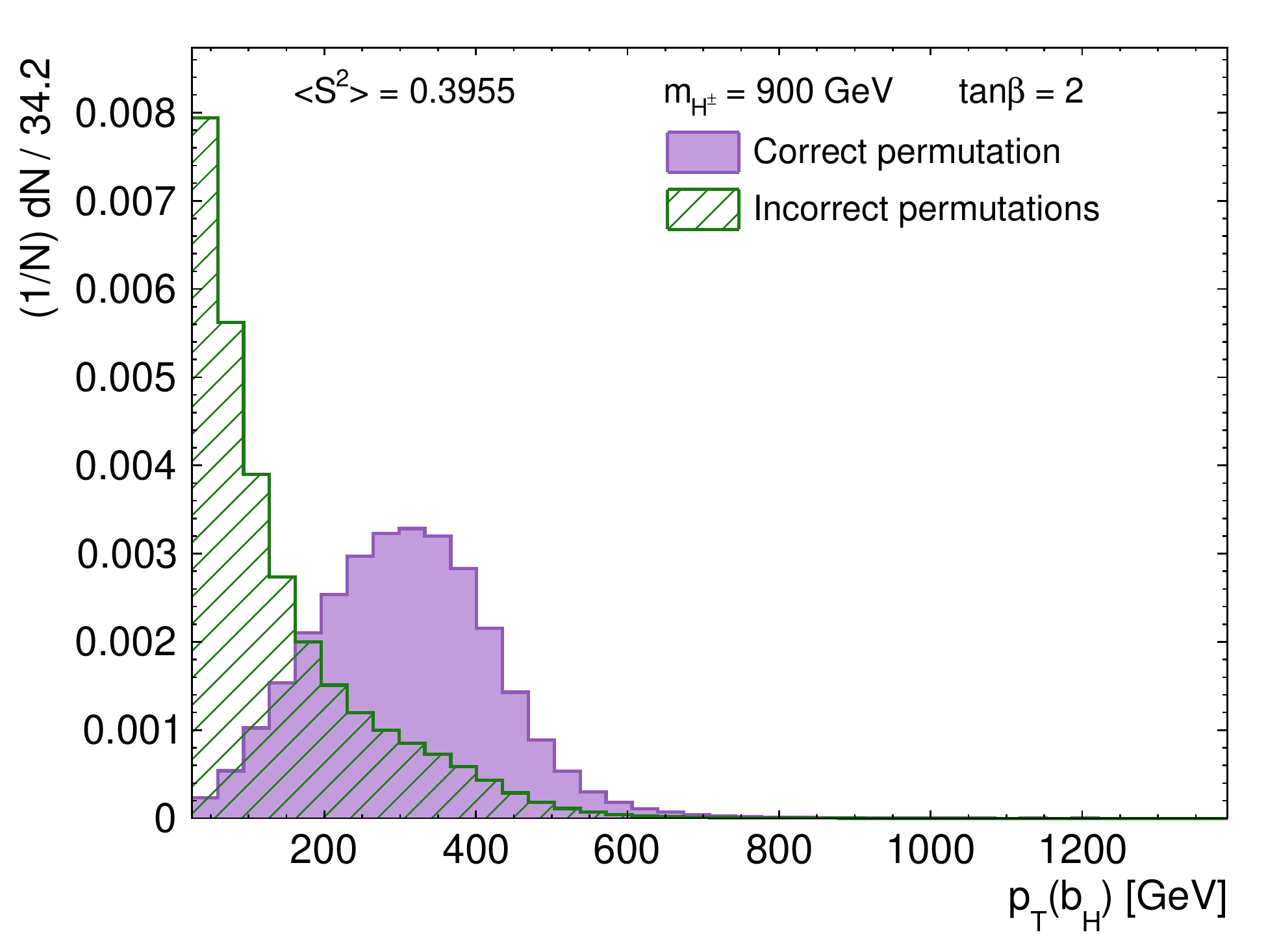}\\
\includegraphics[width=0.32\textwidth]{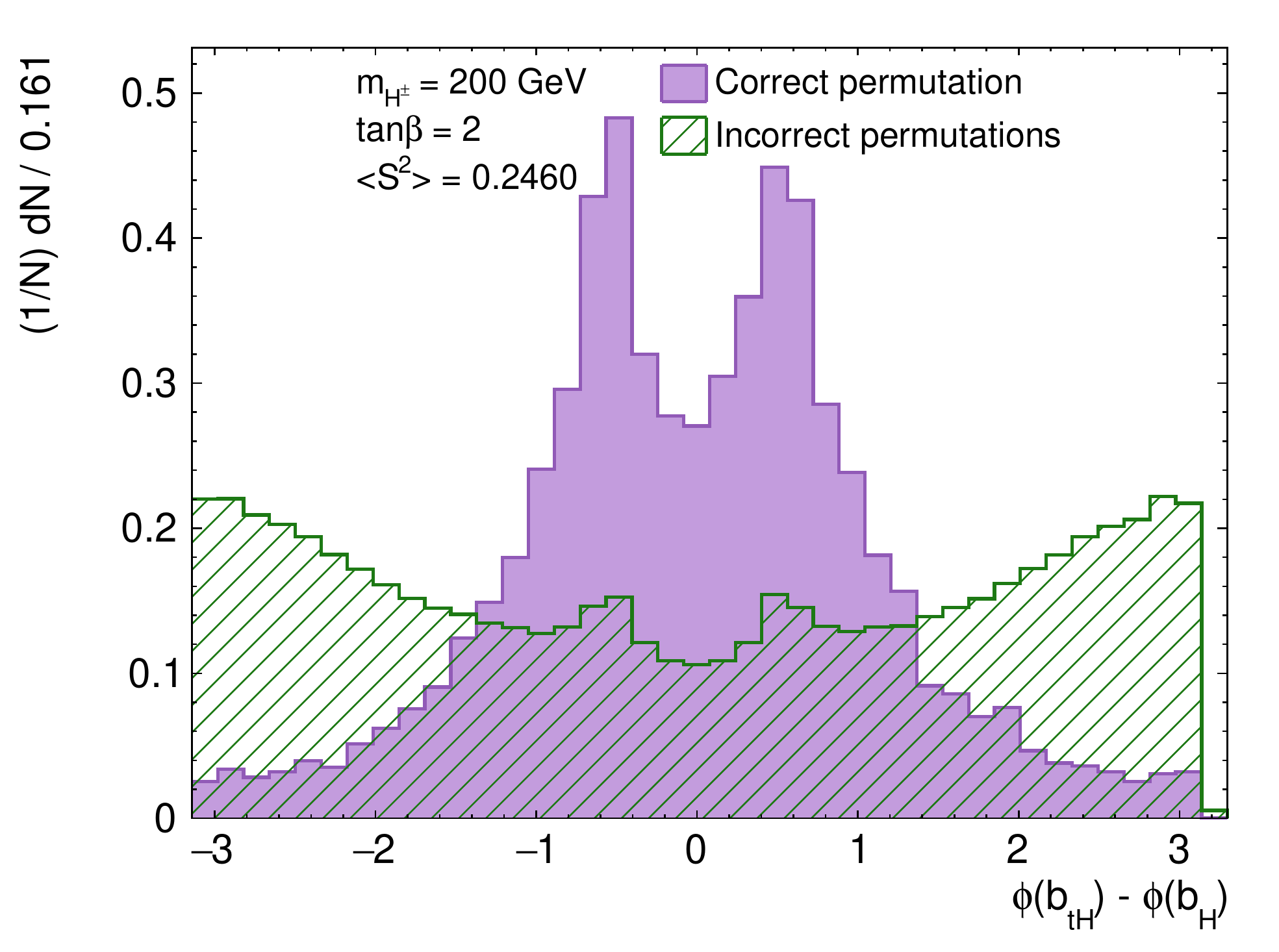}
\includegraphics[width=0.32\textwidth]{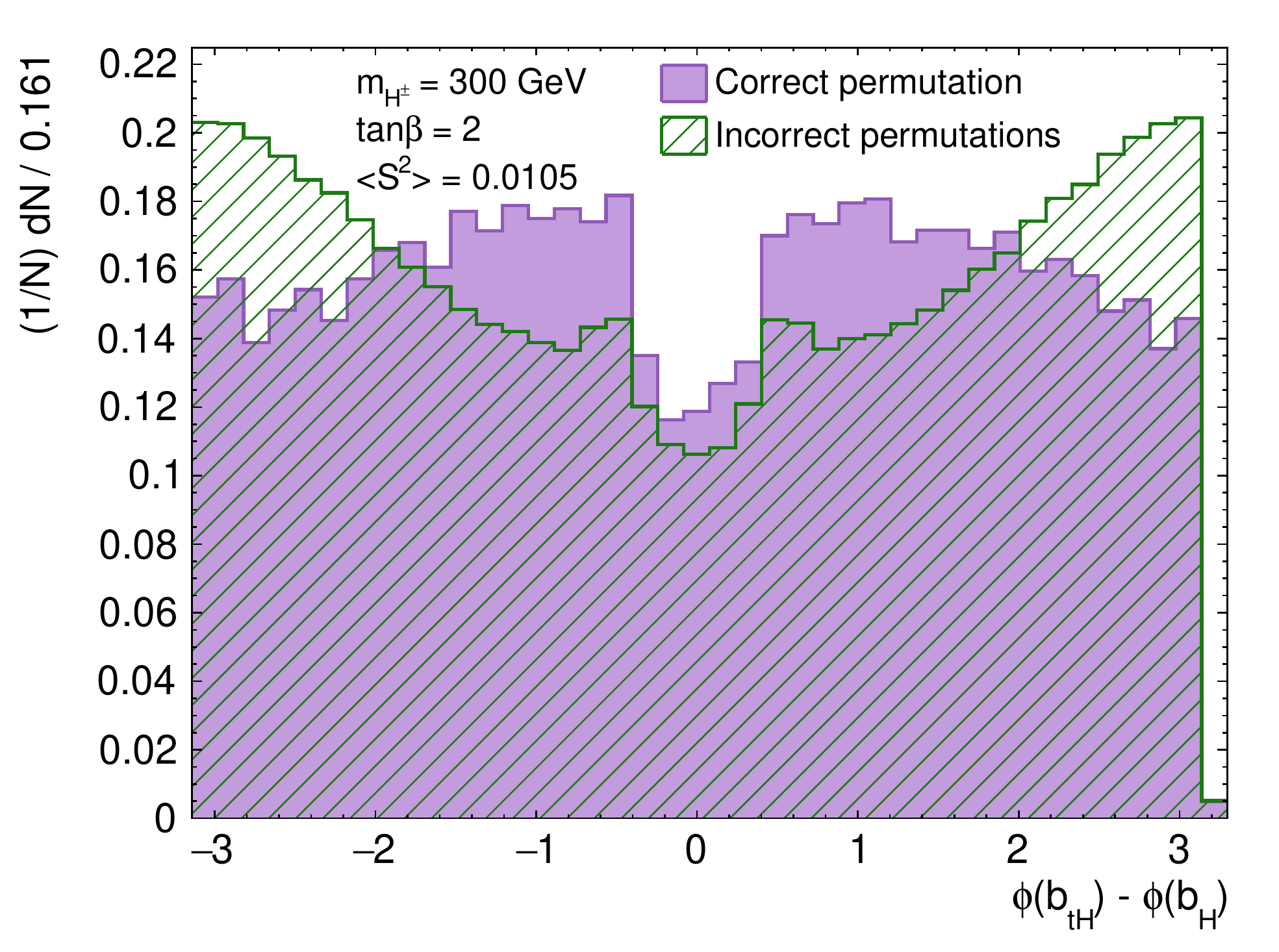}
\includegraphics[width=0.32\textwidth]{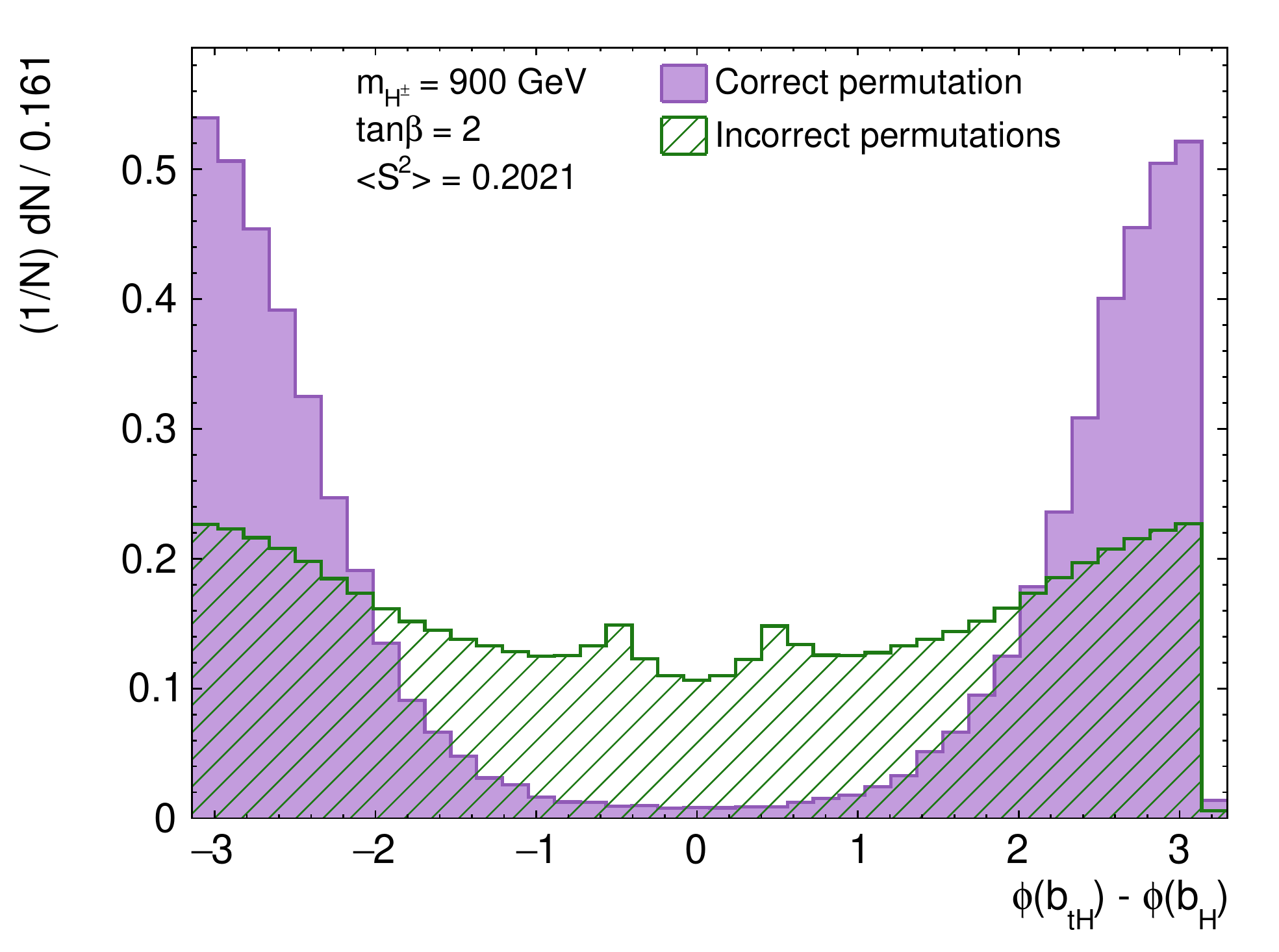}
\caption{Distributions of some of the most important observables in the reconstruction BDT: $m(b_{tH},b_H)$, $p_T^{b_H}$, and $\phi(b_{tH})-\phi(b_H)$. Each of these observables is built entirely on charged Higgs decay products and exhibits greater separation between correct and incorrect combinations at low or high \mhpm.}
\label{fig:reco_BDT_vars}
\end{figure}

At low \mhpm, we find that the most important BDT input\footnote{Here the relative importance of a observable is determined by how often it is used to split a node, weighted by the number of events in the node and the squared separation gain achieved, as defined in TMVA~\cite{Hocker:2007ht}.} is $m(b_{tH},b_H)$, whereas at large \mhpm, $p_T^{b_H}$ becomes the most important.  The distributions for these observables are shown in Fig.~\ref{fig:reco_BDT_vars}, along with $\Delta\phi(b_{tH},b_H)$. The distributions for the correct permutation vary with mass more strongly than those of the incorrect distributions, which are generally fixed for all \mhpm. This results in a turning point around 300~GeV where the correct and incorrect distributions are very similar. This is evident in the BDT performance metrics in Table~\ref{tab:performance_reco}, which show a sharp dropoff from 200--300~GeV followed by a steady increase towards larger \mhpm.

\begin{table}[!ht]
{\scriptsize
\begin{tabular}{l||c|c|c|c|c}
 & \multicolumn{5}{c}{$\mhpm$}\\
{} & 200 GeV & 300 GeV & 500 GeV & 700 GeV & 900 GeV\\\hline
$b$-jet to quark matching [\%] & 34.51 -- 38.86 & 61.77 -- 65.31 & 60.04 -- 65.25 & 59.57 -- 65.02 & 58.80 - 64.70\\
Reconstruction BDT separation & 0.62--0.69 & 0.30 -- 0.39 & 0.52 -- 0.60 & 0.63 -- 0.70 & 0.70 -- 0.75\\
$(b_t,b_{H},b_{tH})$ correct [\%] & 23.4--25.6 & 19.4 -- 24.2 & 29.8 -- 31.0 & 34.2 -- 36.7 & 37.0 -- 40.5 \\
Neutrino weighting solution exists [\%] & 90.1--91.9 & 95.4 -- 97.3 & 93.3 -- 95.9 & 90.9 -- 94.2 & 88.3 -- 93.3\\
$H^\pm$ charge correct [\%] (all events) & 55.9--58.6 & 56.3 -- 59.7 &   58.3 -- 59.4  &  61.8 -- 63.1  &  65.2 -- 66.4 \\
$H^\pm$ charge correct [\%] (events with correct $(b_t,b_{H},b_{tH})$) & 84.0--85.8 & 80.3 -- 81.3 & 82.3 -- 84.1 & 86.4 -- 87.3 & 88.8 -- 89.3\\
$(b_t,b_{H},b_{tH})$ correct \& $H^\pm$ charge correct [\%] & 19.8--21.9 & 15.6 -- 19.7 &   24.7 -- 26.1  &  29.8 -- 31.8  &  33.1 -- 36.2 \\
\end{tabular} }
\caption{Performance of reconstruction BDT and neutrino weighting procedure for $\tan\beta = 1-60$, with minimum and maximum values.}
\label{tab:performance_reco}
\end{table}

\subsection{Neutrino Weighting}

%Check these
%http://arxiv.org/abs/hep-ex/0602008
%http://arxiv.org/abs/1205.0354
%http://arxiv.org/abs/hep-ex/9706014
%http://arxiv.org/abs/1501.07383
%Some possible references:
%1508.03322 ($m_t$ at D0)
%http://inspirehep.net/record/1085854?ln=en ($m_t$ at D0, earlier)
%0904.3195 ($m_t$ at D0, earliest?)

Once the correct jet permutation is identified, it is still necessary to determine the neutrino momenta in order to fully reconstruct the event.  Since the neutrinos are reconstructed only in the form of $\cancel{E}_T$, their individual momenta are unknown. Each neutrino comes from a top ($t\to W^+ b, W^+ \to \ell^+ \nu_\ell$) or anti-top decay, which means the neutrino momenta can be constrained by the top and $W$ masses.  The two $\cancel{E}_T$ and four mass constraints are in principle sufficient to determine the neutrino momenta, though the quadratic nature of the mass constraints and the uncertain pairing of leptons do not provide a unique solution.  To reconstruct the neutrino momenta, we follow a neutrino weighting procedure. This attempts to find the allowed pair of neutrino momenta which best reproduces the observed missing energy. Neutrino weighting is a procedure originally developed at the D{\O} experiment \cite{Abbott:1997fv,Abbott:1998dn} for top quark mass measurements; it has since been used in other measurements, such as the $t\bar{t}$ differential cross section at ATLAS \cite{Aaboud:2016syx}. To the best of our knowledge, it has never been used before in an analysis of the $t\bar{t}b\bar{b}$ channel. In our implementation of neutrino weighting, we sample values from a Gaussian for the pseudorapidity of the two neutrinos, $\eta_1$ and $\eta_2$, in the range $-5\leq \eta_i \leq 5$.  To account for variation in the invariant masses of the top quarks and increase the likelihood of finding real solutions, we also scan between $171.5~\gev$ and $174.0~\gev$ independently for both $m_t$ and $m_{\bar{t}}$. To account for jet resolution, we similarly iterate over several energies of the $b$-jets, sampling from a Gaussian.  For each set of values considered, we solve for the momentum of each neutrino using
\begin{eqnarray}
\nonumber
(p_b+p_\ell+p_\nu)^2 &=& m_t^2\;, \\
(p_\ell+p_\nu)^2 &=& m_W^2\;.
\end{eqnarray}
This reduces to a quadratic constraint for each neutrino, producing up to four real solutions overall.  Additionally, there are two possible ways to pair the leptons and $b$-jets.  For each solution, we calculate a weight

\begin{equation}
w=\exp\left(-\frac{(\cancel{E}_T^{\textrm{calc}}-\cancel{E}_T^{\textrm{obs}})^2}{2\sigma^2_{\cancel{E}_T}}\right)
\exp\left(-\frac{(\cancel{\phi}^{\textrm{calc}}-\cancel{\phi}^{\textrm{obs}})^2}{2\sigma^2_{\cancel{\phi}}}\right),
\end{equation}
where $\cancel{\phi}$ is the azimuthal angle of the missing energy. The resolution of $\cancel{E}_T^{\textrm{obs}}$ is given by $\sigma_{\cancel{E}_T} = 0.2\cancel{E}_T$, following the ATLAS resolution~\cite{ATL-PHYS-PUB-2015-027}. The resolution of $\cancel{\phi}^{\textrm{obs}}$ is given by $\sigma_{\cancel{\phi}}=0.05$, a fixed-value based on the MET resolution in studies from Z boson events during ATLAS Run-1. Out of all the combinations and solutions, we choose the one with the highest weight and take the corresponding neutrino momenta and $\ell$--$b$ pairings for the remainder of our analysis.  If the neutrino weighting procedure is unable to find a real solution for any configuration, the event is discarded.

\begin{figure}[!ht]
\includegraphics[width=0.38\textwidth]{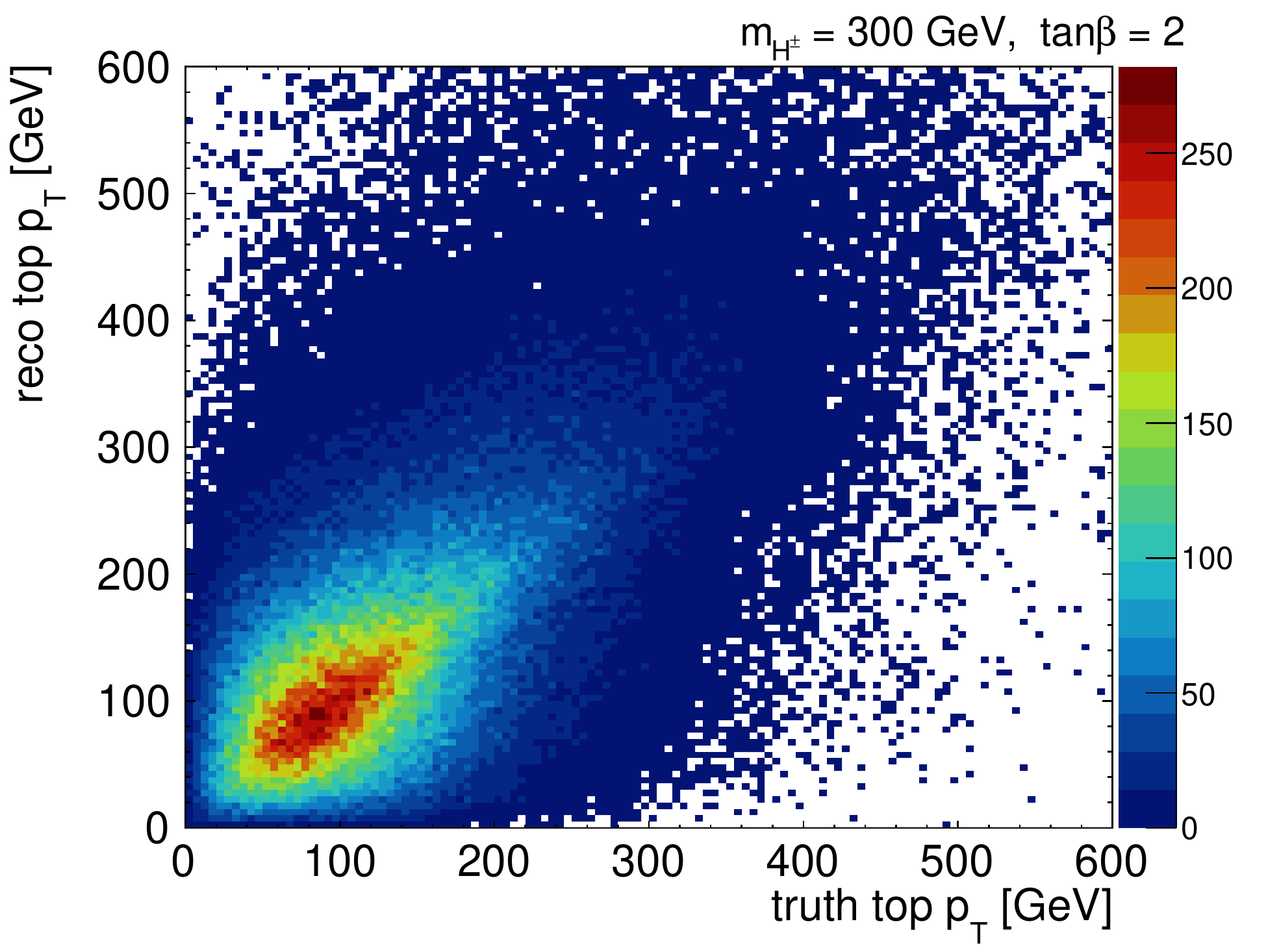}
\includegraphics[width=0.38\textwidth]{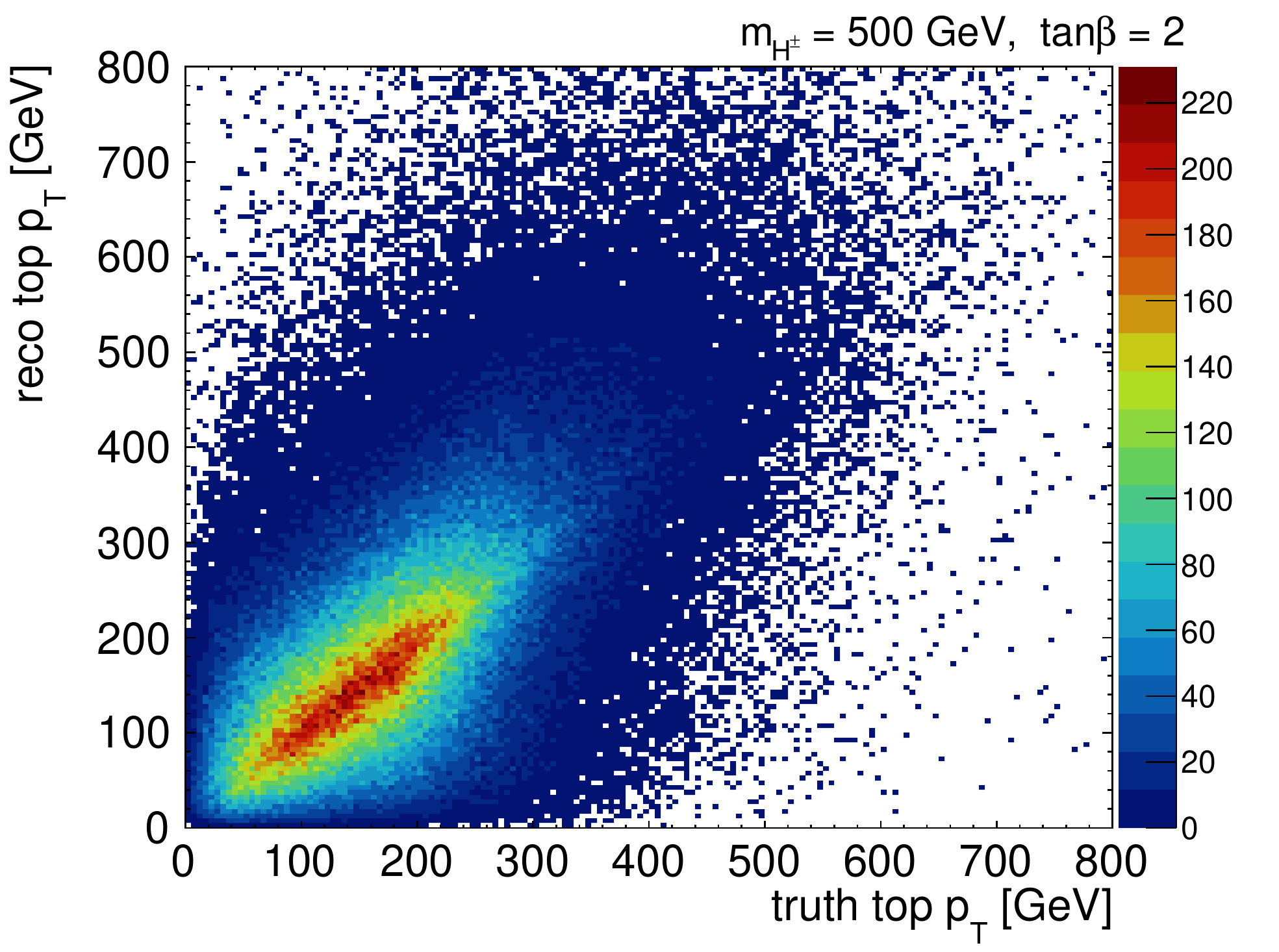}
\includegraphics[width=0.38\textwidth]{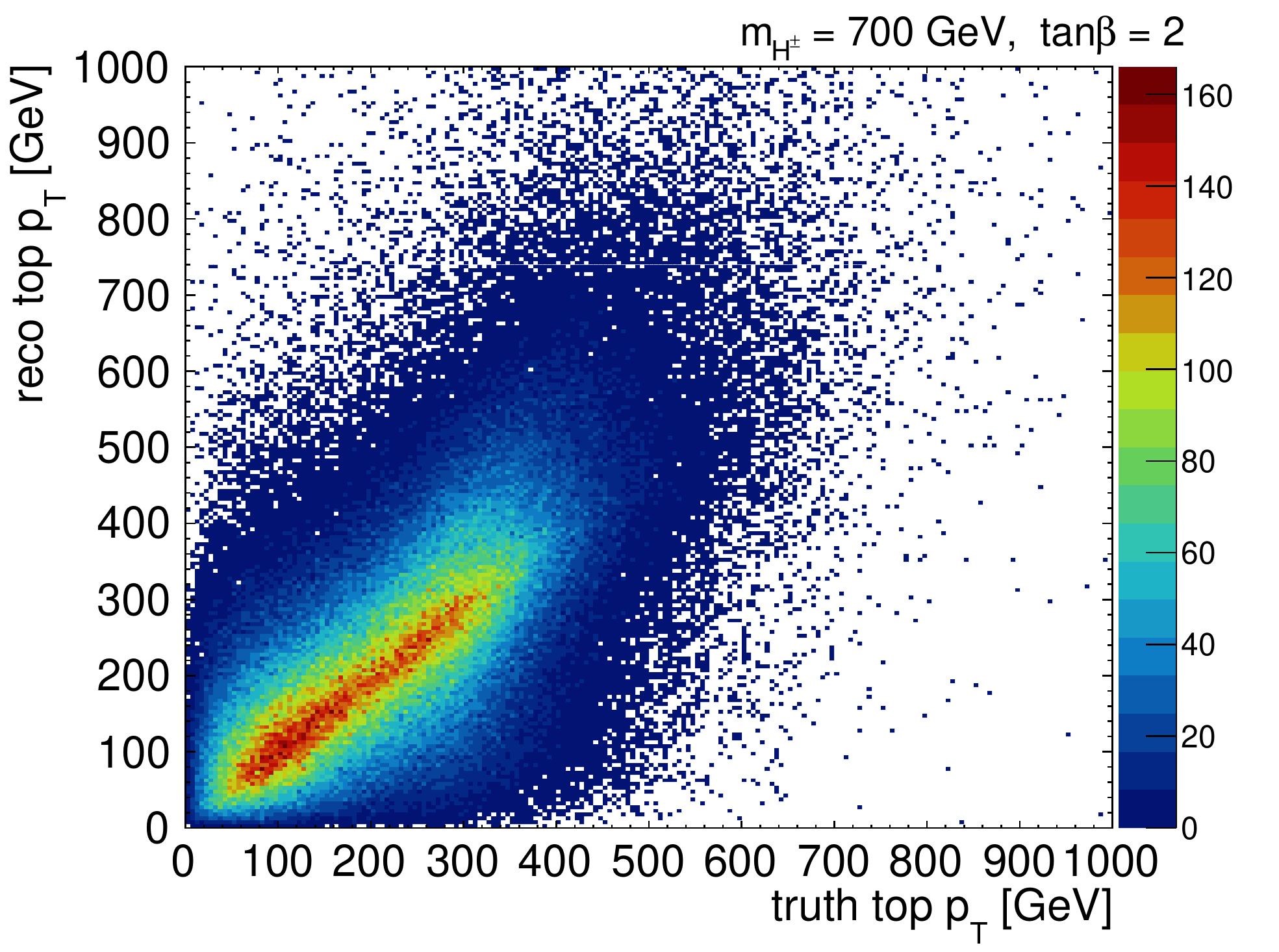}
\includegraphics[width=0.38\textwidth]{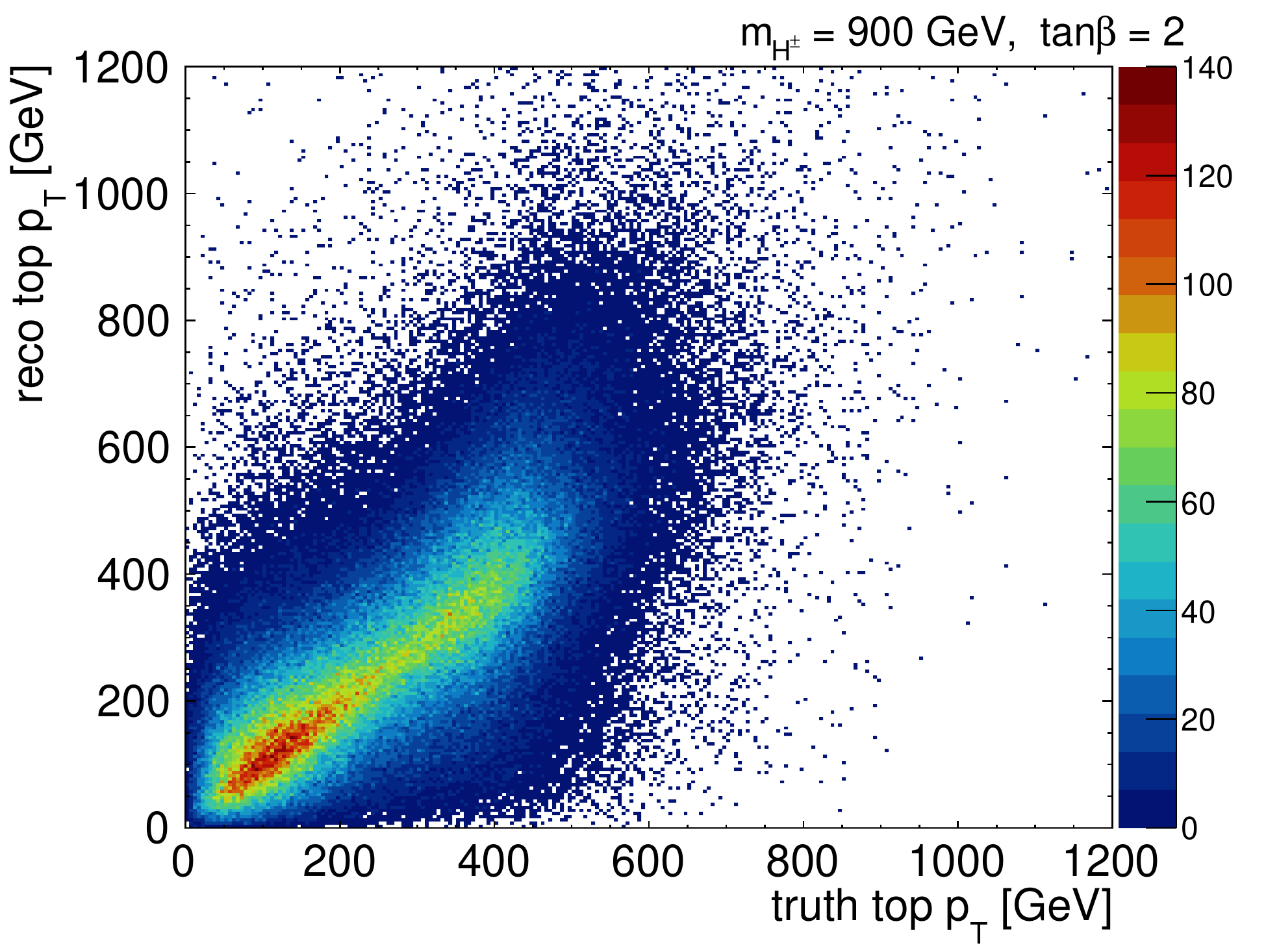}
\caption{Comparison of $p_{T}$ for the truth top and the reconstructed top using the reconstruction BDT and neutrino weighting procedure, for $\mhpm = 300,500,700,900~\gev$ and $\tan\beta=2$.}
\label{fig:neutrinoweight_pT}
\end{figure}

The performance of the neutrino weighting procedure is shown in Table~\ref{tab:performance_reco}. We see that a solution is found a high percentage of the time, and the $\ell$--$b$ pair coming from the charged Higgs decay, which indicates the charge of the Higgs boson, is identified correctly more often than not.  Since this procedure uses the $b$-jet assignments from the reconstruction BDT, to isolate the performance of neutrino weighting, we also show in Table~\ref{tab:performance_reco} the fraction of events correctly reconstructing the charge of the Higgs boson when considering only events for which the $b$-jets have been correctly assigned, which we find happens for 80--90\% of such events.

As with the reconstruction BDT, the neutrino weighting procedure sees only a small variation in performance over the range of $\tan\beta$.  The fraction of events for which all $b$-jets and leptons are correctly assigned is shown in Table~\ref{tab:performance_reco}. Despite seemingly low efficiencies, this is largely a reflection of the large number of possible $b$-jet permutations. The procedure performs much better than a random choice. Ultimately, we use the reconstructed neutrino momenta to reconstruct the top quarks. In Fig.~\ref{fig:neutrinoweight_pT} the reconstructed (reco) top $p_T$ is compared to the corresponding truth top $p_{T}$. These show strong correlations.

\subsection{Classification}
After the reconstruction BDT and neutrino weighting procedure, we have determined the grouping and momenta for all of the final state particles shown in Fig.~\ref{fig:tbhc_diagram}. This allows us to reconstruct a charged Higgs mass, which can help discriminate between signal and background, especially for large masses. However, a stronger discriminant can be constructed by taking advantage of the full kinematic information with a second BDT. This classification BDT is trained on the charged Higgs signal and the combined SM
backgrounds after reconstructing the events. The classification BDT is trained on 21 observables:
\begin{itemize}[itemsep=1pt]
\item Maximum weight from the reconstruction BDT
\item $H_T = \sum_{i}\left|\vec{p}_{i,T}\right|$ for $i=$all jets and leptons
\item Centrality $= H_T/E$, where $E=\sum_i E_i$ for $i=$all jets and leptons
\item $m(b_{i},b_{j})$ and $m(b_{i},l_{j})$ for $i$, $j$ giving smallest $\Delta R$
\item $\min(m(b_{i},l_{j}))$ and $\max(m(b_{i},l_{j}))$
\item $m(b_{1},b_{2})$, $m(t_{H},b_{H})$
\item $p_{T}^{b_{1}+b_{2}}$, $p_{T}^{b_{H}}$, $p_{T}^{b_{1}+t_{1}}$, $p_{T}^{t_{H}}$, $p_{T}^{t_{\text{other}}}$
\item $\Delta R(b_{1},b_{2})$, $\Delta R(b_{1},t_{1})$, $\Delta R(t_{H},t_{\text{non-H}})$, $\Delta R(t_{H},b_{H})$, $\Delta R(b_{tH},b_{H})$, $\Delta R(b_{t},b_{H})$
\item $\cos\theta(l_{H},b_{H})$, the angle between $b_H$ and $\ell_{tH}$ in the reconstructed $H^\pm$ rest frame
\end{itemize}

\begin{figure}[!ht]
\includegraphics[width=0.32\textwidth]{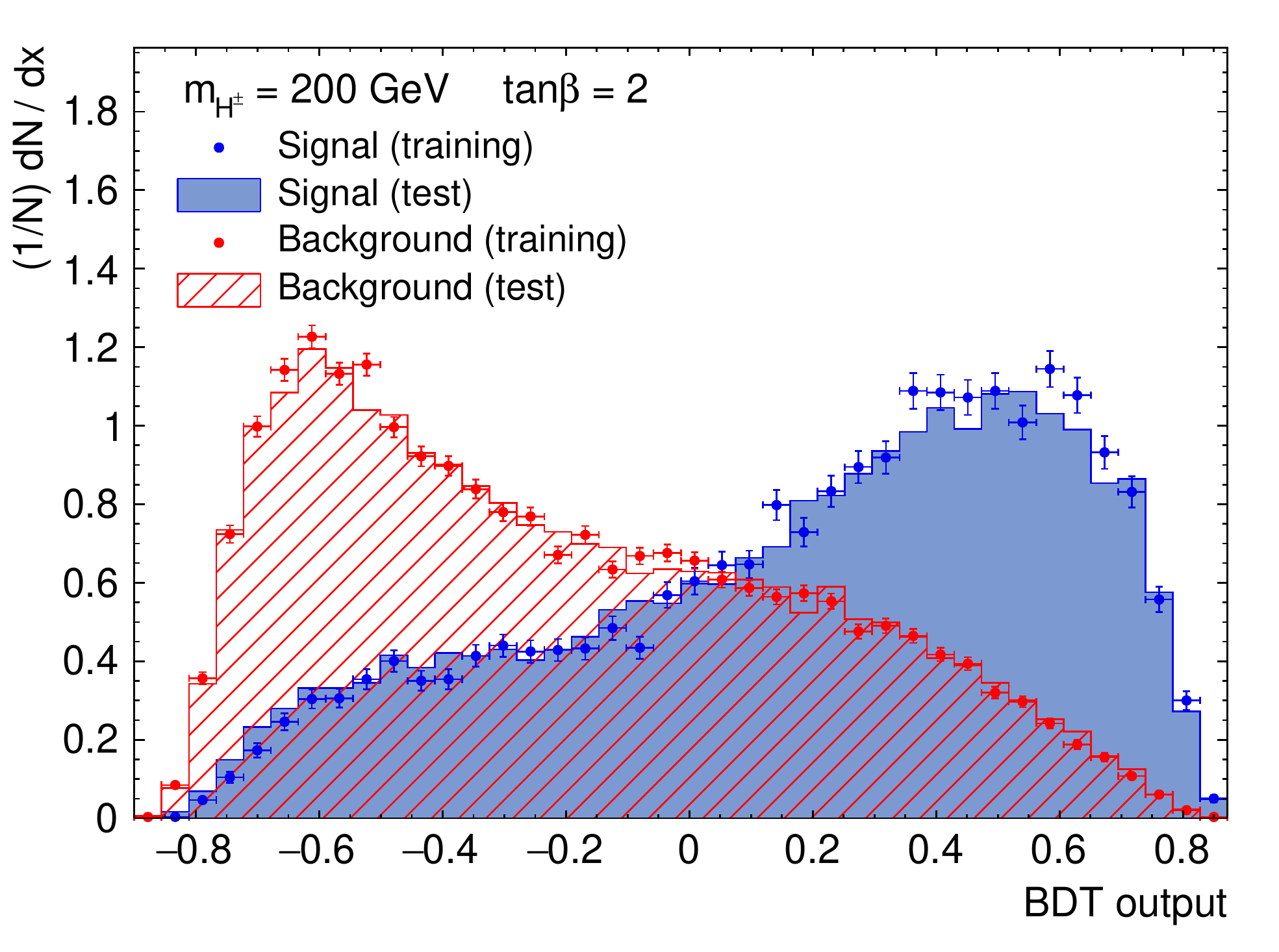}
\includegraphics[width=0.32\textwidth]{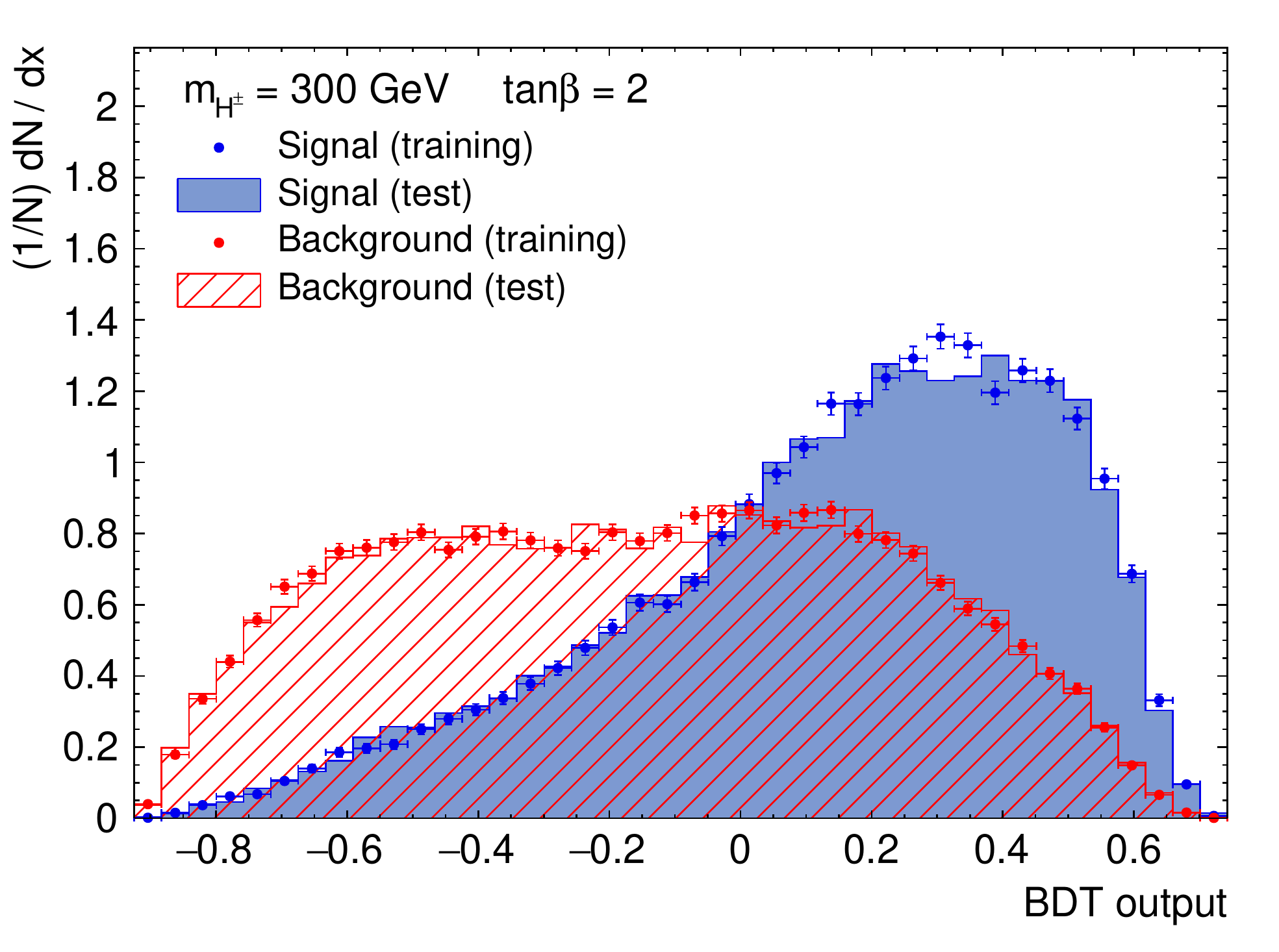}
\includegraphics[width=0.32\textwidth]{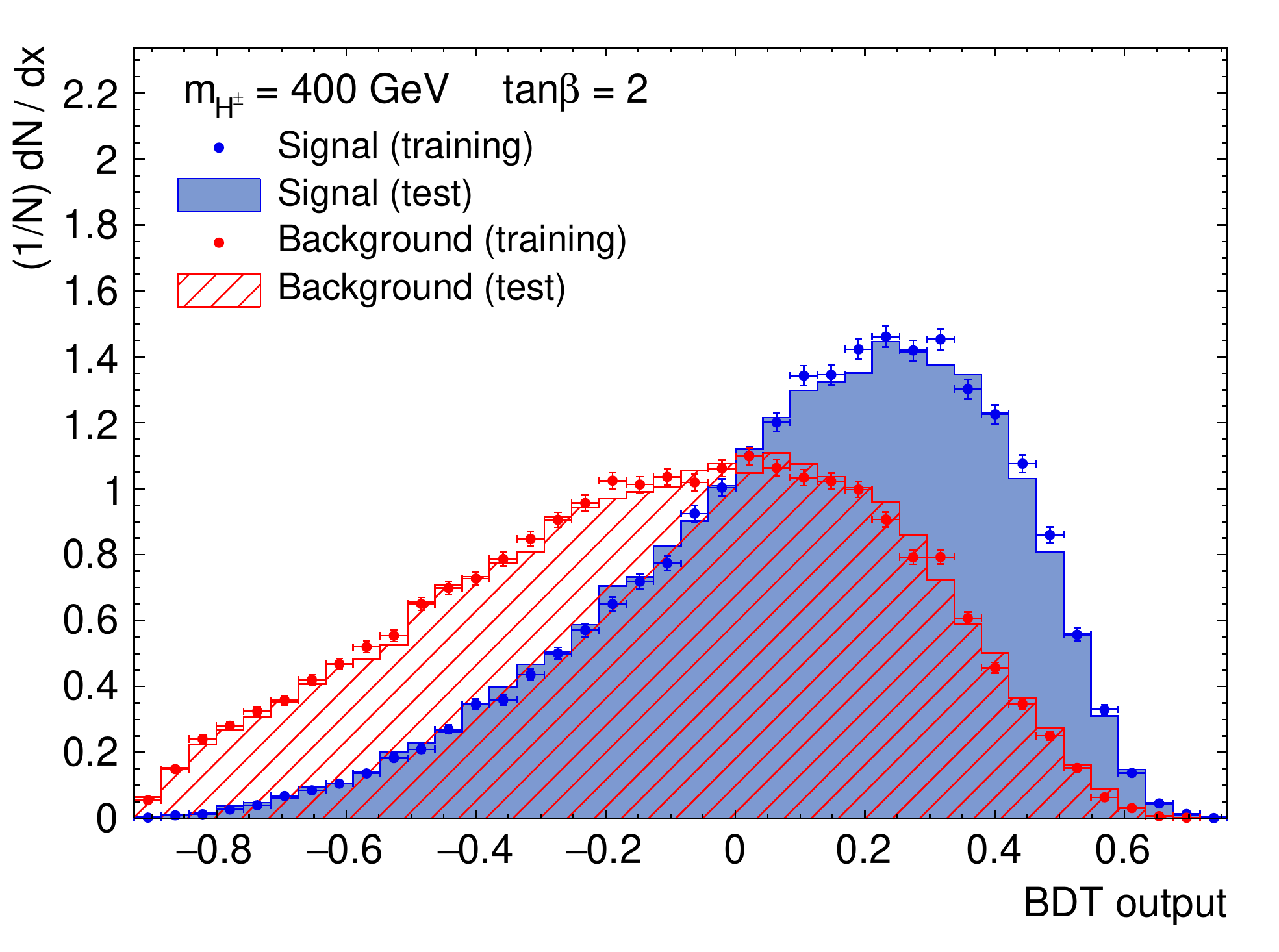}
\includegraphics[width=0.32\textwidth]{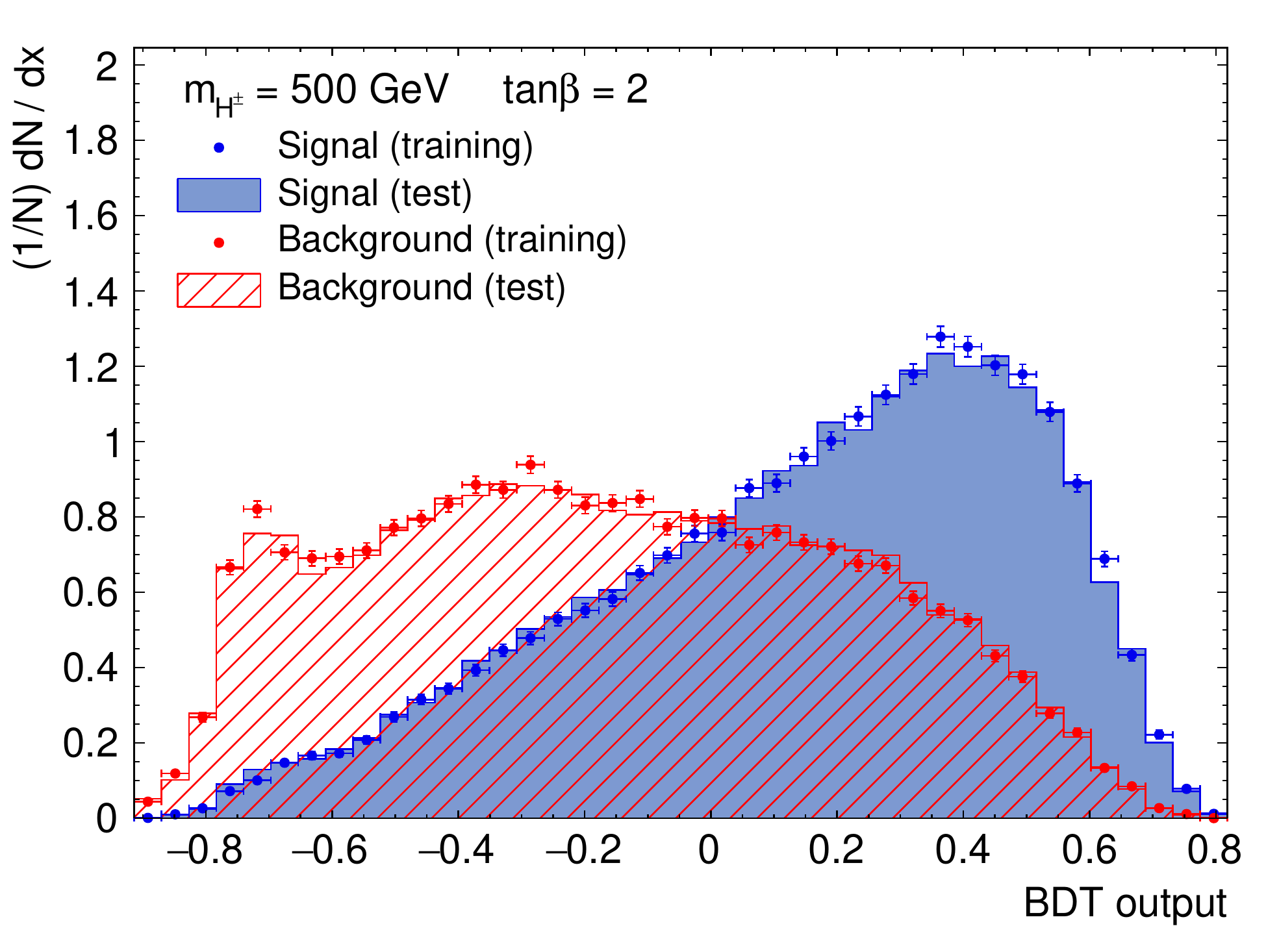}
\includegraphics[width=0.32\textwidth]{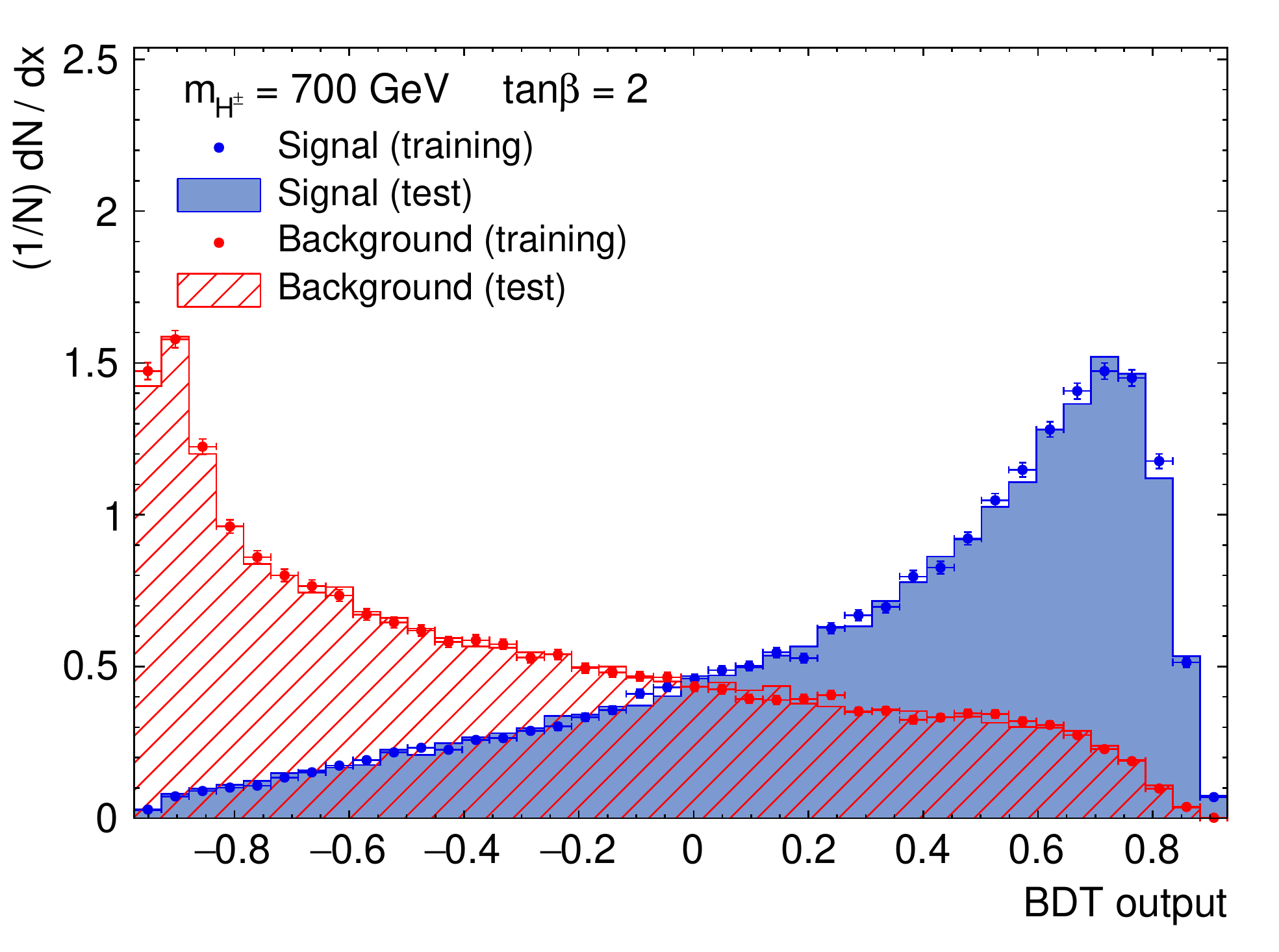}
\includegraphics[width=0.32\textwidth]{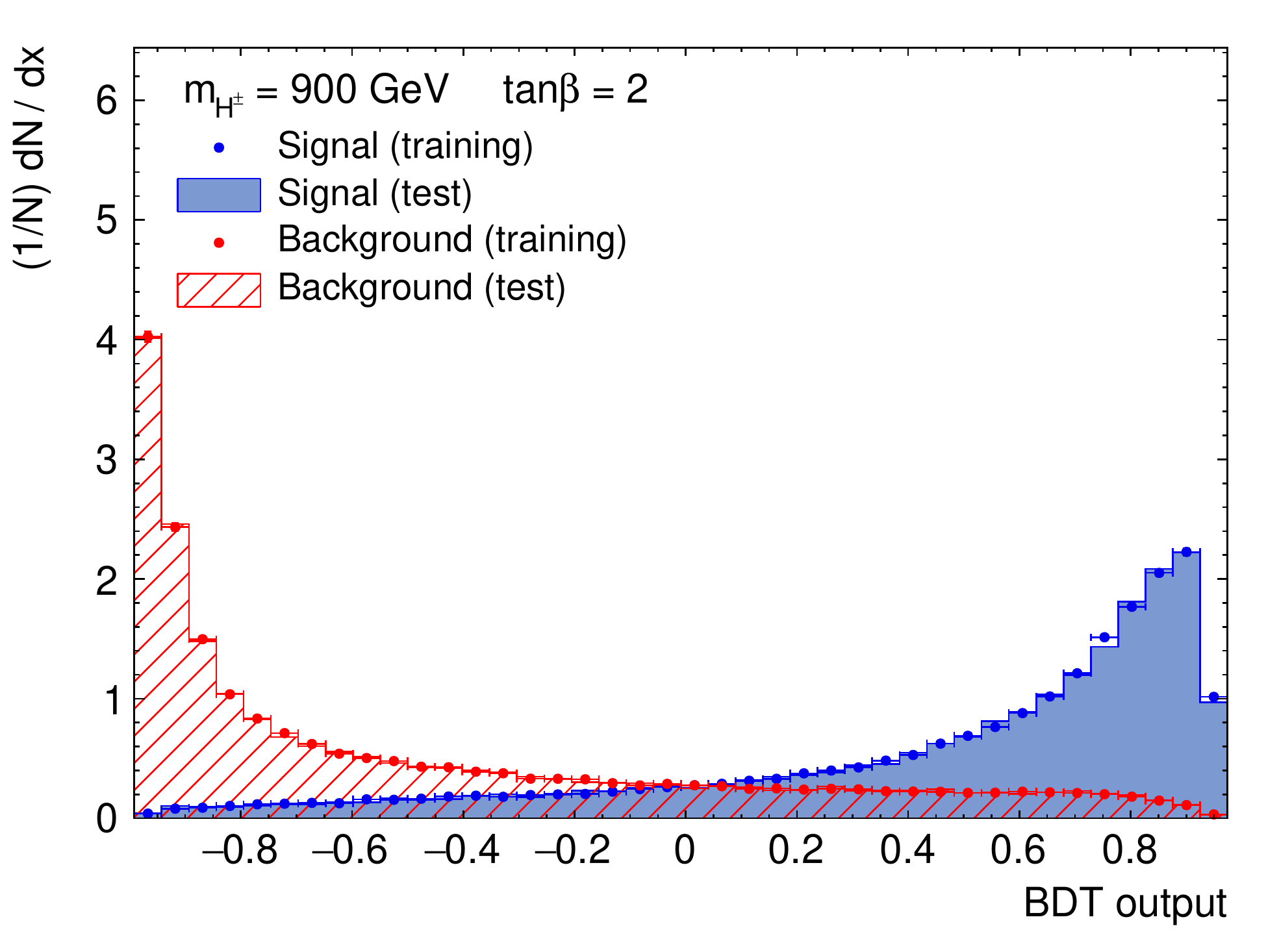}
\caption{Classification BDT response for $\mhpm = 200,300,400,500,700,900~\gev$  and $\tan\beta=2$.}
\label{fig:classification_BDT}
\end{figure}

Here, $b_1$ ($b_2$) is the $b$-jet with the (second) highest $p_T$; $t_1$ is the reconstructed top (including the neutrino) with the highest $p_T$. In general, observables such as $m(j_{1},j_{2})$, the reconstruction BDT weight, and $m(t_{H},b_{H})$ contribute highly to the BDTs. The observables $p_{T}^{b_{H}}$ and $H_{T}$ become important at higher mass.

\begin{figure}
\includegraphics[width=0.32\textwidth]{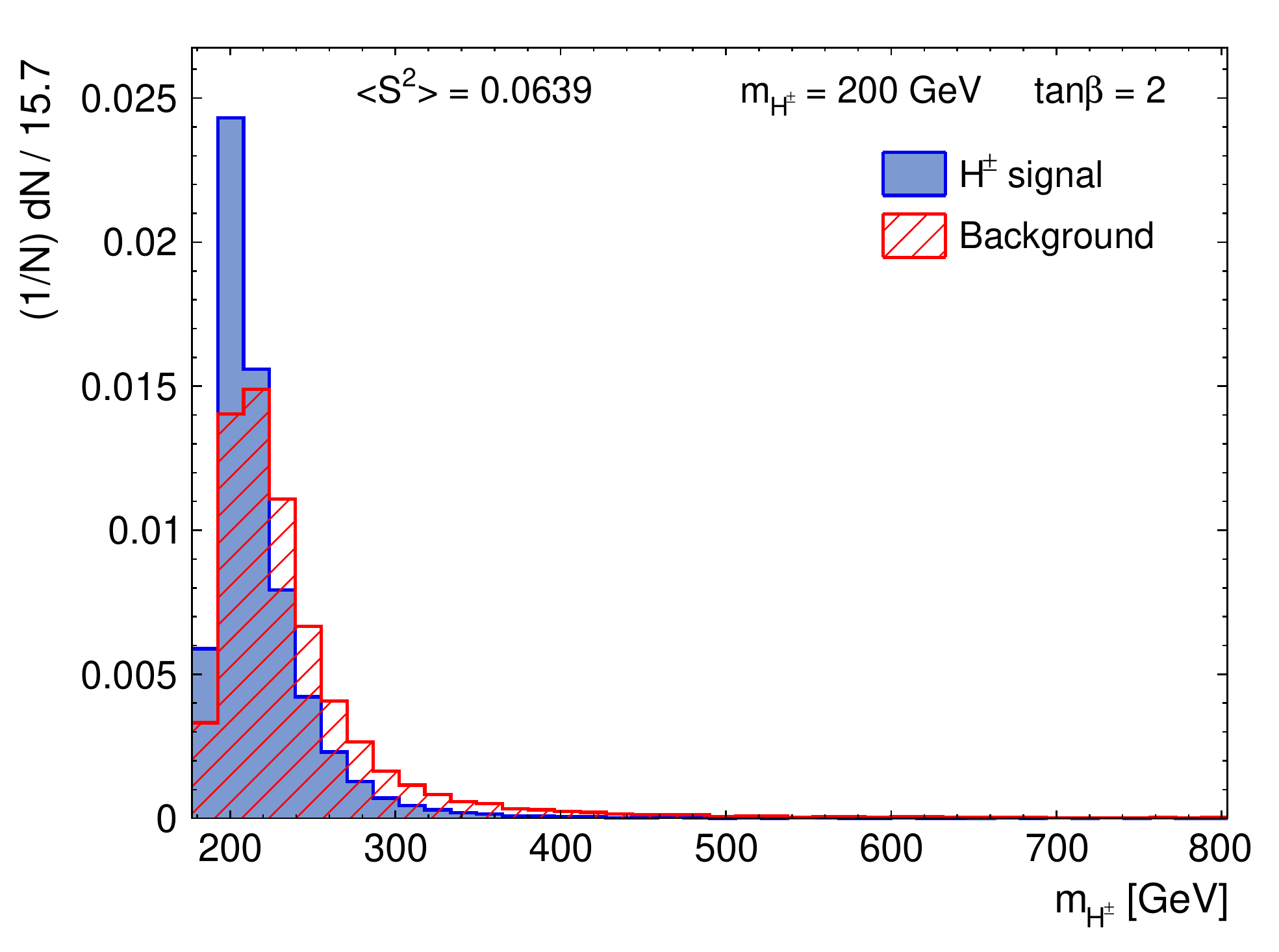}
\includegraphics[width=0.32\textwidth]{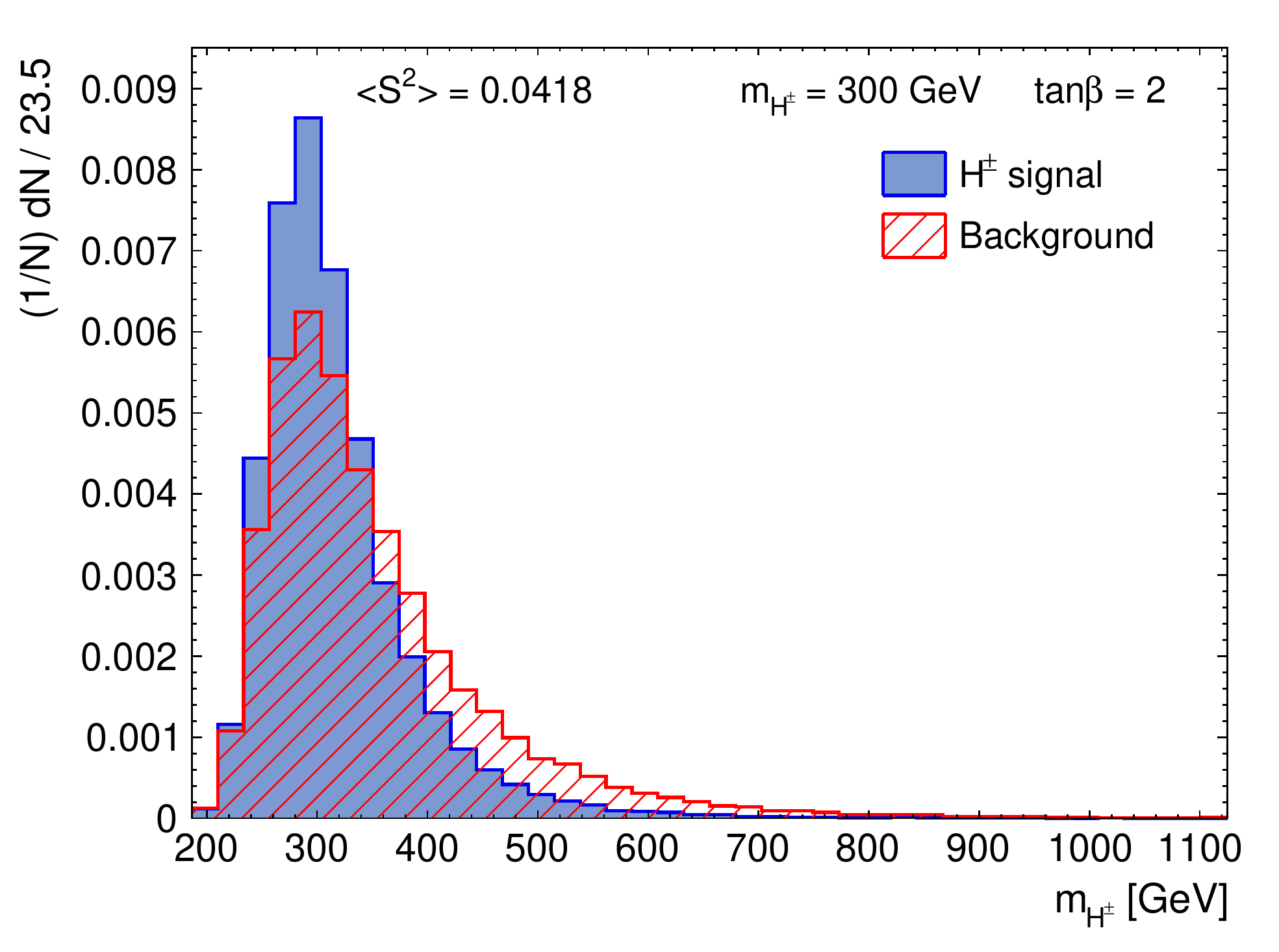}
\includegraphics[width=0.32\textwidth]{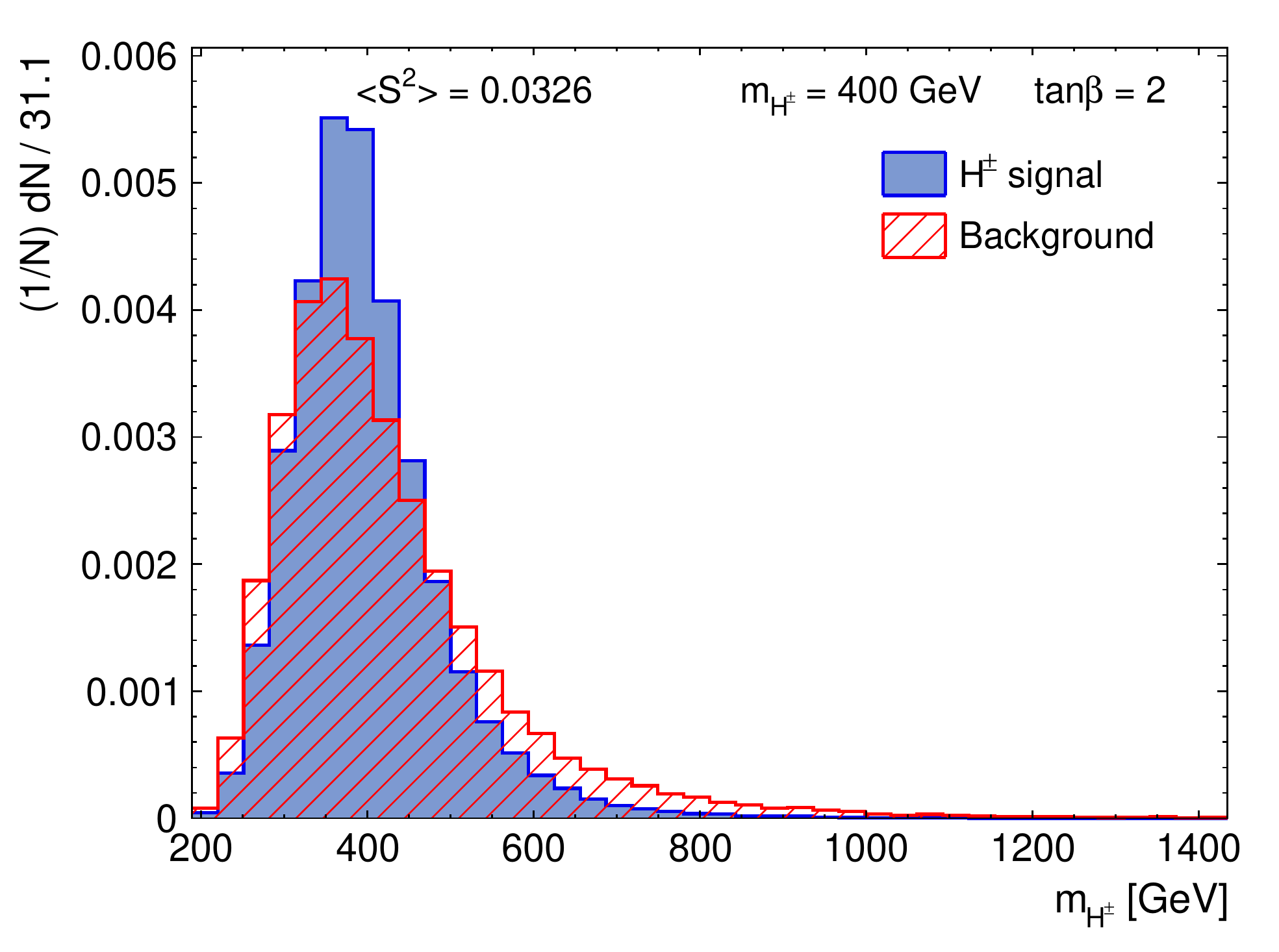}\\
\includegraphics[width=0.32\textwidth]{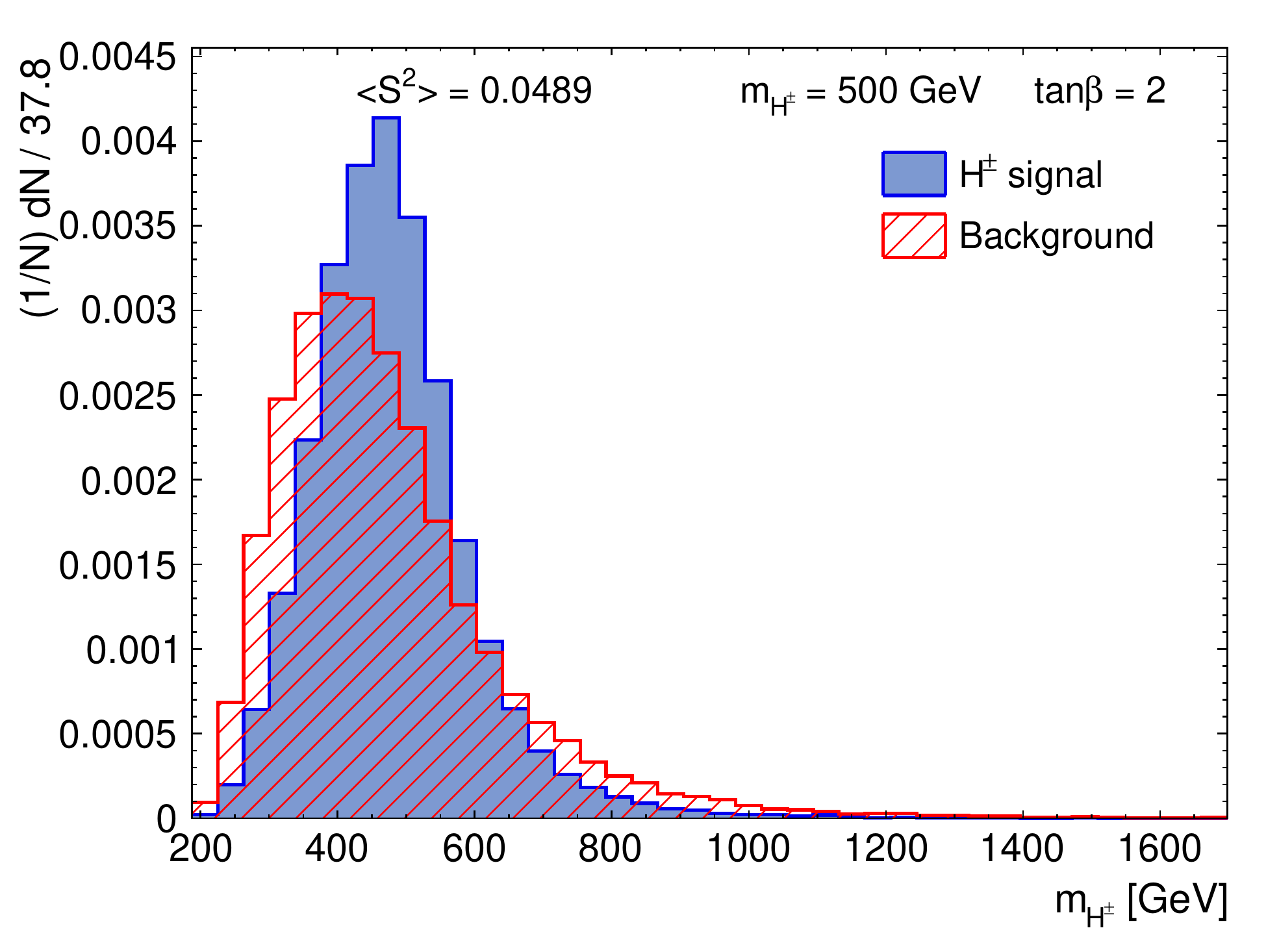}
\includegraphics[width=0.32\textwidth]{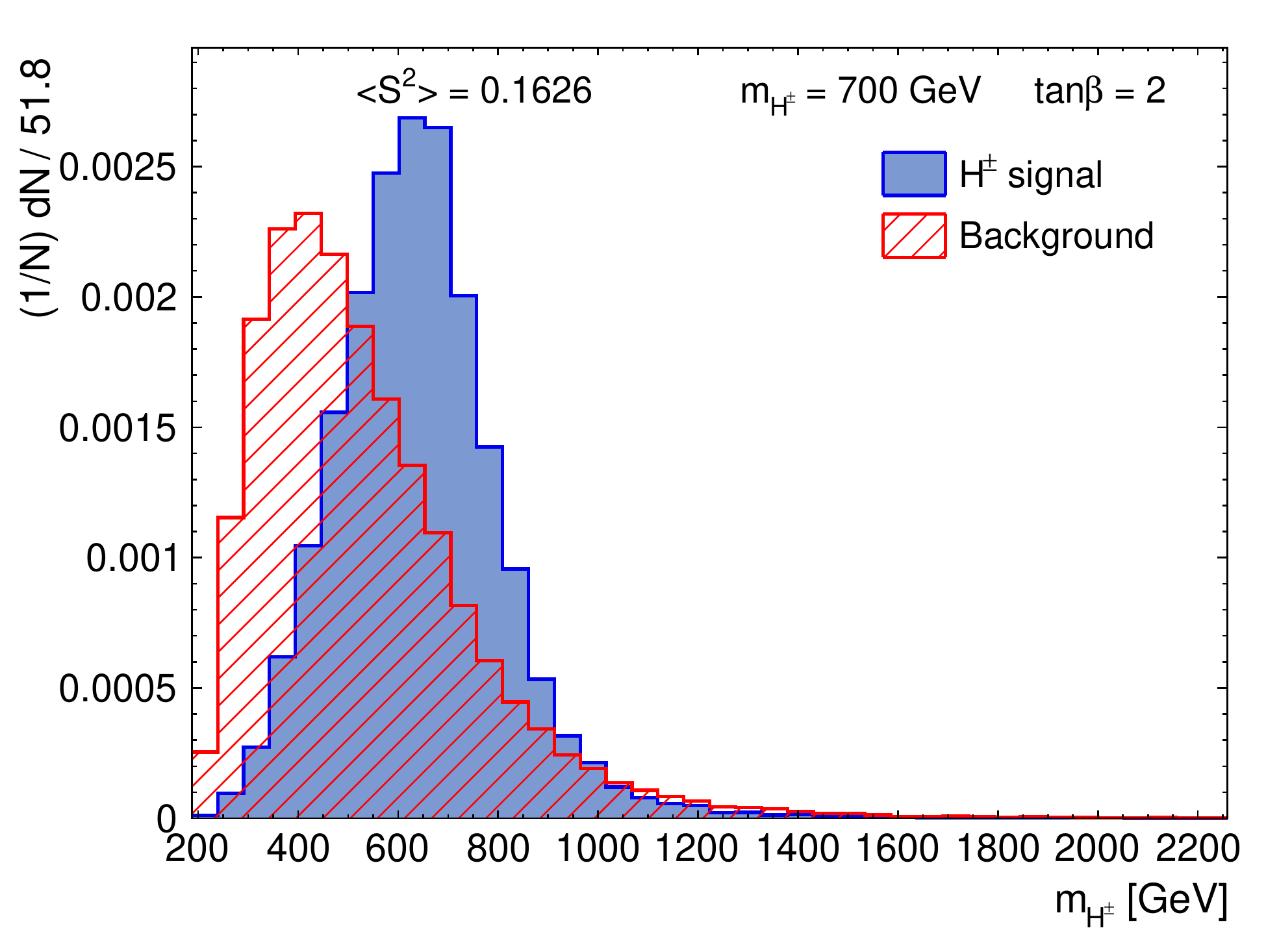}
\includegraphics[width=0.32\textwidth]{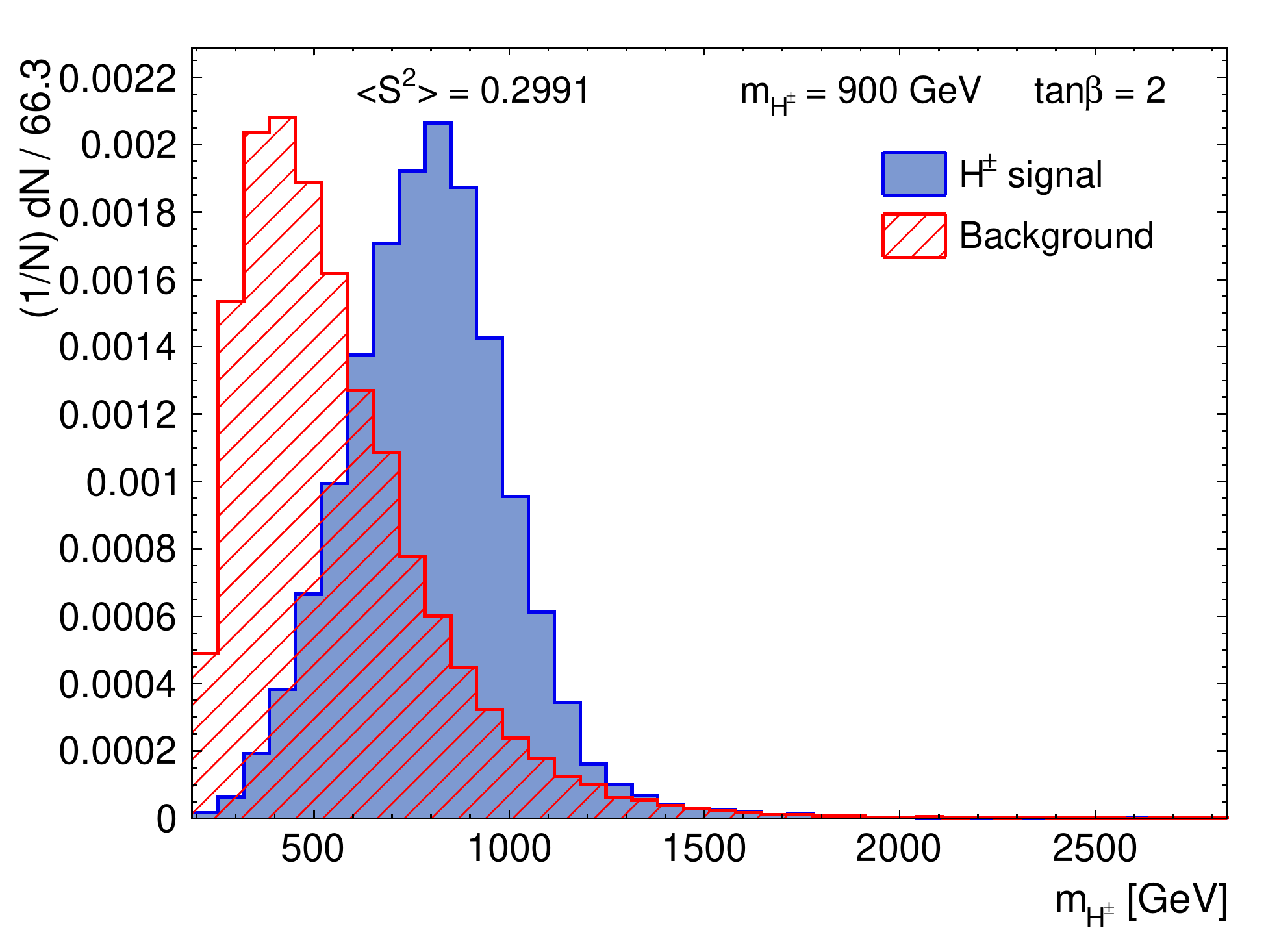}
\caption{Invariant mass of the reconstructed charged Higgs boson at different generated masses for $\tan\beta=2$. The charged Higgs mass is reconstructed well across all masses.}
\label{fig:higgsmass}
\end{figure}

The performance of the classification BDT is shown for several charged Higgs masses in Fig.~\ref{fig:classification_BDT}. While there is only a small separation between signal and background at low $\mhpm$, as $\mhpm$ increases the discriminating power of the BDT increases.  This is also apparent in Table~\ref{tab:performance_class}. As with the reconstruction BDT, there is also an increase in separation at low $\mhpm$, though the effect is smaller here. The cause of the drop in separation at 300 and 400~GeV is because the kinematics of signal and background are most similar at these masses. This can be seen in the invariant mass of the reconstructed charged Higgs, for example, which is shown in Fig.~\ref{fig:higgsmass}. The classification BDT at 200~GeV performs better largely because of the reconstruction BDT weight.

\begin{table}[!ht]
\begin{center}
\begin{tabularx}{0.7\textwidth}{c | Z Z Z Z Z Z Z}
{} & \multicolumn{7}{c}{$\tan\beta$}\\
$\mhpm$[GeV]  ~ &   1  &  2   &    5   &   10  &    15   &   30   &   60 \\\hline
200  & 0.19 & 0.19 & 0.20 & 0.22 & 0.23 & 0.23 & 0.22 \\
300 & 0.15 & 0.15 & 0.15 & 0.16 & 0.17 & 0.17 & 0.16 \\
400 & 0.10 & 0.10 & 0.10 & 0.11 & 0.12 & 0.12 & 0.11 \\
500 & 0.15 & 0.15 & 0.14 & 0.14 & 0.15 & 0.15 & 0.14 \\
600 &  0.23 & 0.24 & 0.23 & 0.21 & 0.22 & 0.22 & 0.21 \\
700 &  0.31  & 0.33 & 0.32 & 0.30 & 0.30 & 0.30 & 0.29 \\
800 &  0.40  & 0.41 & 0.40 & 0.38 & 0.38 & 0.39 & 0.36 \\
900 &  0.46  & 0.49 & 0.48 & 0.46 & 0.46 & 0.46 & 0.43 \\
1000 & 0.52  & 0.55 & 0.54 & 0.52 & 0.52 & 0.52 & 0.49
\end{tabularx}
\caption{Separation $\langle S^2\rangle$ between signal and background in the classification BDT.}
\label{tab:performance_class}
\end{center}
\end{table}

The dependence of separation on $\tan\beta$ is mild for the entire mass range. One of the observables with the largest $\tan\beta$ dependence in the classification BDT is $\cos\theta(l_{H},b_{H})$, shown in Appendix~\ref{app:tanbeta}.

\subsection{LHC sensitivity}

\begin{figure}[!ht]
\includegraphics[width=0.38\textwidth]{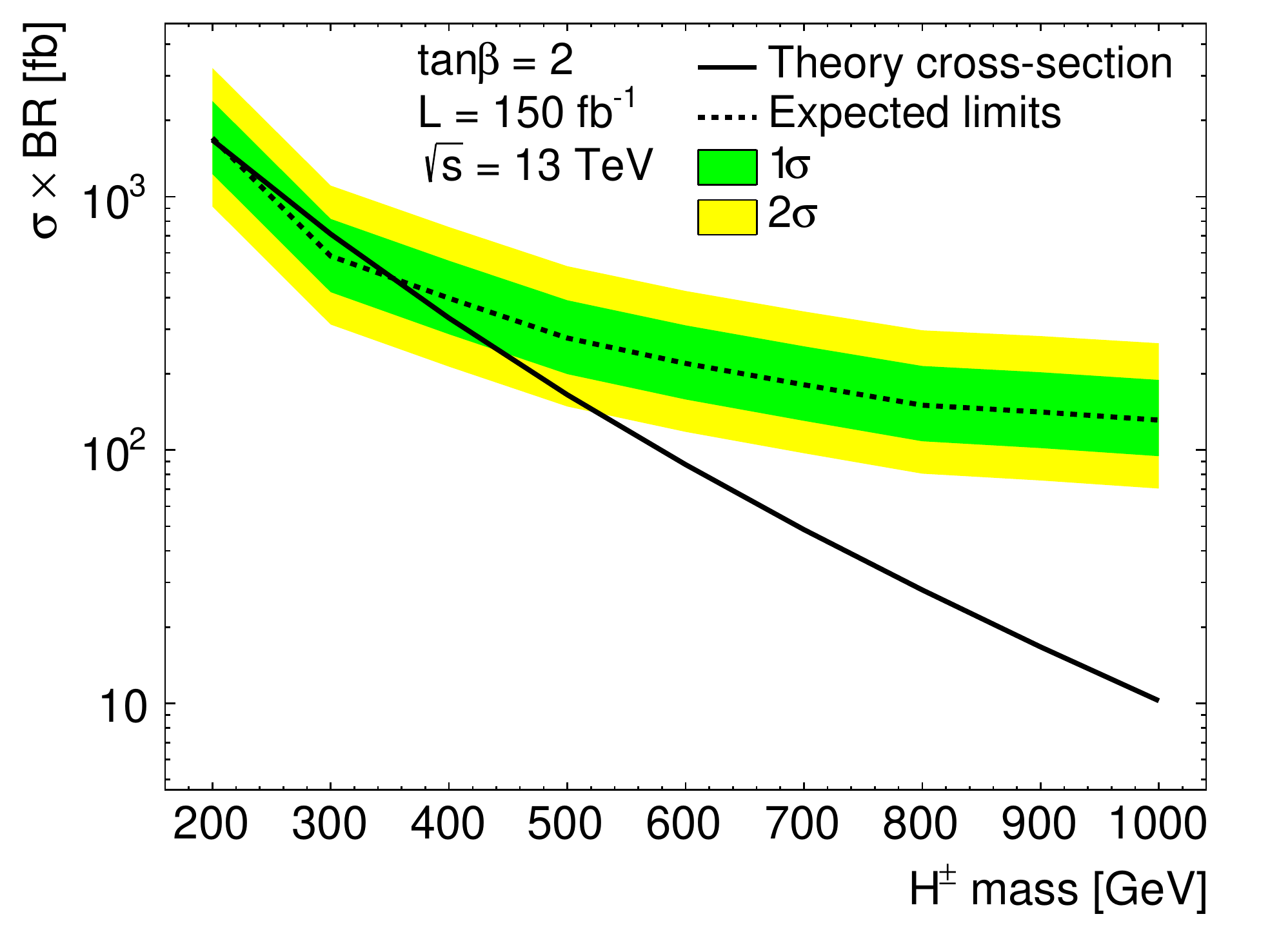}
\includegraphics[width=0.38\textwidth]{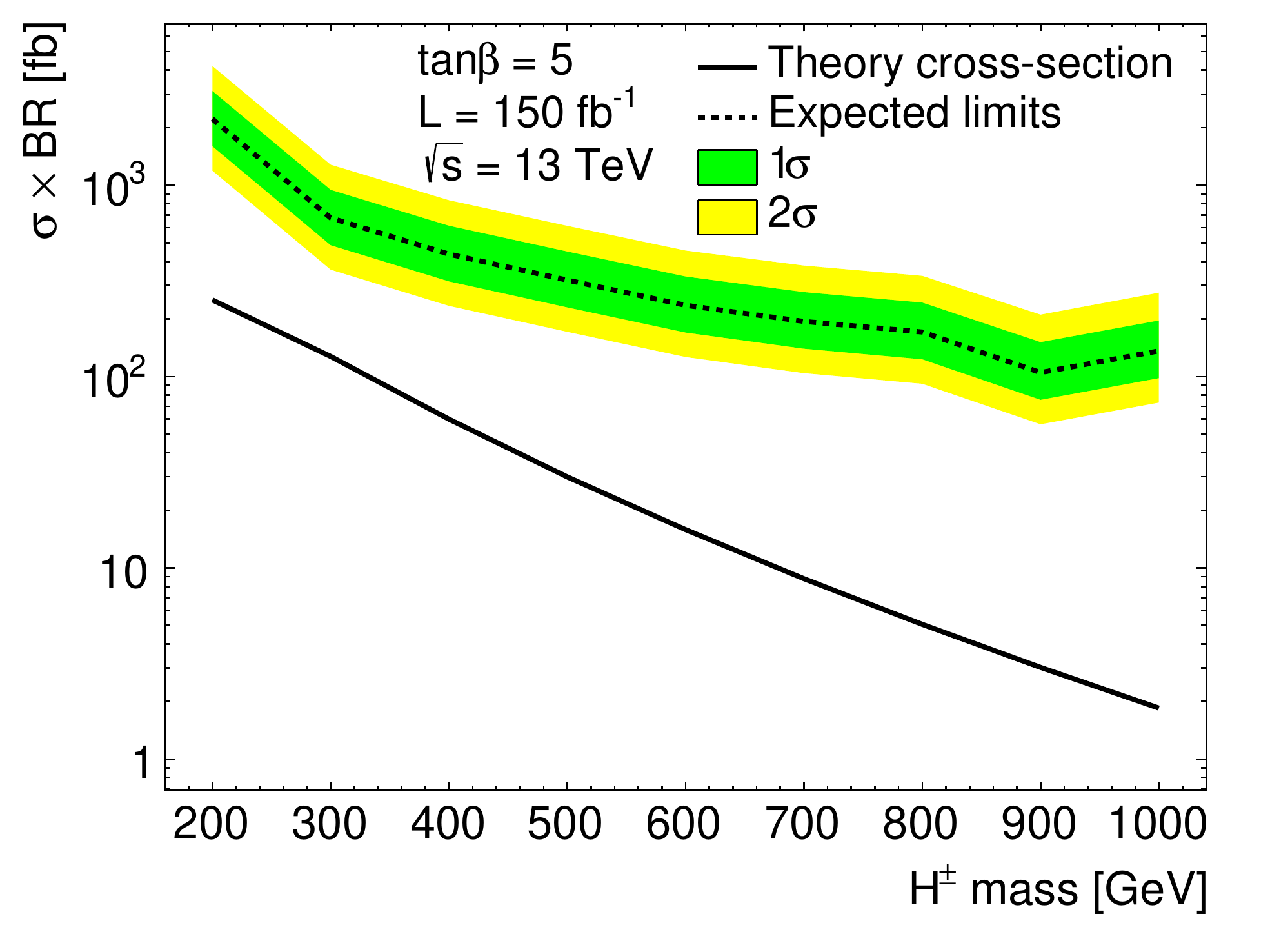}
\includegraphics[width=0.38\textwidth]{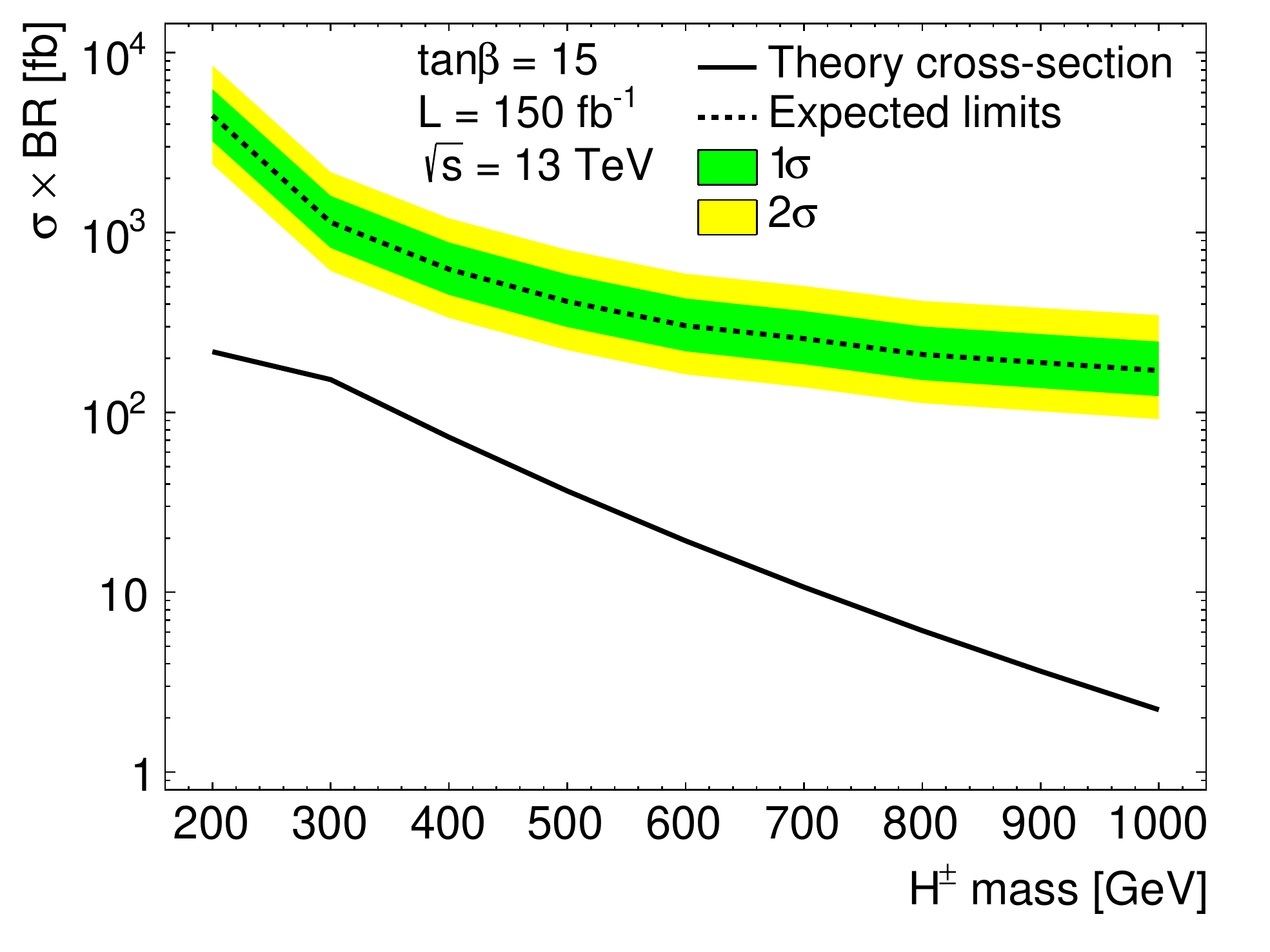}
\includegraphics[width=0.38\textwidth]{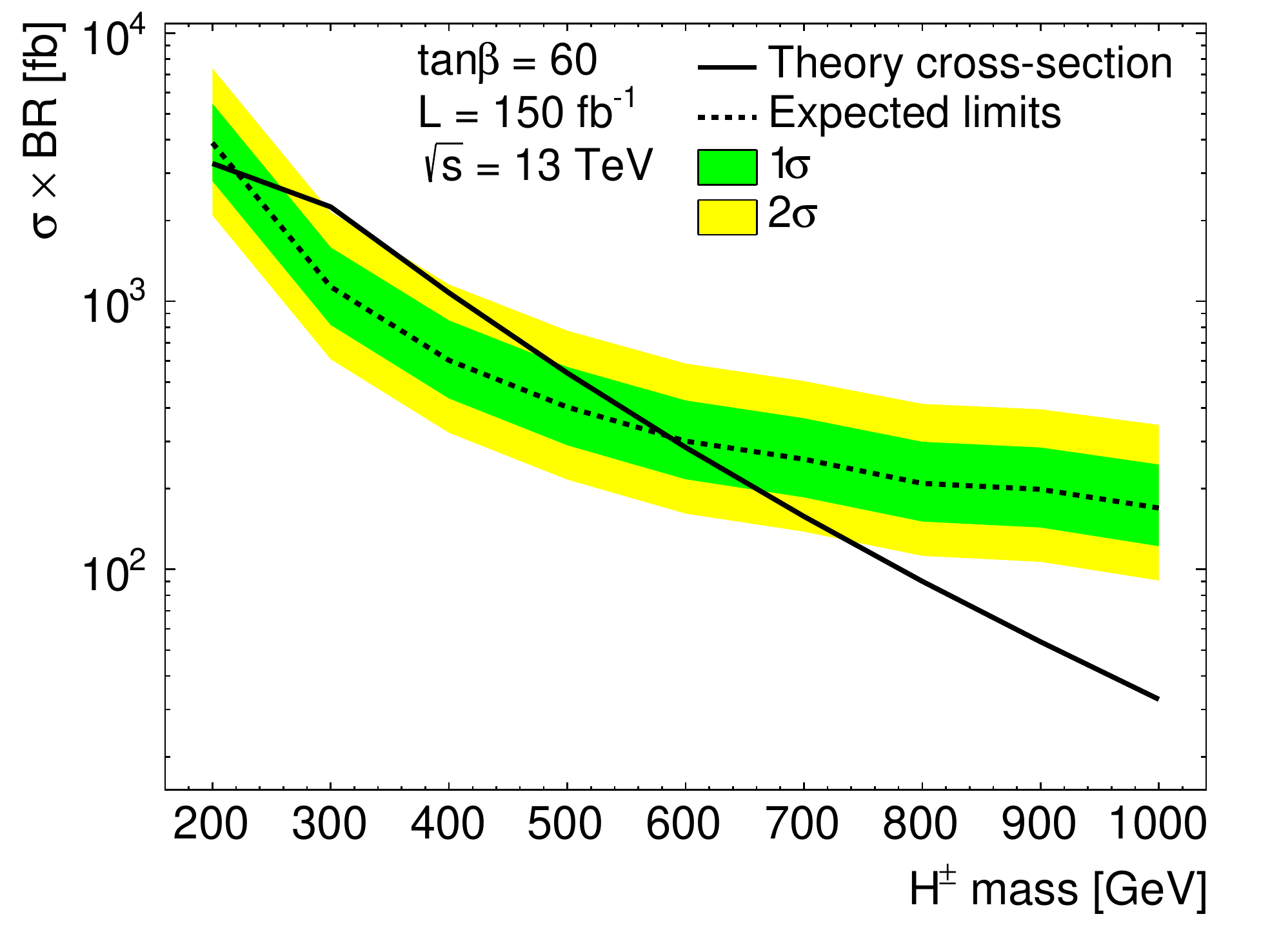}
\caption{Exclusion limits on $\sigma\times BR$ for $\tan\beta = 2,5,15,60$.}
\label{fig:limits}
\end{figure}

\begin{figure}[!ht]
\includegraphics[width=0.45\textwidth]{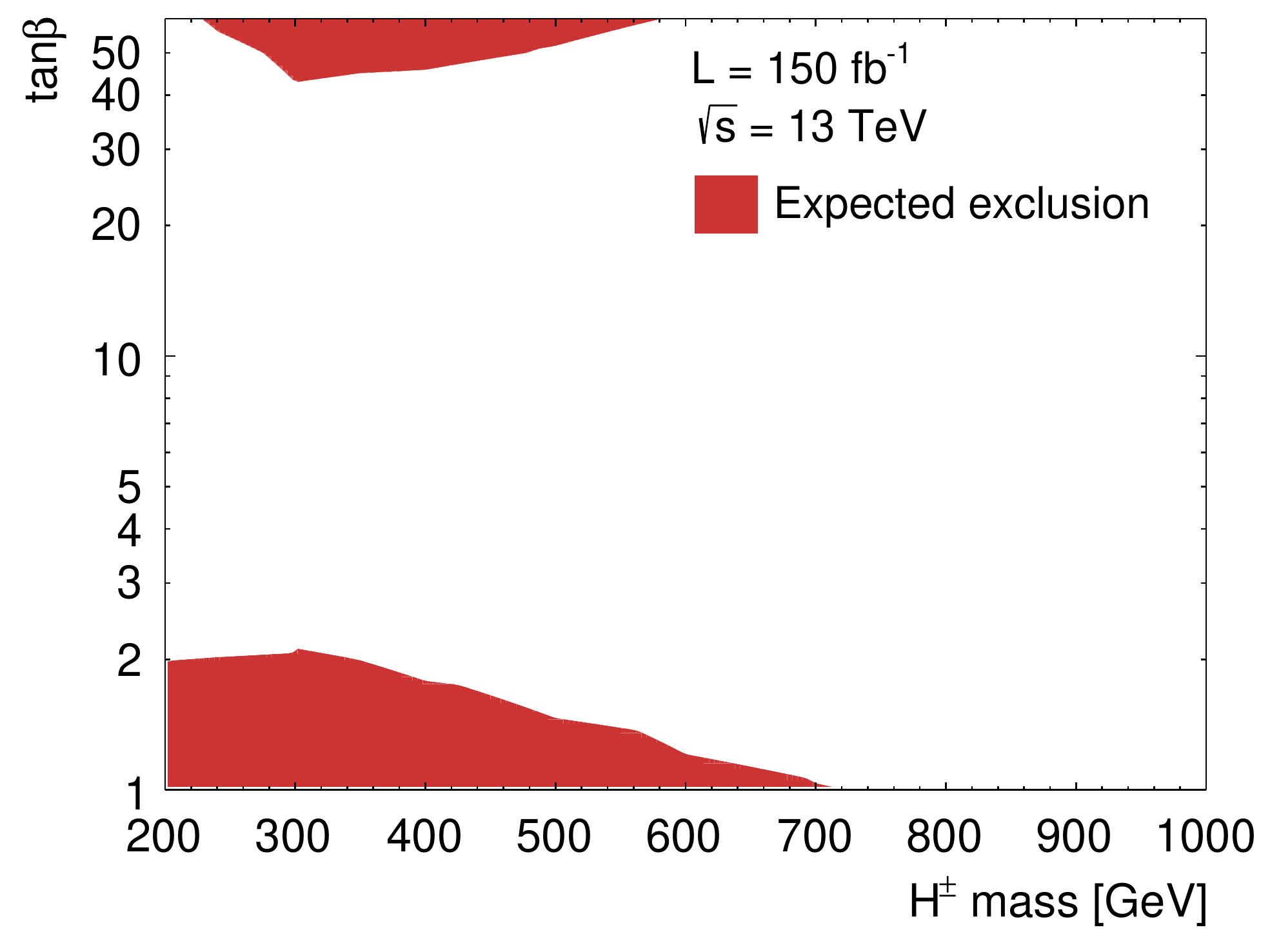}
\caption{Exclusion limits for the MS-2HDM at the 13 TeV LHC with an integrated luminosity of 150 fb$^{-1}$.}
\label{fig:exclusion}
\end{figure}

We determine the sensitivity of this analysis setup at the LHC. As a benchmark, we assume an integrated luminosity of $150~\textrm{fb}^{-1}$, corresponding to LHC Run 2. We derive limits on the $H^\pm$ mediated $pp\to t\bar{t}b\bar{b}$ cross section using the $CL_S$ method~\cite{Read:2002hq}. The samples are split into a set of signal and control regions based on the number of jets and the number of $b$-tagged jets in each event. The signal regions are $\geq 4$j$\ \geq 4b$, $\geq 4$j$3b$ and 3j$3b$; the control regions are 3j$2b$ and 4j$2b$, used to gain a handle on the background in the fit. The derived limits are shown in Fig.~\ref{fig:limits} for several values of $\tan\beta$ and compared with the theoretical signal cross section in the MS-2HDM, given by the LHC Higgs Cross Section Working Group~\cite{Degrande:2015vpa,Flechl:2014wfa,deFlorian:2016spz,PhysRevD.83.055005,PhysRevD.71.115012}.
Figure~\ref{fig:exclusion} shows that, at this luminosity, we can exclude small and large values of $\tan\beta$, for which the $H^\pm t b$ coupling is the largest.

While the theoretical cross section has a strong dependence on $\tan\beta$, the expected limits on $\sigma\times BR$ depend only mildly on $\tan\beta$, as shown in Appendix~\ref{app:tanbeta}.  Consequently, we expect that cross section limits set by this analysis would readily extend to other realisations of the 2HDM, as well as other models with a charged scalar coupling to third-generation quarks.

The limits are compared to those of the most recent ATLAS search in the $H^{\pm}\to tb$ decay channel~\cite{Aaboud:2018cwk}, which does not seek to reconstruct the charged Higgs bosons. This search combines the dilepton and lepton+jets channels. As our study is restricted to the dilepton channel, a fair comparison is made by training the classification BDT described in Ref.~\cite{Aaboud:2018cwk} on our MC samples. The limits obtained from this approach are then compared to the results in Fig.~\ref{fig:limits}. For intermediate masses (400-600~GeV), improvements of 10-15\% may be seen by including the information from reconstruction. As well as improving these limits, a major benefit of the reconstruction is that it provides variables that allow the possibility, in the future, of differential cross-section measurements.

\section{\label{sec:disc} Conclusions}

In this article, we have shown that the charged Higgs boson of the MS-2HDM may be probed in the dileptonic $t\bar{t}$ channel at the LHC for large and small values of $\tan\beta$, and masses as large as $\sim 680~\gev$. This analysis could be extended by including the single lepton and fully hadronic channels. Additionally, there exist contributions to the $t\bar{t}b\bar{b}$ channel from the two neutral MS-2HDM states, $H$ and $A$, as shown in Fig.~\ref{fig:sigma_ttbb_MS2HDM}. Due to differences in their kinematic distributions, it is likely these neutral states would warrant a separate analysis which could be combined with this result. The neutral states could also mediate significant 4$t$ or 4$b$ cross sections, depending on the value of $\tan\beta$.

Furthermore, the analysis presented here generalises to other realisations of the 2HDM and any theory containing a charged scalar with large couplings to third-generation quarks. The small variation in the final limits on $\sigma\times BR$ with $\tan\beta$ suggest that they are dominated by kinematics, rather than other model parameters. This means that they can be readily applied to similar scenarios with little loss of accuracy.  In conclusion, the success of the neutrino weighting procedure, which we implemented in the $t\bar{t}b\bar{b}$ channel
for the first time in this article, give us renewed impetus to apply this procedure to similar channels in the near future.

\section*{Acknowledgements}

\noindent
The authors would like to thank Arghya Choudhury and Bhupal Dev for collaboration
at the early stages of this project. The work of Emily Hanson and Yvonne Peters are supported partially by the ERC research grant 335696 - COLORTTH, and partially by the STFC research grant ST/N504178/1.
Finally, the work of Apostolos Pilaftsis is supported in part by the
Lancaster--Manchester--Sheffield Consortium${}$ for Fundamental
Physics, under STFC research grant ST/L000520/1.

\newpage
\appendix
\section{Dependence on $\tan\beta$} \label{app:tanbeta}
\begin{figure}[!ht]
\includegraphics[width=0.36\textwidth]{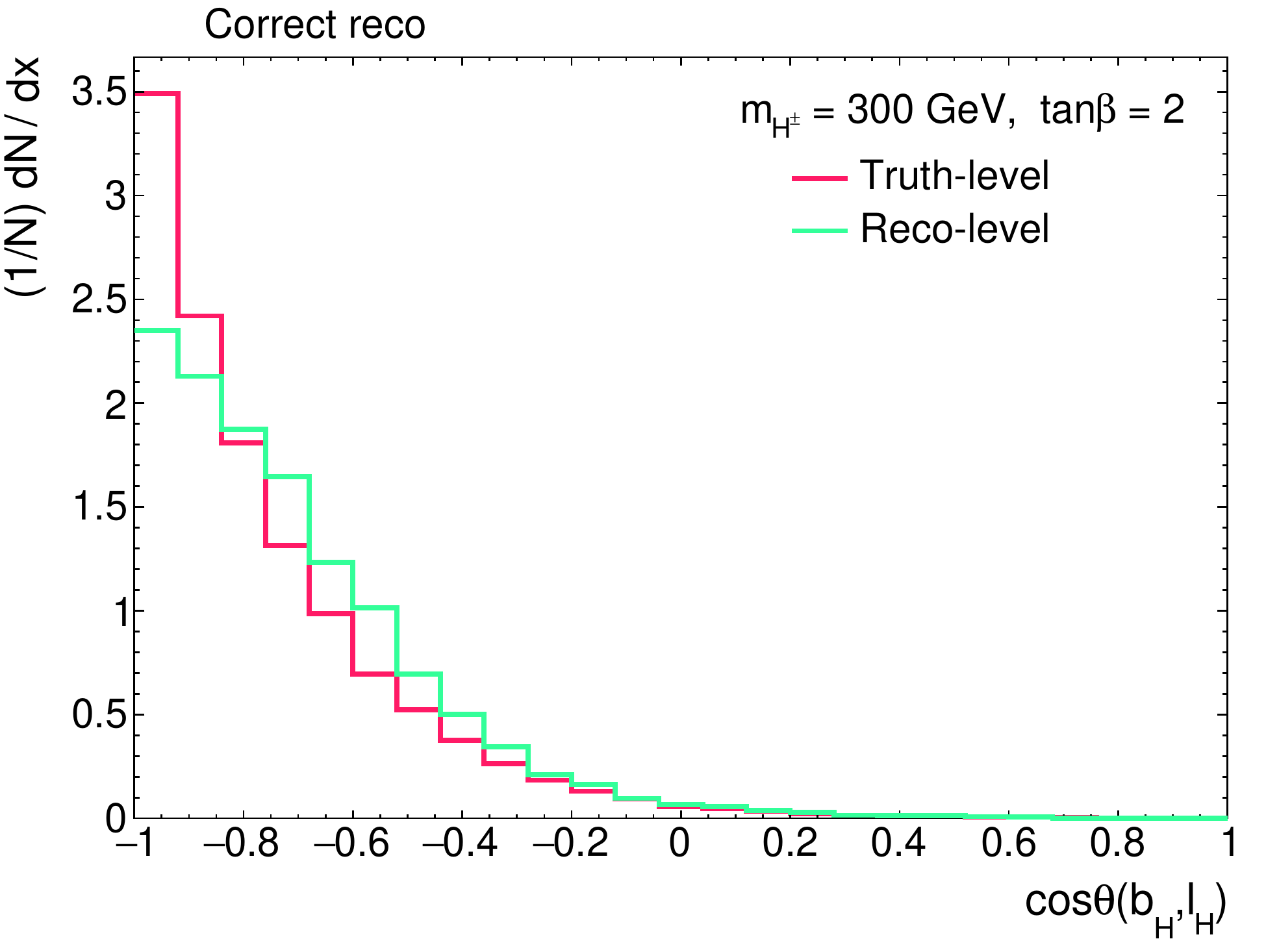}
\includegraphics[width=0.36\textwidth]{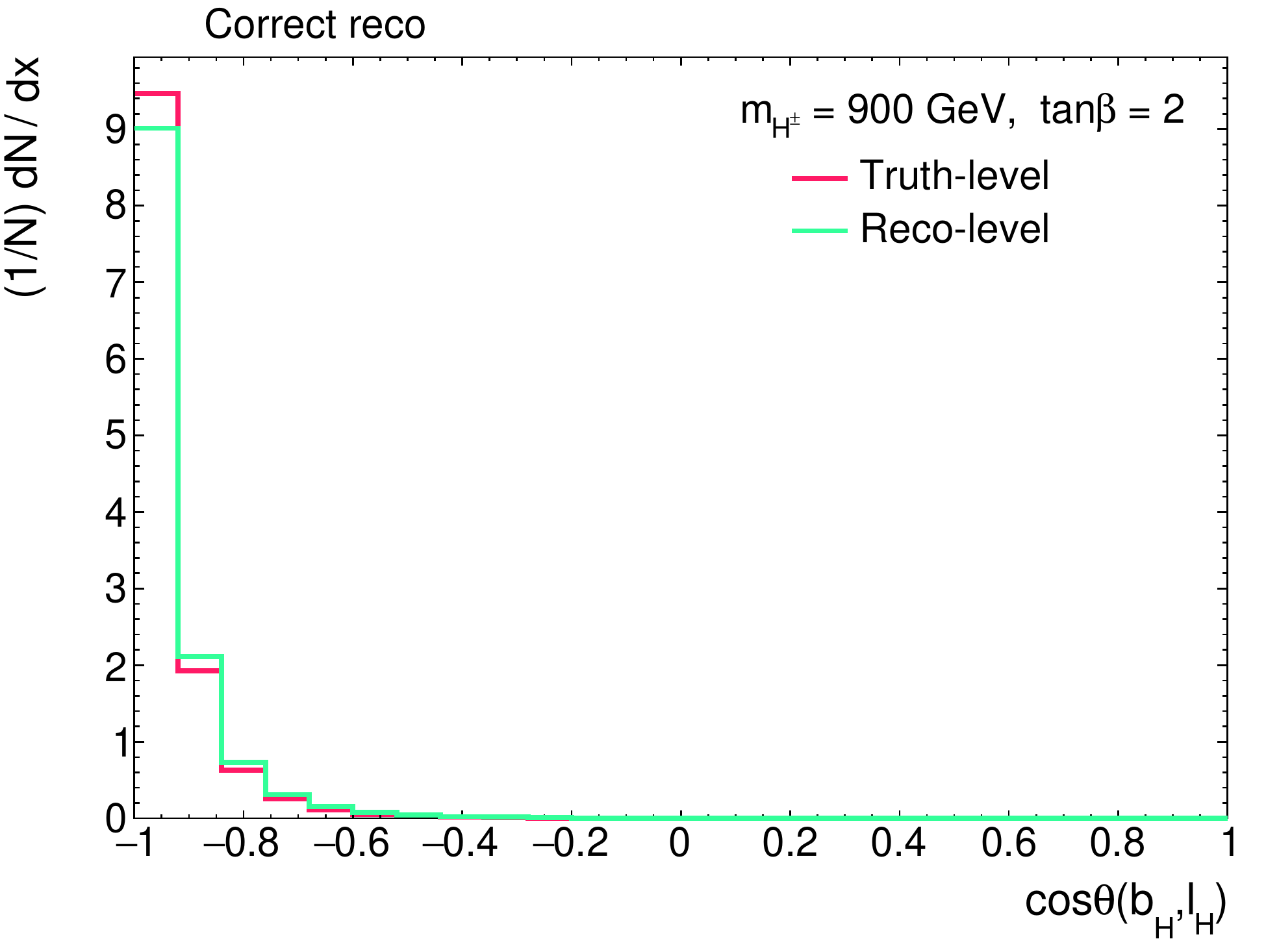}\\
\includegraphics[width=0.36\textwidth]{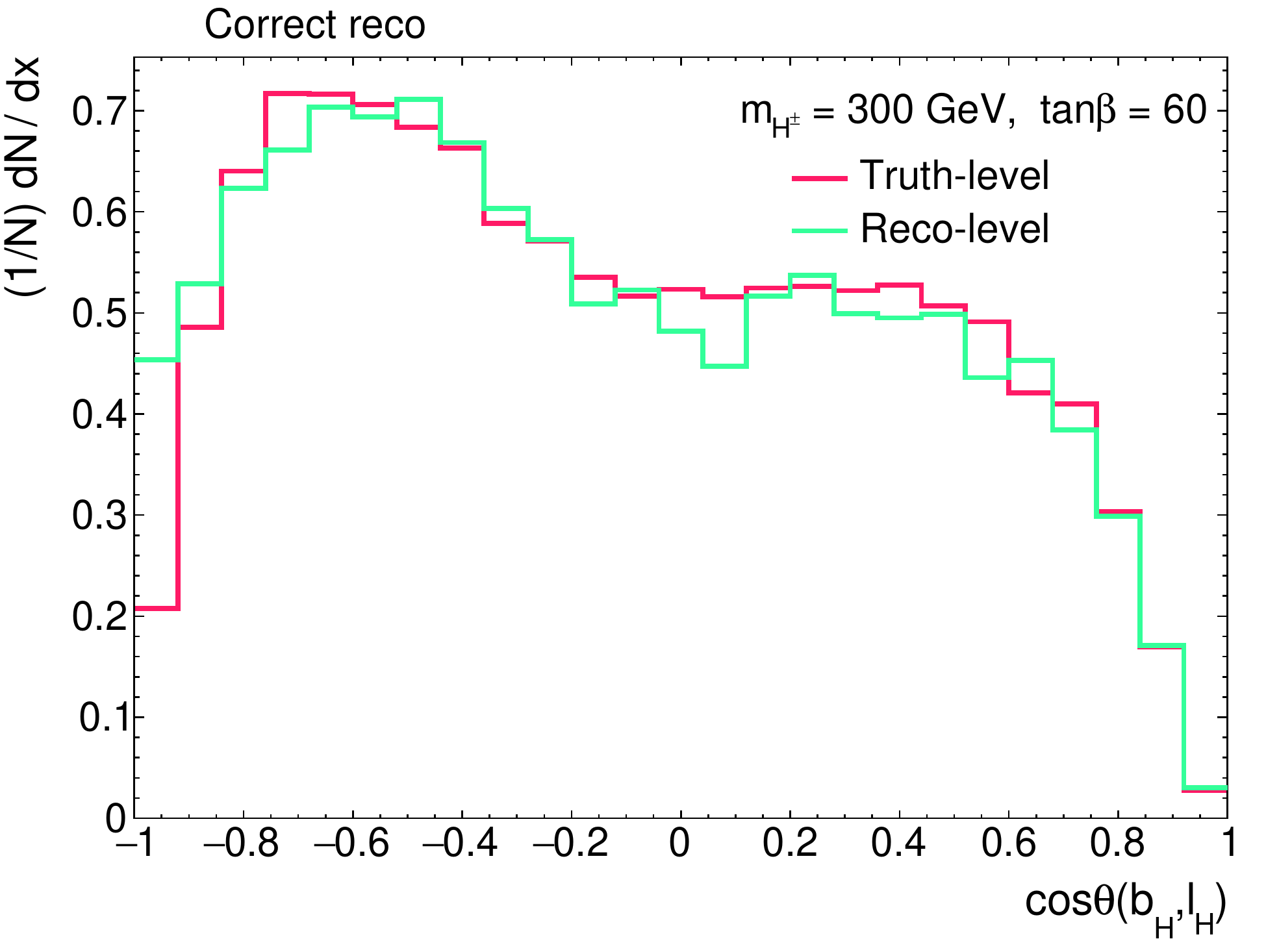}
\includegraphics[width=0.36\textwidth]{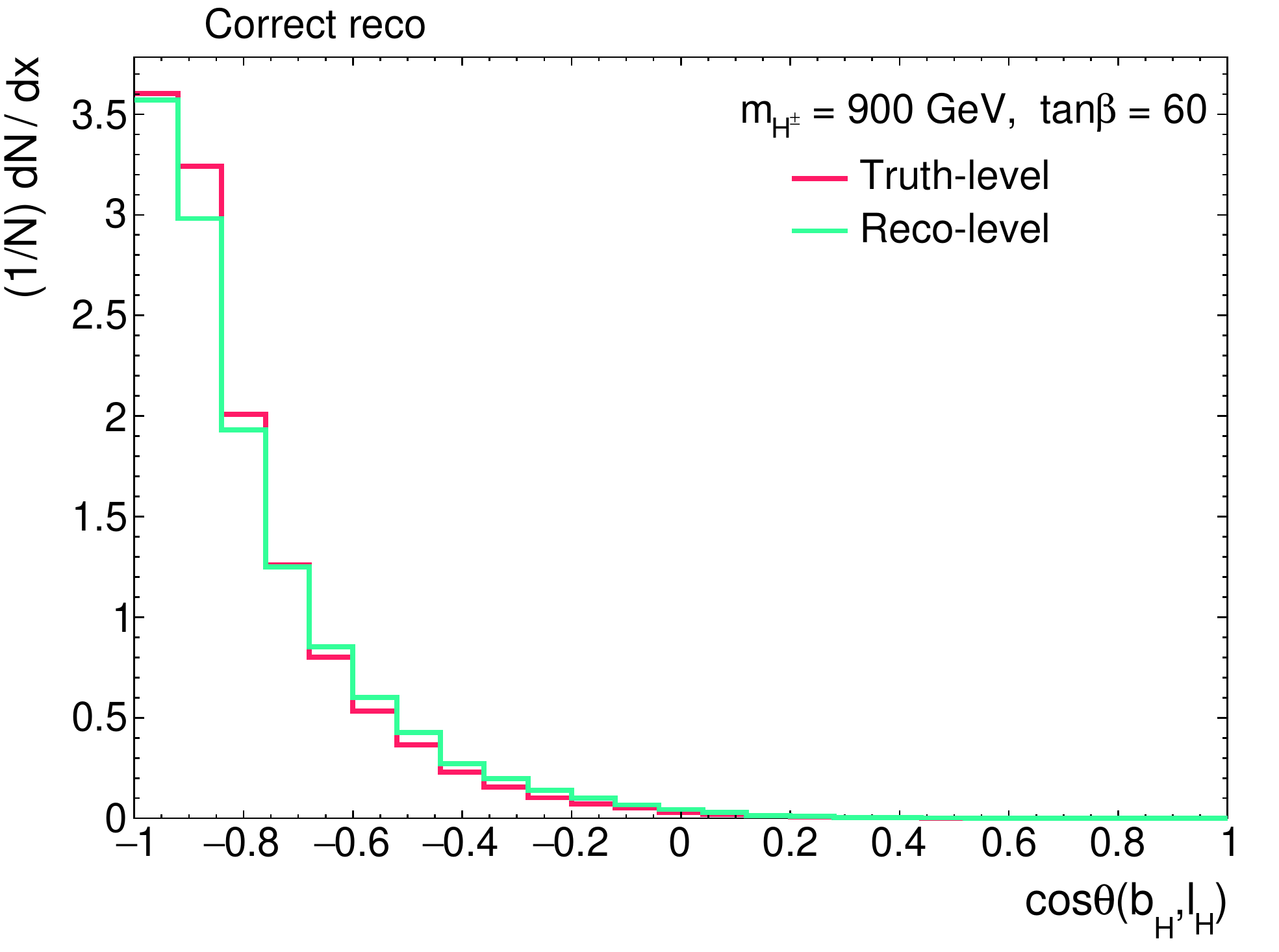}
\caption{The observable $\cos\theta(l_{H},b_{H})$ for the truth partons and for reconstructed jets at $\tan\beta=2,60$ and $\mhpm = 300,900~\gev$, for events where the charged Higgs is correctly reconstructed.}
\label{fig:costheta}
\end{figure}

Figure~\ref{fig:costheta} is an example observable that demonstrates the effect of $\tan\beta$ on the kinematics of the signal. Good agreement is seen in general between the truth $b$-partons and the reconstructed jets (from the reconstruction BDT), improving at higher mass where the reconstruction method has better performance. From the figure, we can see that $l_{H}$ and $b_{H})$ are generally produced back-to-back. At larger $\tan\beta$, the angular spread increases.

\begin{figure}[!ht]
\includegraphics[width=0.36\textwidth]{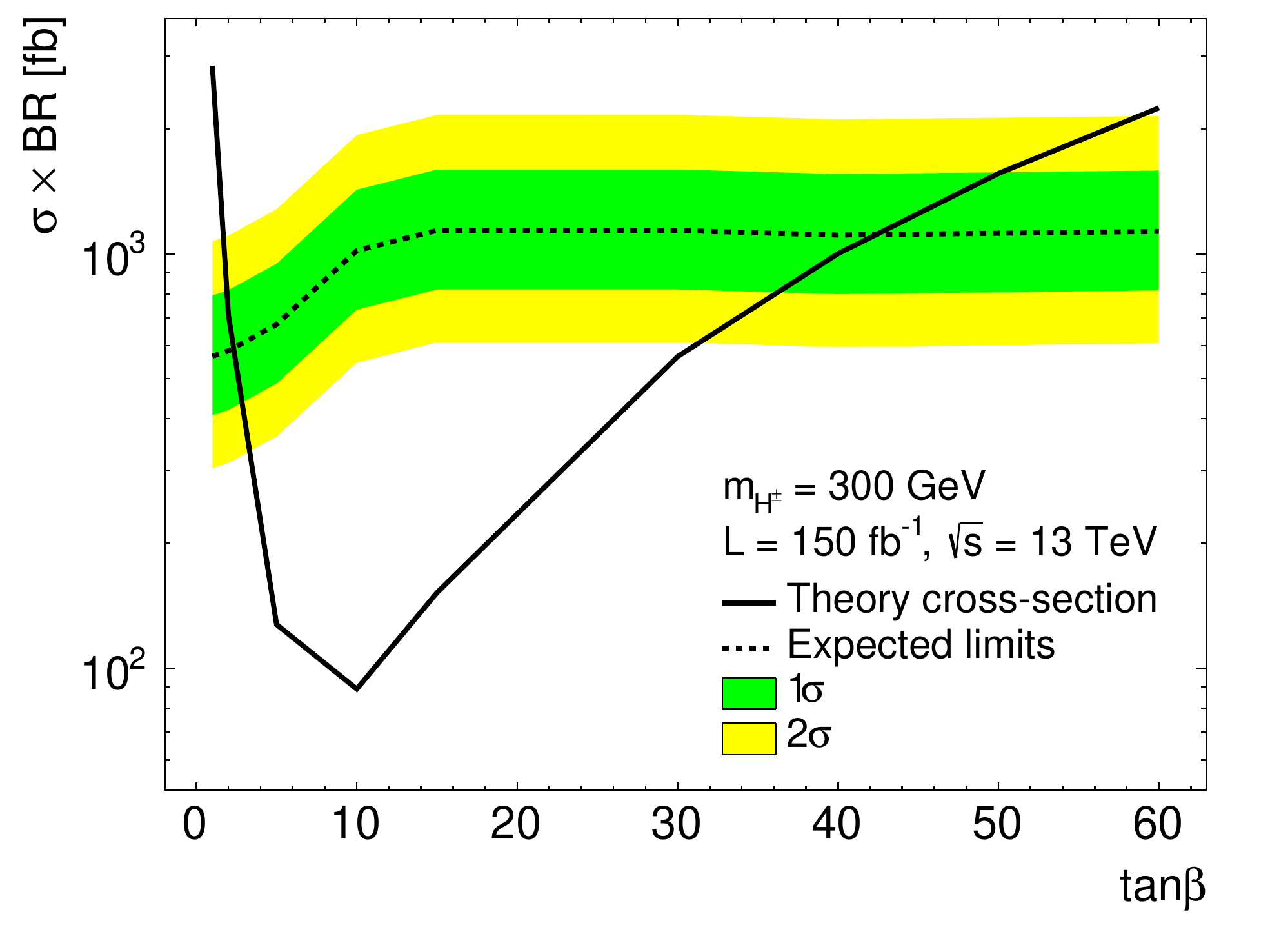}
\includegraphics[width=0.36\textwidth]{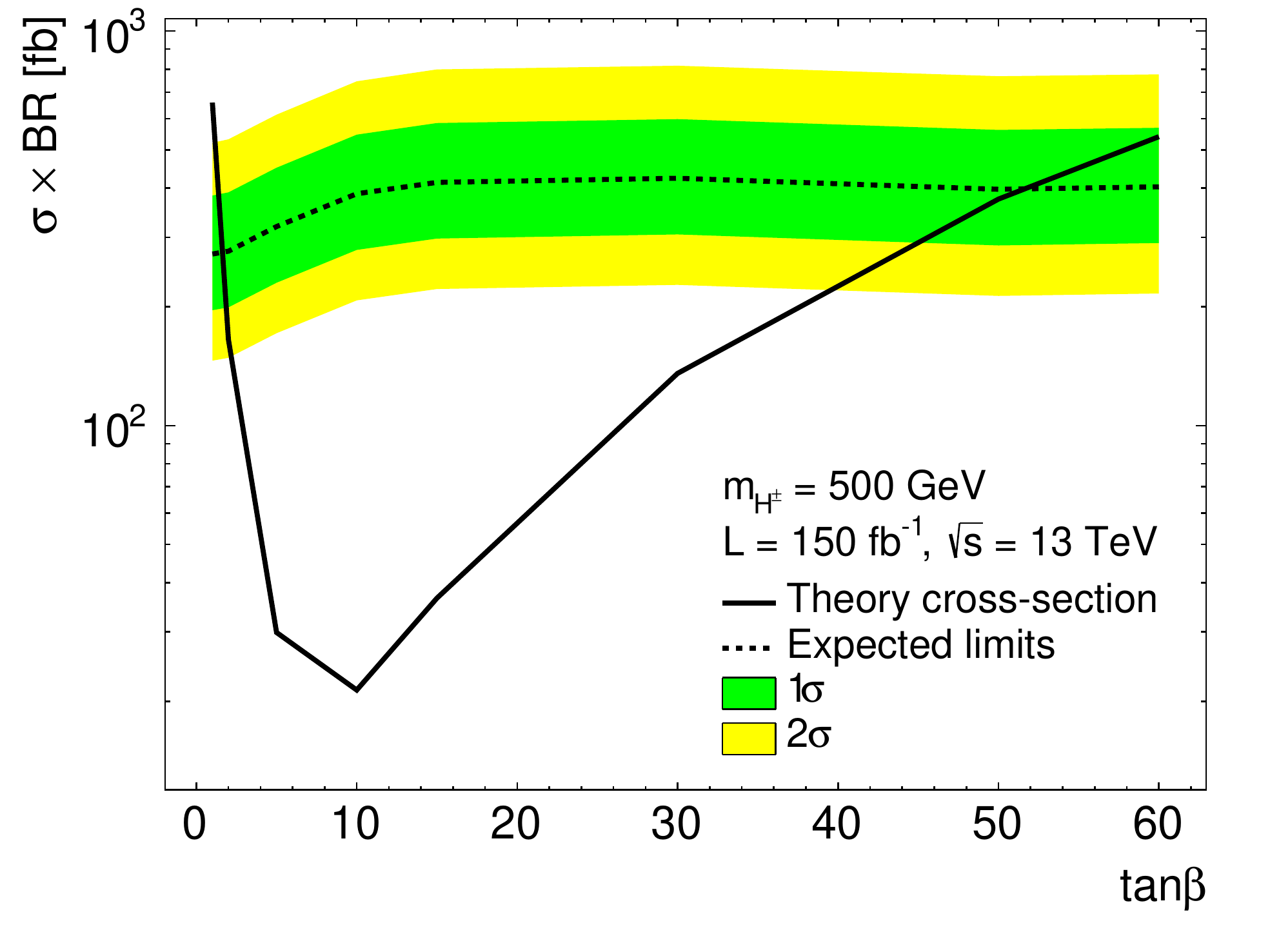}
\includegraphics[width=0.36\textwidth]{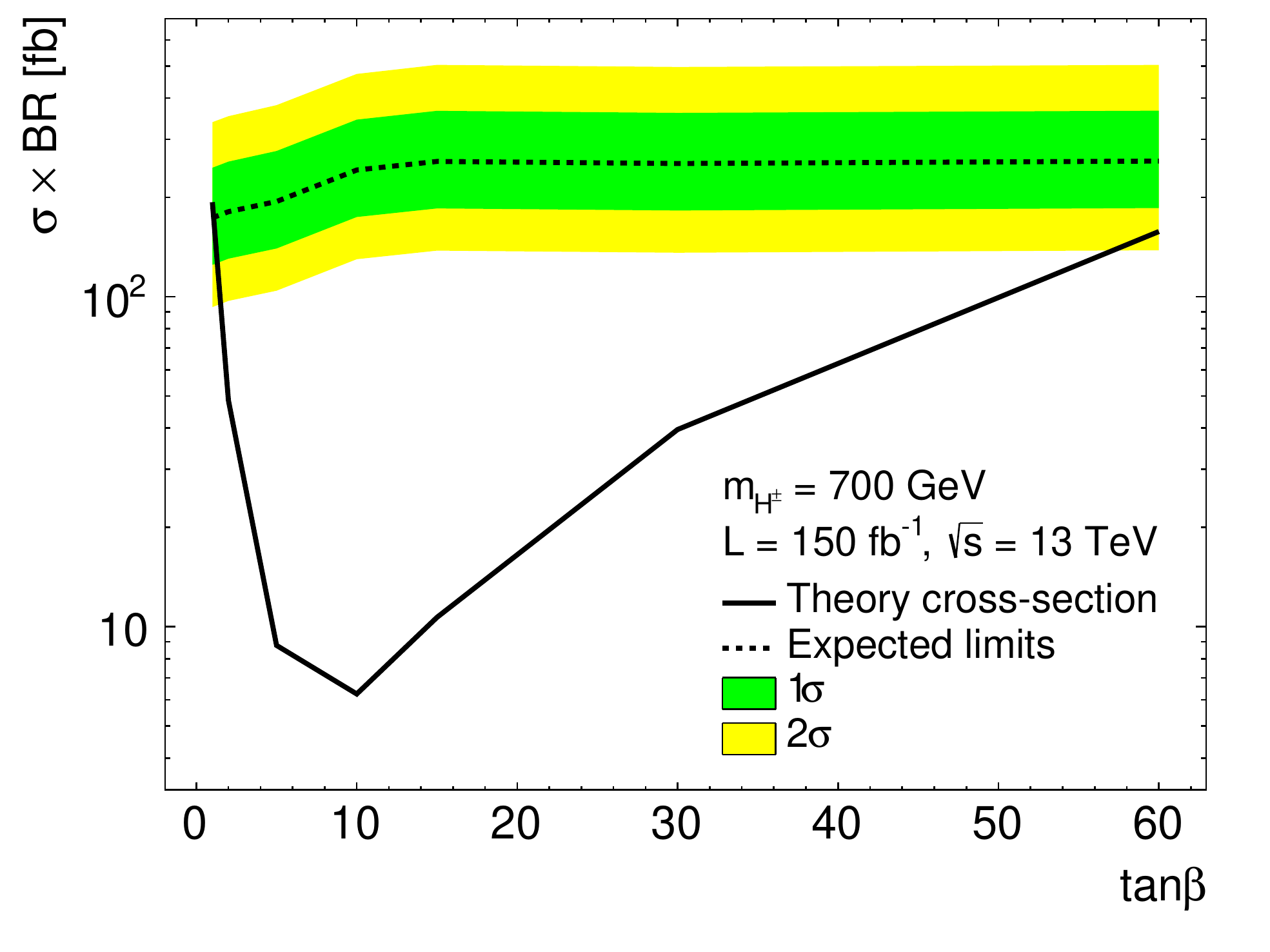}
\includegraphics[width=0.36\textwidth]{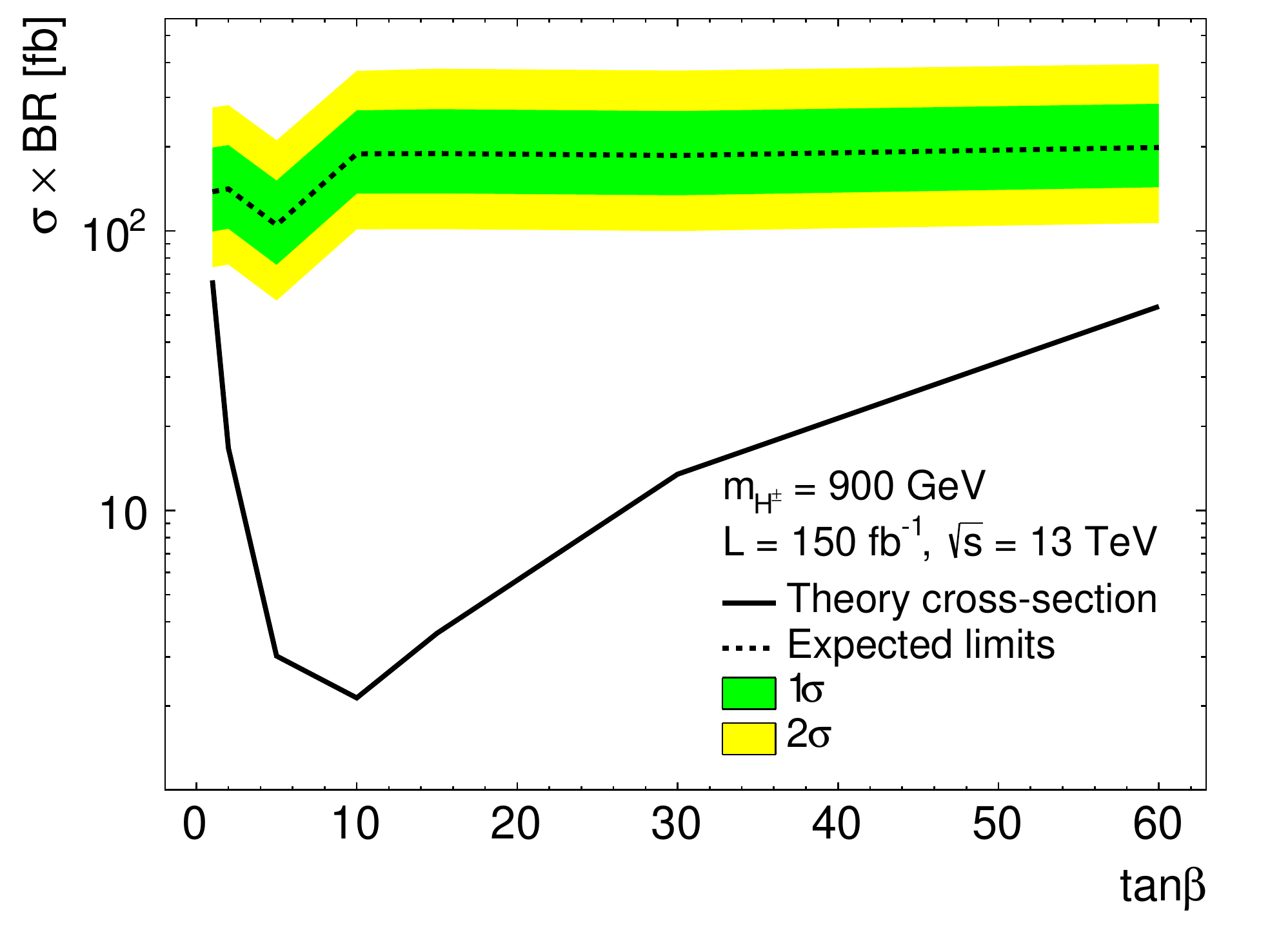}
\caption{Exclusion limits on $\sigma\times BR$ vs $\tan\beta$ for $\mhpm = 300,500,700,900~\gev$.}
\label{fig:limits_tanb}
\end{figure}

Figure~\ref{fig:limits_tanb} shows the dependence of $\tan\beta$ on the limits, shown by the dashed line, for given masses. In general, the $\tan\beta$ dependence is small. The solid line is the theoretical cross-section.

\vfill\eject
%%%%%%%%%%%%%%%%%%%%%%%%%%%%%%%%%%%%%%%%%%%%%%%%%
\newpage
\section*{References}
\bibliography{Charged_Higgs_Paper}

\begin{thebibliography}{58}%
\makeatletter
\providecommand \@ifxundefined [1]{%
 \@ifx{#1\undefined}
}%
\providecommand \@ifnum [1]{%
 \ifnum #1\expandafter \@firstoftwo
 \else \expandafter \@secondoftwo
 \fi
}%
\providecommand \@ifx [1]{%
 \ifx #1\expandafter \@firstoftwo
 \else \expandafter \@secondoftwo
 \fi
}%
\providecommand \natexlab [1]{#1}%
\providecommand \enquote  [1]{``#1''}%
\providecommand \bibnamefont  [1]{#1}%
\providecommand \bibfnamefont [1]{#1}%
\providecommand \citenamefont [1]{#1}%
\providecommand \href@noop [0]{\@secondoftwo}%
\providecommand \href [0]{\begingroup \@sanitize@url \@href}%
\providecommand \@href[1]{\@@startlink{#1}\@@href}%
\providecommand \@@href[1]{\endgroup#1\@@endlink}%
\providecommand \@sanitize@url [0]{\catcode `\\12\catcode `\$12\catcode
  `\&12\catcode `\#12\catcode `\^12\catcode `\_12\catcode `\%12\relax}%
\providecommand \@@startlink[1]{}%
\providecommand \@@endlink[0]{}%
\providecommand \url  [0]{\begingroup\@sanitize@url \@url }%
\providecommand \@url [1]{\endgroup\@href {#1}{\urlprefix }}%
\providecommand \urlprefix  [0]{URL }%
\providecommand \Eprint [0]{\href }%
\providecommand \doibase [0]{http://dx.doi.org/}%
\providecommand \selectlanguage [0]{\@gobble}%
\providecommand \bibinfo  [0]{\@secondoftwo}%
\providecommand \bibfield  [0]{\@secondoftwo}%
\providecommand \translation [1]{[#1]}%
\providecommand \BibitemOpen [0]{}%
\providecommand \bibitemStop [0]{}%
\providecommand \bibitemNoStop [0]{.\EOS\space}%
\providecommand \EOS [0]{\spacefactor3000\relax}%
\providecommand \BibitemShut  [1]{\csname bibitem#1\endcsname}%
\let\auto@bib@innerbib\@empty
%</preamble>
\bibitem [{\citenamefont {Aad}\ \emph {et~al.}(2012)\citenamefont {Aad} \emph
  {et~al.}}]{Aad:2012tfa}%
  \BibitemOpen
  \bibfield  {author} {\bibinfo {author} {\bibfnamefont {G.}~\bibnamefont
  {Aad}} \emph {et~al.} (\bibinfo {collaboration} {ATLAS Collaboration}),\
  }\href {\doibase 10.1016/j.physletb.2012.08.020} {\bibfield  {journal}
  {\bibinfo  {journal} {Phys. Lett.}\ }\textbf {\bibinfo {volume} {B716}},\
  \bibinfo {pages} {1} (\bibinfo {year} {2012})},\ \Eprint
  {http://arxiv.org/abs/1207.7214} {arXiv:1207.7214 [hep-ex]} \BibitemShut
  {NoStop}%
%%CITATION = ARXIV:1207.7214;%%
\bibitem [{\citenamefont {Chatrchyan}\ \emph {et~al.}(2012)\citenamefont
  {Chatrchyan} \emph {et~al.}}]{Chatrchyan:2012xdj}%
  \BibitemOpen
  \bibfield  {author} {\bibinfo {author} {\bibfnamefont {S.}~\bibnamefont
  {Chatrchyan}} \emph {et~al.} (\bibinfo {collaboration} {CMS}),\ }\href
  {\doibase 10.1016/j.physletb.2012.08.021} {\bibfield  {journal} {\bibinfo
  {journal} {Phys. Lett.}\ }\textbf {\bibinfo {volume} {B716}},\ \bibinfo
  {pages} {30} (\bibinfo {year} {2012})},\ \Eprint
  {http://arxiv.org/abs/1207.7235} {arXiv:1207.7235 [hep-ex]} \BibitemShut
  {NoStop}%
%%CITATION = ARXIV:1207.7235;%%
\bibitem [{\citenamefont {Aad}\ \emph {et~al.}(2016)\citenamefont {Aad} \emph
  {et~al.}}]{Khachatryan:2016vau}%
  \BibitemOpen
  \bibfield  {author} {\bibinfo {author} {\bibfnamefont {G.}~\bibnamefont
  {Aad}} \emph {et~al.} (\bibinfo {collaboration} {ATLAS, CMS}),\ }\href
  {\doibase 10.1007/JHEP08(2016)045} {\bibfield  {journal} {\bibinfo  {journal}
  {JHEP}\ }\textbf {\bibinfo {volume} {08}},\ \bibinfo {pages} {045} (\bibinfo
  {year} {2016})},\ \Eprint {http://arxiv.org/abs/1606.02266} {arXiv:1606.02266
  [hep-ex]} \BibitemShut {NoStop}%
%%CITATION = ARXIV:1606.02266;%%
\bibitem [{\citenamefont {Branco}\ \emph {et~al.}(2012)\citenamefont {Branco},
  \citenamefont {Ferreira}, \citenamefont {Lavoura}, \citenamefont {Rebelo},
  \citenamefont {Sher} \emph {et~al.}}]{Branco:2011iw}%
  \BibitemOpen
  \bibfield  {author} {\bibinfo {author} {\bibfnamefont {G.}~\bibnamefont
  {Branco}}, \bibinfo {author} {\bibfnamefont {P.}~\bibnamefont {Ferreira}},
  \bibinfo {author} {\bibfnamefont {L.}~\bibnamefont {Lavoura}}, \bibinfo
  {author} {\bibfnamefont {M.}~\bibnamefont {Rebelo}}, \bibinfo {author}
  {\bibfnamefont {M.}~\bibnamefont {Sher}},  \emph {et~al.},\ }\href {\doibase
  10.1016/j.physrep.2012.02.002} {\bibfield  {journal} {\bibinfo  {journal}
  {Phys. Rept.}\ }\textbf {\bibinfo {volume} {516}},\ \bibinfo {pages} {1}
  (\bibinfo {year} {2012})},\ \Eprint {http://arxiv.org/abs/1106.0034}
  {arXiv:1106.0034 [hep-ph]} \BibitemShut {NoStop}%
%%CITATION = ARXIV:1106.0034;%%
\bibitem [{\citenamefont {Haber}\ and\ \citenamefont
  {Kane}(1985)}]{Haber:1984rc}%
  \BibitemOpen
  \bibfield  {author} {\bibinfo {author} {\bibfnamefont {H.~E.}\ \bibnamefont
  {Haber}}\ and\ \bibinfo {author} {\bibfnamefont {G.~L.}\ \bibnamefont
  {Kane}},\ }\href {\doibase 10.1016/0370-1573(85)90051-1} {\bibfield
  {journal} {\bibinfo  {journal} {Phys. Rept.}\ }\textbf {\bibinfo {volume}
  {117}},\ \bibinfo {pages} {75} (\bibinfo {year} {1985})}\BibitemShut
  {NoStop}%
%%CITATION = PRPLC,117,75;%%
\bibitem [{\citenamefont {Pilaftsis}\ and\ \citenamefont
  {Wagner}(1999)}]{Pilaftsis:1999qt}%
  \BibitemOpen
  \bibfield  {author} {\bibinfo {author} {\bibfnamefont {A.}~\bibnamefont
  {Pilaftsis}}\ and\ \bibinfo {author} {\bibfnamefont {C.~E.~M.}\ \bibnamefont
  {Wagner}},\ }\href {\doibase 10.1016/S0550-3213(99)00261-8} {\bibfield
  {journal} {\bibinfo  {journal} {Nucl. Phys.}\ }\textbf {\bibinfo {volume}
  {B553}},\ \bibinfo {pages} {3} (\bibinfo {year} {1999})},\ \Eprint
  {http://arxiv.org/abs/hep-ph/9902371} {arXiv:hep-ph/9902371 [hep-ph]}
  \BibitemShut {NoStop}%
%%CITATION = HEP-PH/9902371;%%
\bibitem [{\citenamefont {Djouadi}(2008)}]{Djouadi:2005gj}%
  \BibitemOpen
  \bibfield  {author} {\bibinfo {author} {\bibfnamefont {A.}~\bibnamefont
  {Djouadi}},\ }\href {\doibase 10.1016/j.physrep.2007.10.005} {\bibfield
  {journal} {\bibinfo  {journal} {Phys. Rept.}\ }\textbf {\bibinfo {volume}
  {459}},\ \bibinfo {pages} {1} (\bibinfo {year} {2008})},\ \Eprint
  {http://arxiv.org/abs/hep-ph/0503173} {arXiv:hep-ph/0503173 [hep-ph]}
  \BibitemShut {NoStop}%
%%CITATION = HEP-PH/0503173;%%
\bibitem [{\citenamefont {Gunion}\ and\ \citenamefont
  {Haber}(2003)}]{Gunion:2002zf}%
  \BibitemOpen
  \bibfield  {author} {\bibinfo {author} {\bibfnamefont {J.~F.}\ \bibnamefont
  {Gunion}}\ and\ \bibinfo {author} {\bibfnamefont {H.~E.}\ \bibnamefont
  {Haber}},\ }\href {\doibase 10.1103/PhysRevD.67.075019} {\bibfield  {journal}
  {\bibinfo  {journal} {Phys. Rev.}\ }\textbf {\bibinfo {volume} {D67}},\
  \bibinfo {pages} {075019} (\bibinfo {year} {2003})},\ \Eprint
  {http://arxiv.org/abs/hep-ph/0207010} {arXiv:hep-ph/0207010 [hep-ph]}
  \BibitemShut {NoStop}%
%%CITATION = HEP-PH/0207010;%%
\bibitem [{\citenamefont {Carena}\ \emph {et~al.}(2014)\citenamefont {Carena},
  \citenamefont {Low}, \citenamefont {Shah},\ and\ \citenamefont
  {Wagner}}]{Carena:2013ooa}%
  \BibitemOpen
  \bibfield  {author} {\bibinfo {author} {\bibfnamefont {M.}~\bibnamefont
  {Carena}}, \bibinfo {author} {\bibfnamefont {I.}~\bibnamefont {Low}},
  \bibinfo {author} {\bibfnamefont {N.~R.}\ \bibnamefont {Shah}}, \ and\
  \bibinfo {author} {\bibfnamefont {C.~E.~M.}\ \bibnamefont {Wagner}},\ }\href
  {\doibase 10.1007/JHEP04(2014)015} {\bibfield  {journal} {\bibinfo  {journal}
  {JHEP}\ }\textbf {\bibinfo {volume} {04}},\ \bibinfo {pages} {015} (\bibinfo
  {year} {2014})},\ \Eprint {http://arxiv.org/abs/1310.2248} {arXiv:1310.2248
  [hep-ph]} \BibitemShut {NoStop}%
%%CITATION = ARXIV:1310.2248;%%
\bibitem [{\citenamefont {Bhupal~Dev}\ and\ \citenamefont
  {Pilaftsis}(2014)}]{Dev:2014yca}%
  \BibitemOpen
  \bibfield  {author} {\bibinfo {author} {\bibfnamefont {P.~S.}\ \bibnamefont
  {Bhupal~Dev}}\ and\ \bibinfo {author} {\bibfnamefont {A.}~\bibnamefont
  {Pilaftsis}},\ }\href {\doibase 10.1007/JHEP11(2015)147,
  10.1007/JHEP12(2014)024} {\bibfield  {journal} {\bibinfo  {journal} {JHEP}\
  }\textbf {\bibinfo {volume} {12}},\ \bibinfo {pages} {024} (\bibinfo {year}
  {2014})},\ \bibinfo {note} {[Erratum: JHEP11,147(2015)]},\ \Eprint
  {http://arxiv.org/abs/1408.3405} {arXiv:1408.3405 [hep-ph]} \BibitemShut
  {NoStop}%
%%CITATION = ARXIV:1408.3405;%%
\bibitem [{\citenamefont {Bernon}\ \emph {et~al.}(2015)\citenamefont {Bernon},
  \citenamefont {Gunion}, \citenamefont {Haber}, \citenamefont {Jiang},\ and\
  \citenamefont {Kraml}}]{Bernon:2015qea}%
  \BibitemOpen
  \bibfield  {author} {\bibinfo {author} {\bibfnamefont {J.}~\bibnamefont
  {Bernon}}, \bibinfo {author} {\bibfnamefont {J.~F.}\ \bibnamefont {Gunion}},
  \bibinfo {author} {\bibfnamefont {H.~E.}\ \bibnamefont {Haber}}, \bibinfo
  {author} {\bibfnamefont {Y.}~\bibnamefont {Jiang}}, \ and\ \bibinfo {author}
  {\bibfnamefont {S.}~\bibnamefont {Kraml}},\ }\href {\doibase
  10.1103/PhysRevD.92.075004} {\bibfield  {journal} {\bibinfo  {journal} {Phys.
  Rev.}\ }\textbf {\bibinfo {volume} {D92}},\ \bibinfo {pages} {075004}
  (\bibinfo {year} {2015})},\ \Eprint {http://arxiv.org/abs/1507.00933}
  {arXiv:1507.00933 [hep-ph]} \BibitemShut {NoStop}%
%%CITATION = ARXIV:1507.00933;%%
\bibitem [{\citenamefont {Ferreira}\ \emph {et~al.}(2012)\citenamefont
  {Ferreira}, \citenamefont {Santos}, \citenamefont {Sher},\ and\ \citenamefont
  {Silva}}]{Ferreira:2012my}%
  \BibitemOpen
  \bibfield  {author} {\bibinfo {author} {\bibfnamefont {P.~M.}\ \bibnamefont
  {Ferreira}}, \bibinfo {author} {\bibfnamefont {R.}~\bibnamefont {Santos}},
  \bibinfo {author} {\bibfnamefont {M.}~\bibnamefont {Sher}}, \ and\ \bibinfo
  {author} {\bibfnamefont {J.~P.}\ \bibnamefont {Silva}},\ }\href {\doibase
  10.1103/PhysRevD.85.035020} {\bibfield  {journal} {\bibinfo  {journal} {Phys.
  Rev.}\ }\textbf {\bibinfo {volume} {D85}},\ \bibinfo {pages} {035020}
  (\bibinfo {year} {2012})},\ \Eprint {http://arxiv.org/abs/1201.0019}
  {arXiv:1201.0019 [hep-ph]} \BibitemShut {NoStop}%
%%CITATION = ARXIV:1201.0019;%%
\bibitem [{\citenamefont {Bernon}\ \emph {et~al.}(2016)\citenamefont {Bernon},
  \citenamefont {Gunion}, \citenamefont {Haber}, \citenamefont {Jiang},\ and\
  \citenamefont {Kraml}}]{Bernon:2015wef}%
  \BibitemOpen
  \bibfield  {author} {\bibinfo {author} {\bibfnamefont {J.}~\bibnamefont
  {Bernon}}, \bibinfo {author} {\bibfnamefont {J.~F.}\ \bibnamefont {Gunion}},
  \bibinfo {author} {\bibfnamefont {H.~E.}\ \bibnamefont {Haber}}, \bibinfo
  {author} {\bibfnamefont {Y.}~\bibnamefont {Jiang}}, \ and\ \bibinfo {author}
  {\bibfnamefont {S.}~\bibnamefont {Kraml}},\ }\href {\doibase
  10.1103/PhysRevD.93.035027} {\bibfield  {journal} {\bibinfo  {journal} {Phys.
  Rev.}\ }\textbf {\bibinfo {volume} {D93}},\ \bibinfo {pages} {035027}
  (\bibinfo {year} {2016})},\ \Eprint {http://arxiv.org/abs/1511.03682}
  {arXiv:1511.03682 [hep-ph]} \BibitemShut {NoStop}%
%%CITATION = ARXIV:1511.03682;%%
\bibitem [{\citenamefont {Pilaftsis}(2016)}]{Pilaftsis:2016erj}%
  \BibitemOpen
  \bibfield  {author} {\bibinfo {author} {\bibfnamefont {A.}~\bibnamefont
  {Pilaftsis}},\ }\href {\doibase 10.1103/PhysRevD.93.075012} {\bibfield
  {journal} {\bibinfo  {journal} {Phys. Rev.}\ }\textbf {\bibinfo {volume}
  {D93}},\ \bibinfo {pages} {075012} (\bibinfo {year} {2016})},\ \Eprint
  {http://arxiv.org/abs/1602.02017} {arXiv:1602.02017 [hep-ph]} \BibitemShut
  {NoStop}%
%%CITATION = ARXIV:1602.02017;%%
\bibitem [{\citenamefont {Grzadkowski}\ \emph {et~al.}(2018)\citenamefont
  {Grzadkowski}, \citenamefont {Haber}, \citenamefont {Ogreid},\ and\
  \citenamefont {Osland}}]{Grzadkowski:2018ohf}%
  \BibitemOpen
  \bibfield  {author} {\bibinfo {author} {\bibfnamefont {B.}~\bibnamefont
  {Grzadkowski}}, \bibinfo {author} {\bibfnamefont {H.~E.}\ \bibnamefont
  {Haber}}, \bibinfo {author} {\bibfnamefont {O.~M.}\ \bibnamefont {Ogreid}}, \
  and\ \bibinfo {author} {\bibfnamefont {P.}~\bibnamefont {Osland}},\
  }\href@noop {} {\  (\bibinfo {year} {2018})},\ \Eprint
  {http://arxiv.org/abs/1808.01472} {arXiv:1808.01472 [hep-ph]} \BibitemShut
  {NoStop}%
%%CITATION = ARXIV:1808.01472;%%
\bibitem [{\citenamefont {Benakli}\ \emph
  {et~al.}(2018{\natexlab{a}})\citenamefont {Benakli}, \citenamefont
  {Goodsell},\ and\ \citenamefont {Williamson}}]{Benakli:2018vqz}%
  \BibitemOpen
  \bibfield  {author} {\bibinfo {author} {\bibfnamefont {K.}~\bibnamefont
  {Benakli}}, \bibinfo {author} {\bibfnamefont {M.~D.}\ \bibnamefont
  {Goodsell}}, \ and\ \bibinfo {author} {\bibfnamefont {S.~L.}\ \bibnamefont
  {Williamson}},\ }\href {\doibase 10.1140/epjc/s10052-018-6125-1} {\bibfield
  {journal} {\bibinfo  {journal} {Eur. Phys. J.}\ }\textbf {\bibinfo {volume}
  {C78}},\ \bibinfo {pages} {658} (\bibinfo {year} {2018}{\natexlab{a}})},\
  \Eprint {http://arxiv.org/abs/1801.08849} {arXiv:1801.08849 [hep-ph]}
  \BibitemShut {NoStop}%
%%CITATION = ARXIV:1801.08849;%%
\bibitem [{\citenamefont {Benakli}\ \emph
  {et~al.}(2018{\natexlab{b}})\citenamefont {Benakli}, \citenamefont {Chen},\
  and\ \citenamefont {Lafforgue-Marmet}}]{Benakli:2018vjk}%
  \BibitemOpen
  \bibfield  {author} {\bibinfo {author} {\bibfnamefont {K.}~\bibnamefont
  {Benakli}}, \bibinfo {author} {\bibfnamefont {Y.}~\bibnamefont {Chen}}, \
  and\ \bibinfo {author} {\bibfnamefont {G.}~\bibnamefont {Lafforgue-Marmet}},\
  }\href@noop {} {\  (\bibinfo {year} {2018}{\natexlab{b}})},\ \Eprint
  {http://arxiv.org/abs/1811.08435} {arXiv:1811.08435 [hep-ph]} \BibitemShut
  {NoStop}%
%%CITATION = ARXIV:1811.08435;%%
\bibitem [{\citenamefont {Darvishi}\ and\ \citenamefont
  {Pilaftsis}(2019)}]{Darvishi:2019ltl}%
  \BibitemOpen
  \bibfield  {author} {\bibinfo {author} {\bibfnamefont {N.}~\bibnamefont
  {Darvishi}}\ and\ \bibinfo {author} {\bibfnamefont {A.}~\bibnamefont
  {Pilaftsis}},\ }\href@noop {} {\  (\bibinfo {year} {2019})},\ \Eprint
  {http://arxiv.org/abs/1904.06723} {arXiv:1904.06723 [hep-ph]} \BibitemShut
  {NoStop}%
%%CITATION = ARXIV:1904.06723;%%
\bibitem [{\citenamefont {Akeroyd}\ \emph {et~al.}(2017)\citenamefont {Akeroyd}
  \emph {et~al.}}]{Akeroyd:2016ymd}%
  \BibitemOpen
  \bibfield  {author} {\bibinfo {author} {\bibfnamefont {A.~G.}\ \bibnamefont
  {Akeroyd}} \emph {et~al.},\ }\href {\doibase 10.1140/epjc/s10052-017-4829-2}
  {\bibfield  {journal} {\bibinfo  {journal} {Eur. Phys. J.}\ }\textbf
  {\bibinfo {volume} {C77}},\ \bibinfo {pages} {276} (\bibinfo {year}
  {2017})},\ \Eprint {http://arxiv.org/abs/1607.01320} {arXiv:1607.01320
  [hep-ph]} \BibitemShut {NoStop}%
%%CITATION = ARXIV:1607.01320;%%
\bibitem [{\citenamefont {Arbey}\ \emph {et~al.}(2018)\citenamefont {Arbey},
  \citenamefont {Mahmoudi}, \citenamefont {Stal},\ and\ \citenamefont
  {Stefaniak}}]{Arbey:2017gmh}%
  \BibitemOpen
  \bibfield  {author} {\bibinfo {author} {\bibfnamefont {A.}~\bibnamefont
  {Arbey}}, \bibinfo {author} {\bibfnamefont {F.}~\bibnamefont {Mahmoudi}},
  \bibinfo {author} {\bibfnamefont {O.}~\bibnamefont {Stal}}, \ and\ \bibinfo
  {author} {\bibfnamefont {T.}~\bibnamefont {Stefaniak}},\ }\href {\doibase
  10.1140/epjc/s10052-018-5651-1} {\bibfield  {journal} {\bibinfo  {journal}
  {Eur. Phys. J.}\ }\textbf {\bibinfo {volume} {C78}},\ \bibinfo {pages} {182}
  (\bibinfo {year} {2018})},\ \Eprint {http://arxiv.org/abs/1706.07414}
  {arXiv:1706.07414 [hep-ph]} \BibitemShut {NoStop}%
%%CITATION = ARXIV:1706.07414;%%
\bibitem [{\citenamefont {Ginzburg}\ and\ \citenamefont
  {Krawczyk}(2005)}]{Ginzburg:2004vp}%
  \BibitemOpen
  \bibfield  {author} {\bibinfo {author} {\bibfnamefont {I.~F.}\ \bibnamefont
  {Ginzburg}}\ and\ \bibinfo {author} {\bibfnamefont {M.}~\bibnamefont
  {Krawczyk}},\ }\href {\doibase 10.1103/PhysRevD.72.115013} {\bibfield
  {journal} {\bibinfo  {journal} {Phys. Rev.}\ }\textbf {\bibinfo {volume}
  {D72}},\ \bibinfo {pages} {115013} (\bibinfo {year} {2005})},\ \Eprint
  {http://arxiv.org/abs/hep-ph/0408011} {arXiv:hep-ph/0408011 [hep-ph]}
  \BibitemShut {NoStop}%
%%CITATION = HEP-PH/0408011;%%
\bibitem [{\citenamefont {Glashow}\ and\ \citenamefont
  {Weinberg}(1977)}]{Glashow:1976nt}%
  \BibitemOpen
  \bibfield  {author} {\bibinfo {author} {\bibfnamefont {S.~L.}\ \bibnamefont
  {Glashow}}\ and\ \bibinfo {author} {\bibfnamefont {S.}~\bibnamefont
  {Weinberg}},\ }\href {\doibase 10.1103/PhysRevD.15.1958} {\bibfield
  {journal} {\bibinfo  {journal} {Phys. Rev.}\ }\textbf {\bibinfo {volume}
  {D15}},\ \bibinfo {pages} {1958} (\bibinfo {year} {1977})}\BibitemShut
  {NoStop}%
%%CITATION = PHRVA,D15,1958;%%
\bibitem [{\citenamefont {Battye}\ \emph {et~al.}(2011)\citenamefont {Battye},
  \citenamefont {Brawn},\ and\ \citenamefont {Pilaftsis}}]{Battye:2011jj}%
  \BibitemOpen
  \bibfield  {author} {\bibinfo {author} {\bibfnamefont {R.~A.}\ \bibnamefont
  {Battye}}, \bibinfo {author} {\bibfnamefont {G.~D.}\ \bibnamefont {Brawn}}, \
  and\ \bibinfo {author} {\bibfnamefont {A.}~\bibnamefont {Pilaftsis}},\ }\href
  {\doibase 10.1007/JHEP08(2011)020} {\bibfield  {journal} {\bibinfo  {journal}
  {JHEP}\ }\textbf {\bibinfo {volume} {08}},\ \bibinfo {pages} {020} (\bibinfo
  {year} {2011})},\ \Eprint {http://arxiv.org/abs/1106.3482} {arXiv:1106.3482
  [hep-ph]} \BibitemShut {NoStop}%
%%CITATION = ARXIV:1106.3482;%%
\bibitem [{\citenamefont {Pilaftsis}(2012)}]{Pilaftsis:2011ed}%
  \BibitemOpen
  \bibfield  {author} {\bibinfo {author} {\bibfnamefont {A.}~\bibnamefont
  {Pilaftsis}},\ }\href {\doibase 10.1016/j.physletb.2011.11.047} {\bibfield
  {journal} {\bibinfo  {journal} {Phys. Lett.}\ }\textbf {\bibinfo {volume}
  {B706}},\ \bibinfo {pages} {465} (\bibinfo {year} {2012})},\ \Eprint
  {http://arxiv.org/abs/1109.3787} {arXiv:1109.3787 [hep-ph]} \BibitemShut
  {NoStop}%
%%CITATION = ARXIV:1109.3787;%%
\bibitem [{\citenamefont {Maniatis}\ \emph {et~al.}(2008)\citenamefont
  {Maniatis}, \citenamefont {von Manteuffel},\ and\ \citenamefont
  {Nachtmann}}]{Maniatis:2007vn}%
  \BibitemOpen
  \bibfield  {author} {\bibinfo {author} {\bibfnamefont {M.}~\bibnamefont
  {Maniatis}}, \bibinfo {author} {\bibfnamefont {A.}~\bibnamefont {von
  Manteuffel}}, \ and\ \bibinfo {author} {\bibfnamefont {O.}~\bibnamefont
  {Nachtmann}},\ }\href {\doibase 10.1140/epjc/s10052-008-0712-5} {\bibfield
  {journal} {\bibinfo  {journal} {Eur. Phys. J.}\ }\textbf {\bibinfo {volume}
  {C57}},\ \bibinfo {pages} {719} (\bibinfo {year} {2008})},\ \Eprint
  {http://arxiv.org/abs/0707.3344} {arXiv:0707.3344 [hep-ph]} \BibitemShut
  {NoStop}%
%%CITATION = ARXIV:0707.3344;%%
\bibitem [{\citenamefont {Ivanov}(2008)}]{Ivanov:2007de}%
  \BibitemOpen
  \bibfield  {author} {\bibinfo {author} {\bibfnamefont {I.~P.}\ \bibnamefont
  {Ivanov}},\ }\href {\doibase 10.1103/PhysRevD.77.015017} {\bibfield
  {journal} {\bibinfo  {journal} {Phys. Rev.}\ }\textbf {\bibinfo {volume}
  {D77}},\ \bibinfo {pages} {015017} (\bibinfo {year} {2008})},\ \Eprint
  {http://arxiv.org/abs/0710.3490} {arXiv:0710.3490 [hep-ph]} \BibitemShut
  {NoStop}%
%%CITATION = ARXIV:0710.3490;%%
\bibitem [{\citenamefont {Nishi}(2008)}]{Nishi:2007dv}%
  \BibitemOpen
  \bibfield  {author} {\bibinfo {author} {\bibfnamefont {C.~C.}\ \bibnamefont
  {Nishi}},\ }\href {\doibase 10.1103/PhysRevD.77.055009} {\bibfield  {journal}
  {\bibinfo  {journal} {Phys. Rev.}\ }\textbf {\bibinfo {volume} {D77}},\
  \bibinfo {pages} {055009} (\bibinfo {year} {2008})},\ \Eprint
  {http://arxiv.org/abs/0712.4260} {arXiv:0712.4260 [hep-ph]} \BibitemShut
  {NoStop}%
%%CITATION = ARXIV:0712.4260;%%
\bibitem [{\citenamefont {Ferreira}\ \emph {et~al.}(2009)\citenamefont
  {Ferreira}, \citenamefont {Haber},\ and\ \citenamefont
  {Silva}}]{Ferreira:2009wh}%
  \BibitemOpen
  \bibfield  {author} {\bibinfo {author} {\bibfnamefont {P.~M.}\ \bibnamefont
  {Ferreira}}, \bibinfo {author} {\bibfnamefont {H.~E.}\ \bibnamefont {Haber}},
  \ and\ \bibinfo {author} {\bibfnamefont {J.~P.}\ \bibnamefont {Silva}},\
  }\href {\doibase 10.1103/PhysRevD.79.116004} {\bibfield  {journal} {\bibinfo
  {journal} {Phys. Rev.}\ }\textbf {\bibinfo {volume} {D79}},\ \bibinfo {pages}
  {116004} (\bibinfo {year} {2009})},\ \Eprint {http://arxiv.org/abs/0902.1537}
  {arXiv:0902.1537 [hep-ph]} \BibitemShut {NoStop}%
%%CITATION = ARXIV:0902.1537;%%
\bibitem [{\citenamefont {Ferreira}\ \emph {et~al.}(2011)\citenamefont
  {Ferreira}, \citenamefont {Haber}, \citenamefont {Maniatis}, \citenamefont
  {Nachtmann},\ and\ \citenamefont {Silva}}]{Ferreira:2010yh}%
  \BibitemOpen
  \bibfield  {author} {\bibinfo {author} {\bibfnamefont {P.~M.}\ \bibnamefont
  {Ferreira}}, \bibinfo {author} {\bibfnamefont {H.~E.}\ \bibnamefont {Haber}},
  \bibinfo {author} {\bibfnamefont {M.}~\bibnamefont {Maniatis}}, \bibinfo
  {author} {\bibfnamefont {O.}~\bibnamefont {Nachtmann}}, \ and\ \bibinfo
  {author} {\bibfnamefont {J.~P.}\ \bibnamefont {Silva}},\ }\href {\doibase
  10.1142/S0217751X11051494} {\bibfield  {journal} {\bibinfo  {journal} {Int.
  J. Mod. Phys.}\ }\textbf {\bibinfo {volume} {A26}},\ \bibinfo {pages} {769}
  (\bibinfo {year} {2011})},\ \Eprint {http://arxiv.org/abs/1010.0935}
  {arXiv:1010.0935 [hep-ph]} \BibitemShut {NoStop}%
%%CITATION = ARXIV:1010.0935;%%
\bibitem [{\citenamefont {Abbiendi}\ \emph {et~al.}(2013)\citenamefont
  {Abbiendi} \emph {et~al.}}]{Abbiendi:2013hk}%
  \BibitemOpen
  \bibfield  {author} {\bibinfo {author} {\bibfnamefont {G.}~\bibnamefont
  {Abbiendi}} \emph {et~al.} (\bibinfo {collaboration} {LEP, DELPHI, OPAL,
  ALEPH, L3}),\ }\href {\doibase 10.1140/epjc/s10052-013-2463-1} {\bibfield
  {journal} {\bibinfo  {journal} {Eur. Phys. J.}\ }\textbf {\bibinfo {volume}
  {C73}},\ \bibinfo {pages} {2463} (\bibinfo {year} {2013})},\ \Eprint
  {http://arxiv.org/abs/1301.6065} {arXiv:1301.6065 [hep-ex]} \BibitemShut
  {NoStop}%
%%CITATION = ARXIV:1301.6065;%%
\bibitem [{\citenamefont {Aaltonen}\ \emph {et~al.}(2009)\citenamefont
  {Aaltonen} \emph {et~al.}}]{Aaltonen:2009ke}%
  \BibitemOpen
  \bibfield  {author} {\bibinfo {author} {\bibfnamefont {T.}~\bibnamefont
  {Aaltonen}} \emph {et~al.} (\bibinfo {collaboration} {CDF}),\ }\href
  {\doibase 10.1103/PhysRevLett.103.101803} {\bibfield  {journal} {\bibinfo
  {journal} {Phys. Rev. Lett.}\ }\textbf {\bibinfo {volume} {103}},\ \bibinfo
  {pages} {101803} (\bibinfo {year} {2009})},\ \Eprint
  {http://arxiv.org/abs/0907.1269} {arXiv:0907.1269 [hep-ex]} \BibitemShut
  {NoStop}%
%%CITATION = ARXIV:0907.1269;%%
\bibitem [{\citenamefont {Abazov}\ \emph {et~al.}(2009)\citenamefont {Abazov}
  \emph {et~al.}}]{Abazov:2009aa}%
  \BibitemOpen
  \bibfield  {author} {\bibinfo {author} {\bibfnamefont {V.~M.}\ \bibnamefont
  {Abazov}} \emph {et~al.} (\bibinfo {collaboration} {D0}),\ }\href {\doibase
  10.1016/j.physletb.2009.11.016} {\bibfield  {journal} {\bibinfo  {journal}
  {Phys. Lett.}\ }\textbf {\bibinfo {volume} {B682}},\ \bibinfo {pages} {278}
  (\bibinfo {year} {2009})},\ \Eprint {http://arxiv.org/abs/0908.1811}
  {arXiv:0908.1811 [hep-ex]} \BibitemShut {NoStop}%
%%CITATION = ARXIV:0908.1811;%%
\bibitem [{\citenamefont {Gutierrez}(2010)}]{Gutierrez:2010zz}%
  \BibitemOpen
  \bibfield  {author} {\bibinfo {author} {\bibfnamefont {P.}~\bibnamefont
  {Gutierrez}} (\bibinfo {collaboration} {CDF, D0}),\ }\bibfield  {booktitle}
  {\emph {\bibinfo {booktitle} {{Proceedings, 3rd International Workshop on
  Prospects for charged Higgs discovery at colliders (CHARGED 2010): Uppsala,
  Sweden, September 27-30, 2010}}},\ }\href {\doibase 10.22323/1.114.0004}
  {\bibfield  {journal} {\bibinfo  {journal} {PoS}\ }\textbf {\bibinfo {volume}
  {CHARGED2010}},\ \bibinfo {pages} {004} (\bibinfo {year} {2010})}\BibitemShut
  {NoStop}%
%%CITATION = POSCI,CHARGED2010,004;%%
\bibitem [{\citenamefont {Aaboud}\ \emph
  {et~al.}(2018{\natexlab{a}})\citenamefont {Aaboud} \emph
  {et~al.}}]{Aaboud:2018gjj}%
  \BibitemOpen
  \bibfield  {author} {\bibinfo {author} {\bibfnamefont {M.}~\bibnamefont
  {Aaboud}} \emph {et~al.} (\bibinfo {collaboration} {ATLAS}),\ }\href
  {\doibase 10.1007/JHEP09(2018)139} {\bibfield  {journal} {\bibinfo  {journal}
  {JHEP}\ }\textbf {\bibinfo {volume} {09}},\ \bibinfo {pages} {139} (\bibinfo
  {year} {2018}{\natexlab{a}})},\ \Eprint {http://arxiv.org/abs/1807.07915}
  {arXiv:1807.07915 [hep-ex]} \BibitemShut {NoStop}%
%%CITATION = ARXIV:1807.07915;%%
\bibitem [{\citenamefont {Khachatryan}\ \emph
  {et~al.}(2015{\natexlab{a}})\citenamefont {Khachatryan} \emph
  {et~al.}}]{Khachatryan:2015qxa}%
  \BibitemOpen
  \bibfield  {author} {\bibinfo {author} {\bibfnamefont {V.}~\bibnamefont
  {Khachatryan}} \emph {et~al.} (\bibinfo {collaboration} {CMS}),\ }\href
  {\doibase 10.1007/JHEP11(2015)018} {\bibfield  {journal} {\bibinfo  {journal}
  {JHEP}\ }\textbf {\bibinfo {volume} {11}},\ \bibinfo {pages} {018} (\bibinfo
  {year} {2015}{\natexlab{a}})},\ \Eprint {http://arxiv.org/abs/1508.07774}
  {arXiv:1508.07774 [hep-ex]} \BibitemShut {NoStop}%
%%CITATION = ARXIV:1508.07774;%%
\bibitem [{\citenamefont {Aad}\ \emph {et~al.}(2013)\citenamefont {Aad} \emph
  {et~al.}}]{Aad:2013hla}%
  \BibitemOpen
  \bibfield  {author} {\bibinfo {author} {\bibfnamefont {G.}~\bibnamefont
  {Aad}} \emph {et~al.} (\bibinfo {collaboration} {ATLAS}),\ }\href {\doibase
  10.1140/epjc/s10052-013-2465-z} {\bibfield  {journal} {\bibinfo  {journal}
  {Eur. Phys. J.}\ }\textbf {\bibinfo {volume} {C73}},\ \bibinfo {pages} {2465}
  (\bibinfo {year} {2013})},\ \Eprint {http://arxiv.org/abs/1302.3694}
  {arXiv:1302.3694 [hep-ex]} \BibitemShut {NoStop}%
%%CITATION = ARXIV:1302.3694;%%
\bibitem [{\citenamefont {Khachatryan}\ \emph
  {et~al.}(2015{\natexlab{b}})\citenamefont {Khachatryan} \emph
  {et~al.}}]{Khachatryan:2015uua}%
  \BibitemOpen
  \bibfield  {author} {\bibinfo {author} {\bibfnamefont {V.}~\bibnamefont
  {Khachatryan}} \emph {et~al.} (\bibinfo {collaboration} {CMS}),\ }\href
  {\doibase 10.1007/JHEP12(2015)178} {\bibfield  {journal} {\bibinfo  {journal}
  {JHEP}\ }\textbf {\bibinfo {volume} {12}},\ \bibinfo {pages} {178} (\bibinfo
  {year} {2015}{\natexlab{b}})},\ \Eprint {http://arxiv.org/abs/1510.04252}
  {arXiv:1510.04252 [hep-ex]} \BibitemShut {NoStop}%
%%CITATION = ARXIV:1510.04252;%%
\bibitem [{\citenamefont {Aaboud}\ \emph
  {et~al.}(2018{\natexlab{b}})\citenamefont {Aaboud} \emph
  {et~al.}}]{Aaboud:2018cwk}%
  \BibitemOpen
  \bibfield  {author} {\bibinfo {author} {\bibfnamefont {M.}~\bibnamefont
  {Aaboud}} \emph {et~al.},\ }\href {\doibase 10.1007/JHEP11(2018)085}
  {\bibfield  {journal} {\bibinfo  {journal} {Journal of High Energy Physics}\
  }\textbf {\bibinfo {volume} {2018}},\ \bibinfo {pages} {85} (\bibinfo {year}
  {2018}{\natexlab{b}})}\BibitemShut {NoStop}%
\bibitem [{\citenamefont {Enberg}\ \emph {et~al.}(2015)\citenamefont {Enberg},
  \citenamefont {Klemm}, \citenamefont {Moretti}, \citenamefont {Munir},\ and\
  \citenamefont {Wouda}}]{Enberg:2014pua}%
  \BibitemOpen
  \bibfield  {author} {\bibinfo {author} {\bibfnamefont {R.}~\bibnamefont
  {Enberg}}, \bibinfo {author} {\bibfnamefont {W.}~\bibnamefont {Klemm}},
  \bibinfo {author} {\bibfnamefont {S.}~\bibnamefont {Moretti}}, \bibinfo
  {author} {\bibfnamefont {S.}~\bibnamefont {Munir}}, \ and\ \bibinfo {author}
  {\bibfnamefont {G.}~\bibnamefont {Wouda}},\ }\href {\doibase
  10.1016/j.nuclphysb.2015.02.001} {\bibfield  {journal} {\bibinfo  {journal}
  {Nucl. Phys.}\ }\textbf {\bibinfo {volume} {B893}},\ \bibinfo {pages} {420}
  (\bibinfo {year} {2015})},\ \Eprint {http://arxiv.org/abs/1412.5814}
  {arXiv:1412.5814 [hep-ph]} \BibitemShut {NoStop}%
%%CITATION = ARXIV:1412.5814;%%
\bibitem [{\citenamefont {Alwall}\ \emph {et~al.}(2014)\citenamefont {Alwall},
  \citenamefont {Frederix}, \citenamefont {Frixione}, \citenamefont {Hirschi},
  \citenamefont {Maltoni}, \citenamefont {Mattelaer}, \citenamefont {Shao},
  \citenamefont {Stelzer}, \citenamefont {Torrielli},\ and\ \citenamefont
  {Zaro}}]{Alwall:2014hca}%
  \BibitemOpen
  \bibfield  {author} {\bibinfo {author} {\bibfnamefont {J.}~\bibnamefont
  {Alwall}}, \bibinfo {author} {\bibfnamefont {R.}~\bibnamefont {Frederix}},
  \bibinfo {author} {\bibfnamefont {S.}~\bibnamefont {Frixione}}, \bibinfo
  {author} {\bibfnamefont {V.}~\bibnamefont {Hirschi}}, \bibinfo {author}
  {\bibfnamefont {F.}~\bibnamefont {Maltoni}}, \bibinfo {author} {\bibfnamefont
  {O.}~\bibnamefont {Mattelaer}}, \bibinfo {author} {\bibfnamefont {H.~S.}\
  \bibnamefont {Shao}}, \bibinfo {author} {\bibfnamefont {T.}~\bibnamefont
  {Stelzer}}, \bibinfo {author} {\bibfnamefont {P.}~\bibnamefont {Torrielli}},
  \ and\ \bibinfo {author} {\bibfnamefont {M.}~\bibnamefont {Zaro}},\ }\href
  {\doibase 10.1007/JHEP07(2014)079} {\bibfield  {journal} {\bibinfo  {journal}
  {JHEP}\ }\textbf {\bibinfo {volume} {07}},\ \bibinfo {pages} {079} (\bibinfo
  {year} {2014})},\ \Eprint {http://arxiv.org/abs/1405.0301} {arXiv:1405.0301
  [hep-ph]} \BibitemShut {NoStop}%
%%CITATION = ARXIV:1405.0301;%%
\bibitem [{\citenamefont {Misiak}\ and\ \citenamefont
  {Steinhauser}(2017)}]{Misiak:2017bgg}%
  \BibitemOpen
  \bibfield  {author} {\bibinfo {author} {\bibfnamefont {M.}~\bibnamefont
  {Misiak}}\ and\ \bibinfo {author} {\bibfnamefont {M.}~\bibnamefont
  {Steinhauser}},\ }\href {\doibase 10.1140/epjc/s10052-017-4776-y} {\bibfield
  {journal} {\bibinfo  {journal} {Eur. Phys. J.}\ }\textbf {\bibinfo {volume}
  {C77}},\ \bibinfo {pages} {201} (\bibinfo {year} {2017})},\ \Eprint
  {http://arxiv.org/abs/1702.04571} {arXiv:1702.04571 [hep-ph]} \BibitemShut
  {NoStop}%
%%CITATION = ARXIV:1702.04571;%%
\bibitem [{\citenamefont {Guchait}\ and\ \citenamefont
  {Vijay}(2018)}]{Guchait:2018nkp}%
  \BibitemOpen
  \bibfield  {author} {\bibinfo {author} {\bibfnamefont {M.}~\bibnamefont
  {Guchait}}\ and\ \bibinfo {author} {\bibfnamefont {A.~H.}\ \bibnamefont
  {Vijay}},\ }\href@noop {} {\  (\bibinfo {year} {2018})},\ \Eprint
  {http://arxiv.org/abs/1806.01317} {arXiv:1806.01317 [hep-ph]} \BibitemShut
  {NoStop}%
%%CITATION = ARXIV:1806.01317;%%
\bibitem [{\citenamefont {Sj{\"o}strand}\ \emph {et~al.}(2006)\citenamefont
  {Sj{\"o}strand}, \citenamefont {Mrenna},\ and\ \citenamefont
  {Skands}}]{Sjostrand:2006za}%
  \BibitemOpen
  \bibfield  {author} {\bibinfo {author} {\bibfnamefont {T.}~\bibnamefont
  {Sj{\"o}strand}}, \bibinfo {author} {\bibfnamefont {S.}~\bibnamefont
  {Mrenna}}, \ and\ \bibinfo {author} {\bibfnamefont {P.~Z.}\ \bibnamefont
  {Skands}},\ }\href {\doibase 10.1088/1126-6708/2006/05/026} {\bibfield
  {journal} {\bibinfo  {journal} {JHEP}\ }\textbf {\bibinfo {volume} {0605}},\
  \bibinfo {pages} {026} (\bibinfo {year} {2006})},\ \Eprint
  {http://arxiv.org/abs/hep-ph/0603175} {arXiv:hep-ph/0603175 [hep-ph]}
  \BibitemShut {NoStop}%
%%CITATION = HEP-PH/0603175;%%
\bibitem [{\citenamefont {Cacciari}\ \emph {et~al.}(2012)\citenamefont
  {Cacciari}, \citenamefont {Salam},\ and\ \citenamefont
  {Soyez}}]{Cacciari:2011ma}%
  \BibitemOpen
  \bibfield  {author} {\bibinfo {author} {\bibfnamefont {M.}~\bibnamefont
  {Cacciari}}, \bibinfo {author} {\bibfnamefont {G.~P.}\ \bibnamefont {Salam}},
  \ and\ \bibinfo {author} {\bibfnamefont {G.}~\bibnamefont {Soyez}},\ }\href
  {\doibase 10.1140/epjc/s10052-012-1896-2} {\bibfield  {journal} {\bibinfo
  {journal} {Eur. Phys. J.}\ }\textbf {\bibinfo {volume} {C72}},\ \bibinfo
  {pages} {1896} (\bibinfo {year} {2012})},\ \Eprint
  {http://arxiv.org/abs/1111.6097} {arXiv:1111.6097 [hep-ph]} \BibitemShut
  {NoStop}%
%%CITATION = ARXIV:1111.6097;%%
\bibitem [{\citenamefont {de~Favereau}\ \emph {et~al.}(2014)\citenamefont
  {de~Favereau}, \citenamefont {Delaere}, \citenamefont {Demin}, \citenamefont
  {Giammanco}, \citenamefont {Lemaître}, \citenamefont {Mertens},\ and\
  \citenamefont {Selvaggi}}]{deFavereau:2013fsa}%
  \BibitemOpen
  \bibfield  {author} {\bibinfo {author} {\bibfnamefont {J.}~\bibnamefont
  {de~Favereau}}, \bibinfo {author} {\bibfnamefont {C.}~\bibnamefont
  {Delaere}}, \bibinfo {author} {\bibfnamefont {P.}~\bibnamefont {Demin}},
  \bibinfo {author} {\bibfnamefont {A.}~\bibnamefont {Giammanco}}, \bibinfo
  {author} {\bibfnamefont {V.}~\bibnamefont {Lemaître}}, \bibinfo {author}
  {\bibfnamefont {A.}~\bibnamefont {Mertens}}, \ and\ \bibinfo {author}
  {\bibfnamefont {M.}~\bibnamefont {Selvaggi}} (\bibinfo {collaboration}
  {DELPHES 3}),\ }\href {\doibase 10.1007/JHEP02(2014)057} {\bibfield
  {journal} {\bibinfo  {journal} {JHEP}\ }\textbf {\bibinfo {volume} {02}},\
  \bibinfo {pages} {057} (\bibinfo {year} {2014})},\ \Eprint
  {http://arxiv.org/abs/1307.6346} {arXiv:1307.6346 [hep-ex]} \BibitemShut
  {NoStop}%
%%CITATION = ARXIV:1307.6346;%%
\bibitem [{\citenamefont {Eriksson}\ \emph {et~al.}(2010)\citenamefont
  {Eriksson}, \citenamefont {Rathsman},\ and\ \citenamefont
  {St\aa{}l}}]{Eriksson:2009ws}%
  \BibitemOpen
  \bibfield  {author} {\bibinfo {author} {\bibfnamefont {D.}~\bibnamefont
  {Eriksson}}, \bibinfo {author} {\bibfnamefont {J.}~\bibnamefont {Rathsman}},
  \ and\ \bibinfo {author} {\bibfnamefont {O.}~\bibnamefont {St\aa{}l}},\
  }\href {\doibase 10.1016/j.cpc.2009.09.011} {\bibfield  {journal} {\bibinfo
  {journal} {Comput. Phys. Commun.}\ }\textbf {\bibinfo {volume} {181}},\
  \bibinfo {pages} {189} (\bibinfo {year} {2010})},\ \Eprint
  {http://arxiv.org/abs/0902.0851} {arXiv:0902.0851 [hep-ph]} \BibitemShut
  {NoStop}%
%%CITATION = ARXIV:0902.0851;%%
\bibitem [{\citenamefont {Degrande}\ \emph {et~al.}(2015)\citenamefont
  {Degrande}, \citenamefont {Ubiali}, \citenamefont {Wiesemann},\ and\
  \citenamefont {Zaro}}]{Degrande:2015vpa}%
  \BibitemOpen
  \bibfield  {author} {\bibinfo {author} {\bibfnamefont {C.}~\bibnamefont
  {Degrande}}, \bibinfo {author} {\bibfnamefont {M.}~\bibnamefont {Ubiali}},
  \bibinfo {author} {\bibfnamefont {M.}~\bibnamefont {Wiesemann}}, \ and\
  \bibinfo {author} {\bibfnamefont {M.}~\bibnamefont {Zaro}},\ }\href {\doibase
  10.1007/JHEP10(2015)145} {\bibfield  {journal} {\bibinfo  {journal} {JHEP}\
  }\textbf {\bibinfo {volume} {10}},\ \bibinfo {pages} {145} (\bibinfo {year}
  {2015})},\ \Eprint {http://arxiv.org/abs/1507.02549} {arXiv:1507.02549
  [hep-ph]} \BibitemShut {NoStop}%
%%CITATION = ARXIV:1507.02549;%%
\bibitem [{\citenamefont {Flechl}\ \emph {et~al.}(2015)\citenamefont {Flechl},
  \citenamefont {Klees}, \citenamefont {Kramer}, \citenamefont {Spira},\ and\
  \citenamefont {Ubiali}}]{Flechl:2014wfa}%
  \BibitemOpen
  \bibfield  {author} {\bibinfo {author} {\bibfnamefont {M.}~\bibnamefont
  {Flechl}}, \bibinfo {author} {\bibfnamefont {R.}~\bibnamefont {Klees}},
  \bibinfo {author} {\bibfnamefont {M.}~\bibnamefont {Kramer}}, \bibinfo
  {author} {\bibfnamefont {M.}~\bibnamefont {Spira}}, \ and\ \bibinfo {author}
  {\bibfnamefont {M.}~\bibnamefont {Ubiali}},\ }\href {\doibase
  10.1103/PhysRevD.91.075015} {\bibfield  {journal} {\bibinfo  {journal} {Phys.
  Rev.}\ }\textbf {\bibinfo {volume} {D91}},\ \bibinfo {pages} {075015}
  (\bibinfo {year} {2015})},\ \Eprint {http://arxiv.org/abs/1409.5615}
  {arXiv:1409.5615 [hep-ph]} \BibitemShut {NoStop}%
%%CITATION = ARXIV:1409.5615;%%
\bibitem [{\citenamefont {de~Florian}\ \emph {et~al.}(2016)\citenamefont
  {de~Florian} \emph {et~al.}}]{deFlorian:2016spz}%
  \BibitemOpen
  \bibfield  {author} {\bibinfo {author} {\bibfnamefont {D.}~\bibnamefont
  {de~Florian}} \emph {et~al.} (\bibinfo {collaboration} {LHC Higgs Cross
  Section Working Group}),\ }\href {\doibase 10.23731/CYRM-2017-002} {\
  (\bibinfo {year} {2016}),\ 10.23731/CYRM-2017-002},\ \Eprint
  {http://arxiv.org/abs/1610.07922} {arXiv:1610.07922 [hep-ph]} \BibitemShut
  {NoStop}%
%%CITATION = ARXIV:1610.07922;%%
\bibitem [{\citenamefont {Dittmaier}\ \emph {et~al.}(2011)\citenamefont
  {Dittmaier}, \citenamefont {Kr\"amer}, \citenamefont {Spira},\ and\
  \citenamefont {Walser}}]{PhysRevD.83.055005}%
  \BibitemOpen
  \bibfield  {author} {\bibinfo {author} {\bibfnamefont {S.}~\bibnamefont
  {Dittmaier}}, \bibinfo {author} {\bibfnamefont {M.}~\bibnamefont {Kr\"amer}},
  \bibinfo {author} {\bibfnamefont {M.}~\bibnamefont {Spira}}, \ and\ \bibinfo
  {author} {\bibfnamefont {M.}~\bibnamefont {Walser}},\ }\href {\doibase
  10.1103/PhysRevD.83.055005} {\bibfield  {journal} {\bibinfo  {journal} {Phys.
  Rev. D}\ }\textbf {\bibinfo {volume} {83}},\ \bibinfo {pages} {055005}
  (\bibinfo {year} {2011})}\BibitemShut {NoStop}%
\bibitem [{\citenamefont {Berger}\ \emph {et~al.}(2005)\citenamefont {Berger},
  \citenamefont {Han}, \citenamefont {Jiang},\ and\ \citenamefont
  {Plehn}}]{PhysRevD.71.115012}%
  \BibitemOpen
  \bibfield  {author} {\bibinfo {author} {\bibfnamefont {E.~L.}\ \bibnamefont
  {Berger}}, \bibinfo {author} {\bibfnamefont {T.}~\bibnamefont {Han}},
  \bibinfo {author} {\bibfnamefont {J.}~\bibnamefont {Jiang}}, \ and\ \bibinfo
  {author} {\bibfnamefont {T.}~\bibnamefont {Plehn}},\ }\href {\doibase
  10.1103/PhysRevD.71.115012} {\bibfield  {journal} {\bibinfo  {journal} {Phys.
  Rev. D}\ }\textbf {\bibinfo {volume} {71}},\ \bibinfo {pages} {115012}
  (\bibinfo {year} {2005})}\BibitemShut {NoStop}%
\bibitem [{\citenamefont {Hocker}\ \emph {et~al.}(2007)\citenamefont {Hocker}
  \emph {et~al.}}]{Hocker:2007ht}%
  \BibitemOpen
  \bibfield  {author} {\bibinfo {author} {\bibfnamefont {A.}~\bibnamefont
  {Hocker}} \emph {et~al.},\ }\bibfield  {booktitle} {\emph {\bibinfo
  {booktitle} {{Proceedings, 11th International Workshop on Advanced computing
  and analysis techniques in physics research (ACAT 2007): Amsterdam,
  Netherlands, April 23-27, 2007}}},\ }\href@noop {} {\bibfield  {journal}
  {\bibinfo  {journal} {PoS}\ }\textbf {\bibinfo {volume} {ACAT}},\ \bibinfo
  {pages} {040} (\bibinfo {year} {2007})},\ \Eprint
  {http://arxiv.org/abs/physics/0703039} {arXiv:physics/0703039 [PHYSICS]}
  \BibitemShut {NoStop}%
%%CITATION = PHYSICS/0703039;%%
\bibitem [{\citenamefont {Freund}\ and\ \citenamefont
  {Schapire}(1996)}]{Freund96experimentswith}%
  \BibitemOpen
  \bibfield  {author} {\bibinfo {author} {\bibfnamefont {Y.}~\bibnamefont
  {Freund}}\ and\ \bibinfo {author} {\bibfnamefont {R.~E.}\ \bibnamefont
  {Schapire}},\ }\href@noop {} {\enquote {\bibinfo {title} {Experiments with a
  new boosting algorithm},}\ } (\bibinfo {year} {1996})\BibitemShut {NoStop}%
\bibitem [{\citenamefont {Abbott}\ \emph {et~al.}(1998)\citenamefont {Abbott}
  \emph {et~al.}}]{Abbott:1997fv}%
  \BibitemOpen
  \bibfield  {author} {\bibinfo {author} {\bibfnamefont {B.}~\bibnamefont
  {Abbott}} \emph {et~al.} (\bibinfo {collaboration} {D0}),\ }\href {\doibase
  10.1103/PhysRevLett.80.2063} {\bibfield  {journal} {\bibinfo  {journal}
  {Phys. Rev. Lett.}\ }\textbf {\bibinfo {volume} {80}},\ \bibinfo {pages}
  {2063} (\bibinfo {year} {1998})},\ \Eprint
  {http://arxiv.org/abs/hep-ex/9706014} {arXiv:hep-ex/9706014 [hep-ex]}
  \BibitemShut {NoStop}%
%%CITATION = HEP-EX/9706014;%%
\bibitem [{\citenamefont {Abbott}\ \emph {et~al.}(1999)\citenamefont {Abbott}
  \emph {et~al.}}]{Abbott:1998dn}%
  \BibitemOpen
  \bibfield  {author} {\bibinfo {author} {\bibfnamefont {B.}~\bibnamefont
  {Abbott}} \emph {et~al.} (\bibinfo {collaboration} {D0}),\ }\href {\doibase
  10.1103/PhysRevD.60.052001} {\bibfield  {journal} {\bibinfo  {journal} {Phys.
  Rev.}\ }\textbf {\bibinfo {volume} {D60}},\ \bibinfo {pages} {052001}
  (\bibinfo {year} {1999})},\ \Eprint {http://arxiv.org/abs/hep-ex/9808029}
  {arXiv:hep-ex/9808029 [hep-ex]} \BibitemShut {NoStop}%
%%CITATION = HEP-EX/9808029;%%
\bibitem [{\citenamefont {Aaboud}\ \emph {et~al.}(2017)\citenamefont {Aaboud}
  \emph {et~al.}}]{Aaboud:2016syx}%
  \BibitemOpen
  \bibfield  {author} {\bibinfo {author} {\bibfnamefont {M.}~\bibnamefont
  {Aaboud}} \emph {et~al.} (\bibinfo {collaboration} {ATLAS}),\ }\href
  {\doibase 10.1140/epjc/s10052-017-4821-x} {\bibfield  {journal} {\bibinfo
  {journal} {Eur. Phys. J.}\ }\textbf {\bibinfo {volume} {C77}},\ \bibinfo
  {pages} {292} (\bibinfo {year} {2017})},\ \Eprint
  {http://arxiv.org/abs/1612.05220} {arXiv:1612.05220 [hep-ex]} \BibitemShut
  {NoStop}%
%%CITATION = ARXIV:1612.05220;%%
\bibitem [{ATL(2015)}]{ATL-PHYS-PUB-2015-027}%
  \BibitemOpen
  \href {https://cds.cern.ch/record/2037904} {\emph {\bibinfo {title}
  {{Performance of missing transverse momentum reconstruction for the ATLAS
  detector in the first proton-proton collisions at at $\sqrt s$= 13 TeV}}}},\
  \bibinfo {type} {Tech. Rep.}\ \bibinfo {number} {ATL-PHYS-PUB-2015-027}\
  (\bibinfo  {institution} {CERN},\ \bibinfo {address} {Geneva},\ \bibinfo
  {year} {2015})\BibitemShut {NoStop}%
\bibitem [{\citenamefont {Read}(2002)}]{Read:2002hq}%
  \BibitemOpen
  \bibfield  {author} {\bibinfo {author} {\bibfnamefont {A.~L.}\ \bibnamefont
  {Read}},\ }\bibfield  {booktitle} {\emph {\bibinfo {booktitle} {{Advanced
  Statistical Techniques in Particle Physics. Proceedings, Conference, Durham,
  UK, March 18-22, 2002}}},\ }\href {\doibase 10.1088/0954-3899/28/10/313}
  {\bibfield  {journal} {\bibinfo  {journal} {J. Phys.}\ }\textbf {\bibinfo
  {volume} {G28}},\ \bibinfo {pages} {2693} (\bibinfo {year} {2002})},\
  \bibinfo {note} {[,11(2002)]}\BibitemShut {NoStop}%
%%CITATION = JPAGA,G28,2693;%%
\end{thebibliography}%

\end{document}